\documentclass[%
 reprint,
 amsmath,amssymb,
 aps,
]{revtex4-2}

\usepackage{graphicx}
\usepackage{dcolumn}
\usepackage{bm}
\usepackage{siunitx}
\usepackage{float}
\usepackage{color} 
\usepackage{xcolor}
\usepackage[hidelinks]{hyperref}
\hypersetup{
  colorlinks   = true, 
  urlcolor     = blue, 
  linkcolor    = black, 
  citecolor   = blue 
}
\newcommand{\fundmode}{\text{TEM}_{0,0}}

\begin{document}

\preprint{APS/123-QED}

\title{First Results of the Laser-Interferometric Detector for Axions (LIDA)}

\author{Joscha Heinze}
  \email{j.heinze@bham.ac.uk}
  \affiliation{University of Birmingham, School of Physics and Astronomy, Birmingham B15 2TT, United Kingdom.}%

\author{Alex Gill}
\affiliation{University of Birmingham, School of Physics and Astronomy, Birmingham B15 2TT, United Kingdom.}%

\author{Artemiy Dmitriev}
\affiliation{University of Birmingham, School of Physics and Astronomy, Birmingham B15 2TT, United Kingdom.}%

\author{Ji\v{r}\'i Smetana}
\affiliation{University of Birmingham, School of Physics and Astronomy, Birmingham B15 2TT, United Kingdom.}%

\author{Tianliang Yan}
\affiliation{University of Birmingham, School of Physics and Astronomy, Birmingham B15 2TT, United Kingdom.}%

\author{Vincent Boyer}
\affiliation{University of Birmingham, School of Physics and Astronomy, Birmingham B15 2TT, United Kingdom.}%

\author{Matthew Evans}
\affiliation{LIGO, Massachusetts Institute of Technology, Cambridge, MA 02139, USA.}%

\author{Denis Martynov}
\affiliation{University of Birmingham, School of Physics and Astronomy, Birmingham B15 2TT, United Kingdom.}%

\date{\today}

\begin{abstract}
We present the operating principle and the first observing run of a novel kind of direct detector for axions and axion-like particles in the galactic halo. Sensitive to the polarisation rotation of linearly polarised laser light induced by an axion field, our experiment is the first detector of its kind collecting scientific data. We discuss our peak sensitivity of \SI{1.51e-10}{GeV^{-1}} (\SI{95}{\%} confidence level) to the axion-photon coupling strength in the axion mass range of 1.97-2.01\,neV which is, for instance, motivated by supersymmetric grand-unified theories. We also report on effects that arise in our high-finesse in-vacuum cavity at an unprecedented optical continuous-wave intensity of \SI{4.7}{MW\per\centi\meter\squared}. Our detector already belongs to the most sensitive direct searches within its measurement band, and our results pave the way towards surpassing the current sensitivity limits even of astrophysical observations in the mass range from \SI{e-8}{eV} down to \SI{e-16}{eV} via quantum-enhanced laser interferometry, especially with the potential of scaling our detector up to kilometre-length.

\end{abstract}

\maketitle

\textit{Introduction}.---The existence of axions and axion-like particles (ALPs) is well-motivated in a variety of theoretical models. The axion was first introduced in 1977 as a promising candidate to resolve the strong charge-parity problem in quantum chromodynamics \cite{Peccei_1977,Weinberg_1978,Wilczek_1978,ChadhaDay_axionDarkMatterWhatIsIt_2022}. Here, it appears as a field-like Nambu-Goldstone boson in a spontaneously broken Peccei-Quinn symmetry and relaxes to a value which allows the electric dipole moment of the neutron to vanish. After this first proposal, axions as well as ALPs proved to arise generically from many extensions of the Standard Model, e.g.~from string theory and supergravity \cite{Svrcek_2006,Graham_2013, Ringwald_2012,Ringwald_2014,Farina_2017}. Finally, they have also become a leading candidate for dark matter \cite{ABBOTT1983133,PRESKILL1983127,DINE1983137}. This is due to the aforementioned theoretical support, evidence from astronomical observations like gravitational lensing \cite{Amruth_2023}, and since other dark matter candidates like weakly interacting massive particles have not been detected in a variety of attempts \cite{Akerib_Lux_2013,Aprile_2018,Zhang_PandaX_2018}.

In light of the growing significance, various experimental approaches have been proposed, or already employed, to directly measure a signature of axions and ALPs, e.g.~axion haloscopes (MADMAX \cite{Caldwell_2017} and DMRadio \cite{Brouwer_2022}), axion helioscopes (CAST \cite{Anastassopoulos_2017} and IAXO \cite{Armengaud_2014}), ``light shining though a wall'' experiments (ALPS \cite{Baehre_2013} and CROWS \cite{Betz_2013}) and magnetometers (ABRACADABRA \cite{PhysRevLett.127.081801}). However, no signature has been found yet which makes it essential to further diversify the search. 

In this Letter, we present LIDA, a laser-interferometric detector for axions based on Ref.~\cite{Martynov_2020} and related to the studies in Refs.~\cite{DeRocco_2018,Obata_2018,Liu_2019,FirstDanceResults_2023,Nagano_2019}. LIDA uses the coupling of axions to photons, though not their conversion as in several other experiments, and represents a fairly new kind of detector which has not yet contributed to the axion science data. Its general detection approach is also closely related to Ref.\ \cite{Liu_linearlyPolarisedPulsarLight_2020} which, in contrast, utilises pulsar light.

The utilisation of laser-interferometric axion detectors is particularly well motivated in Refs.\ \cite{Obata_2018,Martynov_2020,Michimura_2020} which show the potential of these detectors to even surpass the most stringent constraints from astrophysical observations in almost their entire measurement band. We will first reiterate the operating principle and design, and then discuss LIDA's performance in the first observing run. At the same time, this is the final result in the neV mass range as we will measure at lower axion masses in the future.

\textit{Operating principle}.---If dark matter is made of axions with mass $m_a$, it behaves like a coherent, classical field \cite{Budker_Casper_2014}
\begin{equation}\label{eq:alps_field}
    a(t) = a_0 \sin\left[\Omega_a t + \delta(t)\right]
\end{equation}
with angular frequency $\Omega_a = 2\pi f_a = m_ac^2/\hbar$, field amplitude $a_0^2 = 2\rho_\text{DM}\hbar^2/m_a^2$, the local density of dark matter $\rho_{\rm DM}$, and the phase of the field $\delta (t)$. The interaction Lagrangian for the axion-photon coupling reads \cite{Anastassopoulos_2017}
\begin{equation}\label{eqn:interactionLagrangian}
\mathcal{L}_{a\gamma}=-\frac{g_{a\gamma}}{4}aF^{\mu\nu}\tilde{F}_{\mu\nu}\ \text{,}
\end{equation}
where $a$ is the axion field, $F$ is the electro-magnetic field-strength tensor and $g_{a\gamma}$ is the coupling coefficient. This coupling leads to a phase difference \cite{DeRocco_2018}
\begin{equation}\label{eqn:phaseDifference}
    \Delta\phi(t,\tau)=g_{a\gamma}\left[a(t)-a(t-\tau)\right]
\end{equation}
which accumulates between left- and right-handed circularly polarised light over a time period of $\tau$. Equivalently, the polarisation axis of linearly polarised light is rotated with a period that corresponds to the axion frequency; this rotation is measurable with our detector.

\begin{figure*}
    \centering
    \includegraphics[trim=0mm 54mm 0mm 50mm,clip,width=0.9\linewidth]{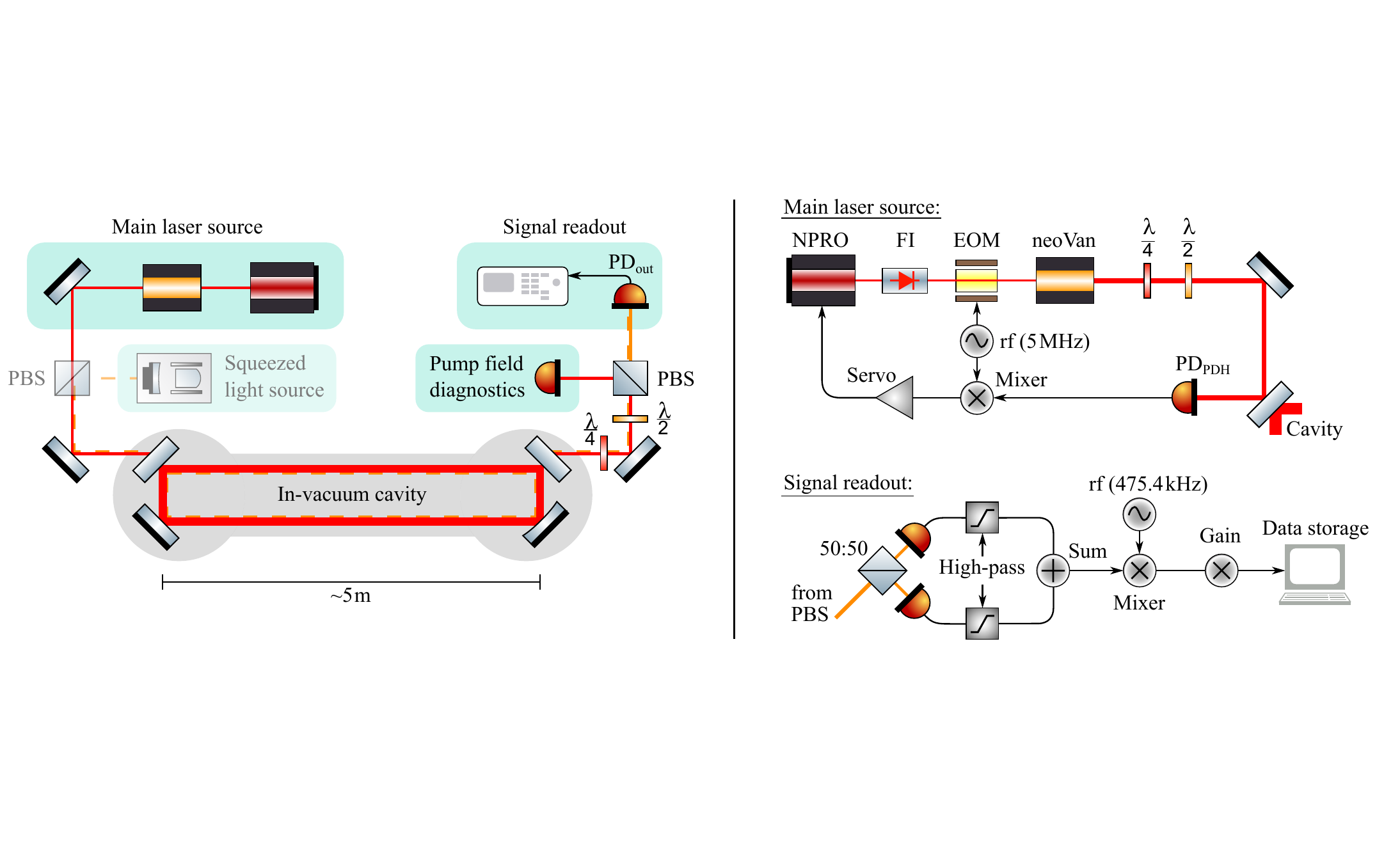}
    \caption{Simplified schematic of the experimental setup on the left, details of the main laser source and of the signal readout on the right. Red beam: pump field, orange beam: signal field, orange-dashed beam: planned squeezed field, EOM: electro-optic modulator, FI: Faraday isolator, NPRO: non-planar ring laser, PBS: polarising beamsplitter, PD: photodetector, RF: radio frequency generator. The installation of a squeezed light source as indicated is planned.}
    \label{fig:setup}
\end{figure*}
Hence, LIDA utilises a laser beam at optical angular frequency $\omega_\text{pmp}$ which is linearly polarised along the vertical axis (S-polarisation) as a \textit{pump field}. As shown in Figure~\ref{fig:setup}, this pump field is kept on resonance with a high-finesse cavity to amplify its optical power. If an axion field periodically rotates the polarisation axis of the circulating intra-cavity pump field, it excites two coherent light fields (\textit{sidebands}) in the orthogonal P-polarisation at frequencies $\omega_\text{pmp}\pm\Omega_a$ (\textit{signal field}). These sidebands build up inside the cavity according to \cite{Martynov_2020}
\begin{align}\label{eqn:signalBuildupInCavity}
\begin{split}
    &E_\text{sig,cav}(\pm\Omega_a)=-\frac{E_\text{pmp,cav}\exp\left(i\frac{\beta\mp\Omega_a\tau}{2}+\delta\right)}{1-\sqrt{1-2T_\text{sig}-l_\text{rt}}\exp\left[i\left(\beta\mp\Omega_a\tau\right)\right]}\\
    & \qquad\ \ \times g_{a\gamma}\frac{\tau}{4}\text{sinc}\left(\frac{\Omega_a\tau}{4}\right)\cos\left(\frac{2\beta\mp\Omega_a\tau}{4}\right)\sqrt{2\tau_a\rho_\text{DM}}
\end{split}
\end{align}

\noindent
where we assume the ``rotating frame'' by setting ${\omega_\text{pmp}=0}$. $E_\text{pmp,cav}$ is the circulating pump field, $\beta$ is an extra cavity roundtrip phase which the signal field accumulates relative to the pump field, $\tau$ is the cavity roundtrip time, $T_\text{sig}$ is the power transmissivity of the cavity input and output couplers for the signal field polarisation, $l_\text{rt}$ is the cavity roundtrip power loss and $\tau_a$ is the coherence time of the axion field. $\beta$ results from the cumulative effect of the four cavity mirrors and their coatings and leads to a non-degeneracy of the cavity's S- and P-eigenmodes (detuning). Hence, each sideband in the signal field is only resonantly enhanced if $\pm\Omega_a$ is sufficiently close to the detuning frequency.

In transmission of the cavity, we separate the signal field from the pump field via a polarising beamsplitter. In addition, a half-wave plate shifts a small constant fraction of the pump field into the signal polarisation to serve as a local oscillator $E_\text{LO}=i\xi\sqrt{T_\text{pmp}}E_\text{pmp,cav}$, where $\xi$ is twice the rotation angle of the half-wave plate and $T_\text{pmp}$ is the power transmissivity of the cavity output coupler for the pump field polarisation. Finally, a photodetector measures the signal as the beat note between the local oscillator and the sidebands, yielding the following amplitude spectral density \cite{Martynov_2020}:
\begin{align}\label{eqn:readoutPower}
\begin{split}
    P_\text{out}(\Omega_a)=&\ (1-l_\text{out})E_\text{LO}\sqrt{T_\text{sig}}\\
    &\times\left[E^*_\text{sig,cav}(-\Omega_a)-E_\text{sig,cav}(\Omega_a)\right]
\end{split}
\end{align}
with the optical loss in the readout beam path $l_\text{out}$. This signal yields a signal-to-noise ratio SNR of
\begin{equation}\label{eqn:SNR}
    \text{SNR}^2=\left|\frac{P_\text{out}(\Omega_a)}{P_\text{N}(\Omega_a)}\right|^2\sqrt{\frac{T_\text{meas}}{\tau_a}}\ \text{,}
\end{equation}
where $P_\text{N}$ is the amplitude spectral density of the total noise and $T_\text{meas}$ is the total measurement time.  

\textit{Experimental setup.}---We now discuss the specifics of our setup as shown in Figure~\ref{fig:setup} and the parameters achieved for the first observing run. The main laser source operates with a \SI{300}{mW} non-planar ring laser (NPRO) which continuously emits linearly polarised light in the $\fundmode$ mode at a wavelength of \SI{1064}{nm}. An electro-optic modulator (EOM) modulates the phase of the light field at a frequency of \SI{5}{MHz}. This enables the stabilisation of the laser frequency to the resonances of the in-vacuum cavity via the Pound-Drever-Hall scheme \cite{Black_PoundDreverHall_2001} using the signal from the photodetector $\text{PD}_\text{PDH}$ in reflection of the cavity. The optical power that is injected into the cavity can be enhanced to about \SI{18}{W} by a neoLASE solid-state laser amplifier. A quarter- and half-wave plate finally tune the pump polarisation. For the first observing run, we injected \SI{12}{W} into the cavity in the S-polarisation.

The rectangular in-vacuum cavity measures about $\SI{4.9}{m}\times\SI{10}{cm}$ in size. The input and output couplers are nominally identical with measured power transmissivities at an angle of incidence of \SI{45}{\degree} of $T_\text{sig}=\SI{0.13}{\%}$ and $T_\text{pmp}=\SI{17}{ppm}$ in the P- and S-polarisation, respectively. We inferred the respective pole frequencies from the cavity's transfer function for power modulations between the input and transmission to be $f_\text{p,P}=\SI{6.76}{kHz}$ and $f_\text{p,S}=\SI{202}{Hz}$. This yields a finesse of $\mathcal{F}_\text{P}=2220$ and $\mathcal{F}_\text{P}=74220$ as well as an intra-cavity roundtrip loss of $l_\text{rt}=\SI{51}{ppm}$. The other two cavity mirrors are highly reflective, the one on the readout side has a radius of curvature of \SI{10.2}{m} setting the beam waist of the cavity eigenmodes to about \SI{1.1}{mm} and \SI{1.5}{mm} on the horizontal and vertical axis, respectively. We measured small phase shifts between the P- and S-polarisation upon reflection off each of the cavity mirrors of \SI{20}{mrad} around an angle of incidence of \SI{45}{\degree} via an ellipsometer. The current detuning between the cavity's P- and S-eigenmodes is \SI{478}{kHz}. This detuning corresponds to a sensitivity peak at an axion mass of about \SI{2}{neV} which is within the range motivated, e.g., by grand unified theories \cite{Co_2016,PhysRevD.98.095011} and observations of the cosmic infrared background \cite{Kohri_2017}. The detuning may be controlled and scanned by an auxiliary cavity in the future \cite{Martynov_2020}.

The signal field was split up by a 50:50 beamsplitter and measured by two photodetectors $\text{PD}_\text{out}$. The two PD signals were high-passed and summed up. The sum was demodulated at \SI{475.4}{kHz}, and, after an amplification by a factor of 50, the demodulated and amplified output signal was logged with a sampling rate of \SI{65.5}{kHz}. The current optical loss in the readout path amounts to $l_\text{out}=\SI{5}{\%}$, mainly due to the two beamsplitters. 

From the optical power in transmission of the cavity, we inferred an average and maximum circulating intra-cavity pump power of \SI{118}{kW} and \SI{124}{kW}, respectively. The latter corresponds to an optical intensity of \SI{4.7}{MW\per\centi\meter\squared} at the waist position. To our knowledge, this level of intensity has not been reached before in any optical continuous-wave experiment \cite{DellaValle_longDecayCavity_2014}. 

\begin{figure}
    \centering
    \includegraphics[trim=9mm 69mm 13mm 62mm,clip,width=0.95\linewidth]{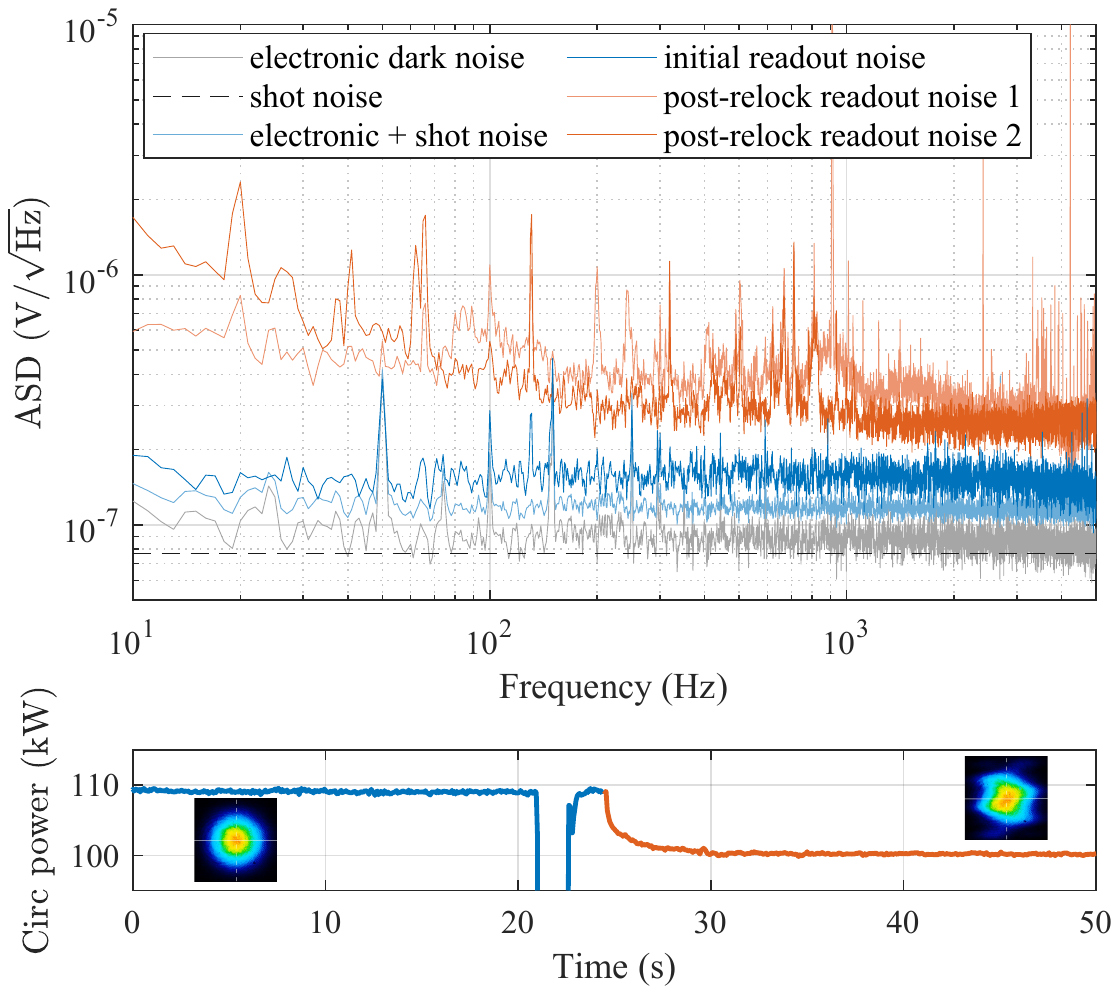}
    \caption{Bottom: time series of the circulating intra-cavity power when the detector is disturbed. It typically relocks automatically, however, often reaching a final state with less circulating power than initially. The CCD camera pictures show the corresponding fields transmitted through the cavity. Top: amplitude spectral densities (ASD) of the electronic dark noise and shot noise in the demodulated output signal. The difference between their incoherent sum and the \textit{initial} readout noise (corresponding to the cavity state of the first \SI{20}{s} of the time series) is caused by technical laser noise. The \textit{post-relock} readout noise (corresponding to the cavity state of the last \SI{20}{s}) is significantly increased over the full measurement band, especially at lower frequencies. 1 and 2 are two examples as the post-relock noise varies.}
    \label{fig:noise}
\end{figure}

\textit{Limiting noise sources}.---Our current signal readout path allows for a measurement frequency band of \SIrange{475.4}{505.1}{kHz}. Within this band, we were limited by electronic dark noise, quantum shot noise and technical laser noise (see Fig.~\ref{fig:noise}). Shot noise is caused by vacuum fluctuations in the signal polarisation that co-propagate with the input pump field, are transmitted through the cavity and reach the readout. The technical laser noise can couple to the signal readout if the input polarisation is not perfectly tuned. In this case, a small fraction of the field that is injected into the cavity is in the signal polarisation and its technical noise is transmitted through the cavity at the detuning frequency. Hence, we had to carefully adjust the tuning of the input waveplates. Coherence measurements with the input intensity noise suggest that laser frequency noise dominates this technical noise coupling channel. 

\begin{figure}
    \centering
    \includegraphics[trim=37mm 86mm 47mm 86mm,clip,width=\linewidth]{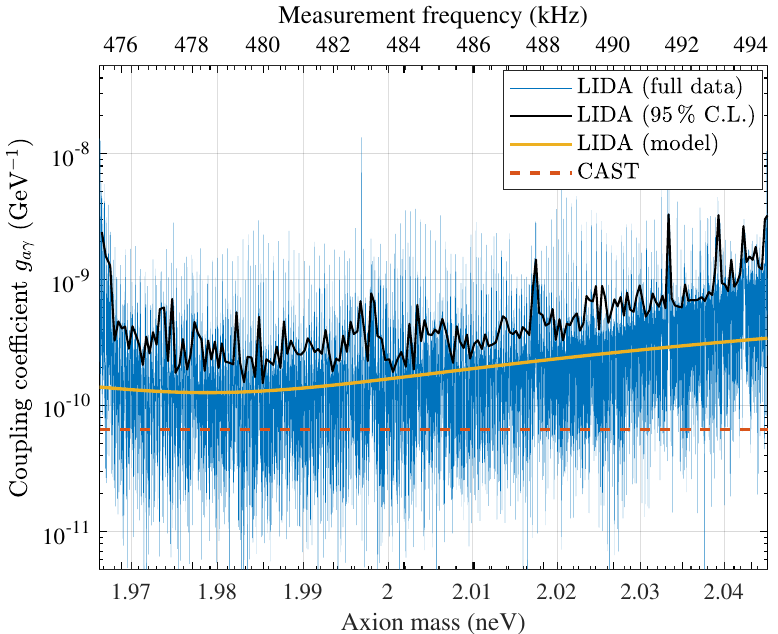}
    \caption{Sensitivity to the axion-photon coupling coefficient $g_{a\gamma}$ that LIDA reached during the first observing run, dependent on the axion mass and measurement frequency, at the \SI{95}{\%} confidence level of the full data. Both curves are compared to the predictions of the shot-noise limited model from Eq.~\ref{eqn:readoutPower}, and to the constraint set by the CAST detector \cite{Anastassopoulos_2017}.}
    \label{fig:results}
\end{figure}

\textit{Results and future prospects}.---Figure~\ref{fig:results} shows our sensitivity at the \SI{95}{\%} confidence level of the full data. The full data is derived by averaging the amplitude spectral density of the readout signal over the total measurement time, subtracting the noise floor and calibrating the result with our theoretical model from Eq.~\ref{eqn:readoutPower} using experimentally determined parameters. The numerous narrow lines originate from the electronic dark noise. For the \SI{95}{\%} confidence level, we first identified the frequencies of the lines in the electronic dark noise and then removed the corresponding lines in the full data. After the removal of 2,106 lines, we could still probe about 19,000 independent axions masses between 1.97-2.01\,neV given their linewidth of $\Delta f/f\sim\num{e-6}$. Within this mass range, we could identify 343 candidates which we excluded from axion or ALP signatures via an analysis of their frequency (and existence) over time and of their linewidth as outlined in Ref.\ \cite{FirstDanceResults_2023}.

LIDA reached a maximum sensitivity of $g_{a\gamma}=\SI{1.51e-10}{GeV^{-1}}$ at \SI{1.985}{neV}, or \SI{480.0}{kHz}, in a measurement time of $T_\text{meas}=\SI{85}{h}$. This is only a factor of \num{2.3} above the constraints set by the most sensitive direct searches at this axion mass, the CAST helioscope \cite{Anastassopoulos_2017} and ABRACADABRA \cite{PhysRevLett.127.081801}, and about a factor of 30 above the most stringent astrophysical constraints from Fermi-LAT (NGC1275) \cite{Fermi-LAT_NGC1275_2016} and magnetic white dwarf polarisations \cite{MWDpolarisation_2022}. The average sensitivity in the range of 1.97-2.01\,neV, which is relatively narrow at these axion masses, was \SI{3.2e-10}{GeV^{-1}}. We have not measured a significant evidence for axions or ALPs.

We predict a significant increase in LIDA's sensitivity from three main improvements in future observing runs. First, an additional cavity in the input beam path will suppress both technical laser intensity and frequency noise above the cavity's pole frequency and reduce its coupling to the readout. Second, we will add a squeezed light source to the input optics to mitigate the readout shot noise similarly to the gravitational-wave detectors Advanced LIGO \cite{LSC_SQUEEZING_2013,Buikema_performanceOfLIGO_2020}, Advanced Virgo \cite{Acernese_2019,Bersanetti_VirgoStatus_2021} and GEO600 \cite{Lough_6dBGEO600_2021}. Thirdly, we will operate LIDA at a detuning of about zero for a year. In combination, these changes will result in a sensitivity of about \SI{e-13}{GeV^{-1}} at \SI{e-14}{eV} as shown in Ref.\ \cite{Martynov_2020}. LIDA will then be able to probe a region of the mass-coupling parameter space which has not been explored yet, directly or via astrophysical observations. At a detuning close to zero, LIDA's spectral sensitivity will moreover be significantly broader than presented here. Finally, a major upscaling of LIDA to kilometre-length would even allow to reach below \SI{e-16}{GeV^{-1}} at \SI{e-14}{eV} \cite{Martynov_2020}.

\textit{Challenges}.---We will now discuss two challenging and not yet completely explained aspects of LIDA which may also become relevant to similar detectors in Tokyo \cite{FirstDanceResults_2023} and at the MIT \cite{Liu_2019}, and to high-intensity and high-finesse experiments, in general. First, if the intra-cavity pump power is sufficiently high, our cavity can assume at least two stable states when the laser frequency is stabilised to the cavity's $\text{TEM}_{0,0}$ eigenmode (\textit{locked}) as shown in Figure~\ref{fig:noise}. Each state is characterised by its circulating power, readout noise pattern and transmitted light field. We obtain the state with the highest circulating power when we lock the detector manually to start an observing run. This state corresponds to the lowest (``initial'') readout noise as well as to the purest transmitted field. However, when the detector relocks automatically after an external disturbance, it typically decays into a state with less circulating power, higher (``post-relock'') readout noise with additional noise peaks and a transmitted field in which the $\text{TEM}_{0,0}$ mode is superimposed with varying higher-order Hermite-Gaussian modes. We have not yet identified the exact mechanism behind this effect, but it is likely to have a thermal origin and limits the effective measurement time. If the mechanism turns out to be related to parametric instabilities \cite{EVANS2010_parametricInstabilitiesInGeneral}, this issue could be solved via acoustic mode dampers \cite{Biscans2019_acousticModeDampers}.


Second, the pump field that is transmitted through the cavity shows a significant amount of light in the signal polarisation, i.e.\ it is elliptically polarised, if linearly polarised light in the S-polarisation is injected. In transmission of the polarising beamsplitter in the readout, we consistently measure contrasts of only \SIrange{65}{70}{\%}. A theoretical model of the cavity shows that this observation can be explained by a slight non-planarity of the cavity geometry which would cause a coupling of the external S- and P-polarisation. The measured contrast only requires a misalignment at the cavity mirrors of about \SI{1}{mrad} which is within a reasonable range. Moreover, we measured that the viewports of our vacuum system convert linearly into elliptically polarised light dependent on the point of tranmission; in general, this effect seems to grow with increasing distance from the viewport centre. Most likely, the reduced contrast in transmission of the PBS arises due to a combination of both effects, and we compensate for it with an additional quarter-wave plate in transmission of the cavity. This waveplate changes the phase relation between the signal and pump field but, since the current cavity detuning of \SI{480}{kHz} is relatively large, only one of the signal sidebands is effectively enhanced and measured. Hence, we can measure the signal in an arbitrary quadrature. In the future, we will try to reduce the cavity non-planarity and might switch to an in-vacuum readout.

\textit{Conclusion}.---We presented the results of the first \SI{85}{h}-long observing run of a laser-interferometric detector for axions and axion-like particles (ALPs) called LIDA. Our current peak sensitivity to the axion-photon coupling coefficient $g_{a\gamma}$ is inside an axion mass range of 1.97-2.01\,neV where we reached up to \SI{1.51e-10}{GeV^{-1}} at a \SI{95}{\%} confidence level. This is only a factor of 2.3 higher than the CAST limit and among the most sensitive direct axion searches. Besides the electronic dark noise, we were limited by quantum shot noise and technical laser noise which will be further reduced by the implementation of a squeezed light source and an input mode cleaner cavity, respectively. From these techniques and an increase in the measurement time to a year, we expect to reach a sensitivity about three orders of magnitude higher if we reduce the frequency separation of the cavity's S- and P-eigenmodes and measure axion masses down to \SI{e-14}{eV} and lower, where the axion field exhibits a larger coherence time. This would allow LIDA to probe a yet unexplored region of the mass-coupling parameter space. These results are a highly promising milestone for advancing direct axion and ALP searches by expanding them to the field of quantum-enhanced laser interferometry. They are furthermore a strong argument to ultimately set LIDA up as a kilometre-scale detector, as done in the gravitational-wave research, which would further boost the sensitivity by several orders of magnitude \cite{Martynov_2020}.

We acknowledge members of the UK Quantum Interferometry collaboration for useful discussions, the support of the Institute for Gravitational Wave Astronomy at the University of Birmingham and STFC Quantum Technology for Fundamental Physics scheme (Grant No. ST/T006609/1 and ST/W006375/1). D.M. is supported by the 2021 Philip Leverhulme Prize.

\bibliography{LIDA1}

\begin{thebibliography}{46}%
\makeatletter
\providecommand \@ifxundefined [1]{%
 \@ifx{#1\undefined}
}%
\providecommand \@ifnum [1]{%
 \ifnum #1\expandafter \@firstoftwo
 \else \expandafter \@secondoftwo
 \fi
}%
\providecommand \@ifx [1]{%
 \ifx #1\expandafter \@firstoftwo
 \else \expandafter \@secondoftwo
 \fi
}%
\providecommand \natexlab [1]{#1}%
\providecommand \enquote  [1]{``#1''}%
\providecommand \bibnamefont  [1]{#1}%
\providecommand \bibfnamefont [1]{#1}%
\providecommand \citenamefont [1]{#1}%
\providecommand \href@noop [0]{\@secondoftwo}%
\providecommand \href [0]{\begingroup \@sanitize@url \@href}%
\providecommand \@href[1]{\@@startlink{#1}\@@href}%
\providecommand \@@href[1]{\endgroup#1\@@endlink}%
\providecommand \@sanitize@url [0]{\catcode `\\12\catcode `\$12\catcode
  `\&12\catcode `\#12\catcode `\^12\catcode `\_12\catcode `\%12\relax}%
\providecommand \@@startlink[1]{}%
\providecommand \@@endlink[0]{}%
\providecommand \url  [0]{\begingroup\@sanitize@url \@url }%
\providecommand \@url [1]{\endgroup\@href {#1}{\urlprefix }}%
\providecommand \urlprefix  [0]{URL }%
\providecommand \Eprint [0]{\href }%
\providecommand \doibase [0]{https://doi.org/}%
\providecommand \selectlanguage [0]{\@gobble}%
\providecommand \bibinfo  [0]{\@secondoftwo}%
\providecommand \bibfield  [0]{\@secondoftwo}%
\providecommand \translation [1]{[#1]}%
\providecommand \BibitemOpen [0]{}%
\providecommand \bibitemStop [0]{}%
\providecommand \bibitemNoStop [0]{.\EOS\space}%
\providecommand \EOS [0]{\spacefactor3000\relax}%
\providecommand \BibitemShut  [1]{\csname bibitem#1\endcsname}%
\let\auto@bib@innerbib\@empty
\bibitem [{\citenamefont {Peccei}\ and\ \citenamefont
  {Quinn}(1977)}]{Peccei_1977}%
  \BibitemOpen
  \bibfield  {author} {\bibinfo {author} {\bibfnamefont {R.~D.}\ \bibnamefont
  {Peccei}}\ and\ \bibinfo {author} {\bibfnamefont {H.~R.}\ \bibnamefont
  {Quinn}},\ }\bibfield  {title} {\bibinfo {title} {$\mathrm{CP}$ conservation
  in the presence of pseudoparticles},\ }\href
  {https://doi.org/10.1103/PhysRevLett.38.1440} {\bibfield  {journal} {\bibinfo
   {journal} {Phys. Rev. Lett.}\ }\textbf {\bibinfo {volume} {38}},\ \bibinfo
  {pages} {1440} (\bibinfo {year} {1977})}\BibitemShut {NoStop}%
\bibitem [{\citenamefont {Weinberg}(1978)}]{Weinberg_1978}%
  \BibitemOpen
  \bibfield  {author} {\bibinfo {author} {\bibfnamefont {S.}~\bibnamefont
  {Weinberg}},\ }\bibfield  {title} {\bibinfo {title} {A new light boson?},\
  }\href {https://doi.org/10.1103/PhysRevLett.40.223} {\bibfield  {journal}
  {\bibinfo  {journal} {Phys. Rev. Lett.}\ }\textbf {\bibinfo {volume} {40}},\
  \bibinfo {pages} {223} (\bibinfo {year} {1978})}\BibitemShut {NoStop}%
\bibitem [{\citenamefont {Wilczek}(1978)}]{Wilczek_1978}%
  \BibitemOpen
  \bibfield  {author} {\bibinfo {author} {\bibfnamefont {F.}~\bibnamefont
  {Wilczek}},\ }\bibfield  {title} {\bibinfo {title} {Problem of strong $p$ and
  $t$ invariance in the presence of instantons},\ }\href
  {https://doi.org/10.1103/PhysRevLett.40.279} {\bibfield  {journal} {\bibinfo
  {journal} {Phys. Rev. Lett.}\ }\textbf {\bibinfo {volume} {40}},\ \bibinfo
  {pages} {279} (\bibinfo {year} {1978})}\BibitemShut {NoStop}%
\bibitem [{\citenamefont {Chadha-Day}\ \emph {et~al.}(2022)\citenamefont
  {Chadha-Day}, \citenamefont {Ellis},\ and\ \citenamefont
  {Marsh}}]{ChadhaDay_axionDarkMatterWhatIsIt_2022}%
  \BibitemOpen
  \bibfield  {author} {\bibinfo {author} {\bibfnamefont {F.}~\bibnamefont
  {Chadha-Day}}, \bibinfo {author} {\bibfnamefont {J.}~\bibnamefont {Ellis}},\
  and\ \bibinfo {author} {\bibfnamefont {D.~J.~E.}\ \bibnamefont {Marsh}},\
  }\bibfield  {title} {\bibinfo {title} {Axion dark matter: What is it and why
  now?},\ }\href {https://doi.org/10.1126/sciadv.abj3618} {\bibfield  {journal}
  {\bibinfo  {journal} {Science Advances}\ }\textbf {\bibinfo {volume} {8}},\
  \bibinfo {pages} {eabj3618} (\bibinfo {year} {2022})}\BibitemShut {NoStop}%
\bibitem [{\citenamefont {{Svrcek}}\ and\ \citenamefont
  {{Witten}}(2006)}]{Svrcek_2006}%
  \BibitemOpen
  \bibfield  {author} {\bibinfo {author} {\bibfnamefont {P.}~\bibnamefont
  {{Svrcek}}}\ and\ \bibinfo {author} {\bibfnamefont {E.}~\bibnamefont
  {{Witten}}},\ }\bibfield  {title} {\bibinfo {title} {{Axions in string
  theory}},\ }\href {https://doi.org/10.1088/1126-6708/2006/06/051} {\bibfield
  {journal} {\bibinfo  {journal} {Journal of High Energy Physics}\ }\textbf
  {\bibinfo {volume} {2006}},\ \bibinfo {eid} {051} (\bibinfo {year}
  {2006})}\BibitemShut {NoStop}%
\bibitem [{\citenamefont {Graham}\ and\ \citenamefont
  {Rajendran}(2013)}]{Graham_2013}%
  \BibitemOpen
  \bibfield  {author} {\bibinfo {author} {\bibfnamefont {P.~W.}\ \bibnamefont
  {Graham}}\ and\ \bibinfo {author} {\bibfnamefont {S.}~\bibnamefont
  {Rajendran}},\ }\bibfield  {title} {\bibinfo {title} {New observables for
  direct detection of axion dark matter},\ }\href
  {https://doi.org/10.1103/PhysRevD.88.035023} {\bibfield  {journal} {\bibinfo
  {journal} {Phys. Rev. D}\ }\textbf {\bibinfo {volume} {88}},\ \bibinfo
  {pages} {035023} (\bibinfo {year} {2013})}\BibitemShut {NoStop}%
\bibitem [{\citenamefont {{Ringwald}}(2012)}]{Ringwald_2012}%
  \BibitemOpen
  \bibfield  {author} {\bibinfo {author} {\bibfnamefont {A.}~\bibnamefont
  {{Ringwald}}},\ }\bibfield  {title} {\bibinfo {title} {{Exploring the role of
  axions and other WISPs in the dark universe}},\ }\href
  {https://doi.org/10.1016/j.dark.2012.10.008} {\bibfield  {journal} {\bibinfo
  {journal} {Physics of the Dark Universe}\ }\textbf {\bibinfo {volume} {1}},\
  \bibinfo {pages} {116} (\bibinfo {year} {2012})},\ \Eprint
  {https://arxiv.org/abs/1210.5081} {arXiv:1210.5081 [hep-ph]} \BibitemShut
  {NoStop}%
\bibitem [{\citenamefont {Ringwald}(2014)}]{Ringwald_2014}%
  \BibitemOpen
  \bibfield  {author} {\bibinfo {author} {\bibfnamefont {A.}~\bibnamefont
  {Ringwald}},\ }\bibfield  {title} {\bibinfo {title} {Searching for axions and
  {ALPs} from string theory},\ }\href
  {https://doi.org/10.1088/1742-6596/485/1/012013} {\bibfield  {journal}
  {\bibinfo  {journal} {Journal of Physics: Conference Series}\ }\textbf
  {\bibinfo {volume} {485}},\ \bibinfo {pages} {012013} (\bibinfo {year}
  {2014})}\BibitemShut {NoStop}%
\bibitem [{\citenamefont {Farina}\ \emph {et~al.}(2017)\citenamefont {Farina},
  \citenamefont {Pappadopulo}, \citenamefont {Rompineve},\ and\ \citenamefont
  {Tesi}}]{Farina_2017}%
  \BibitemOpen
  \bibfield  {author} {\bibinfo {author} {\bibfnamefont {M.}~\bibnamefont
  {Farina}}, \bibinfo {author} {\bibfnamefont {D.}~\bibnamefont {Pappadopulo}},
  \bibinfo {author} {\bibfnamefont {F.}~\bibnamefont {Rompineve}},\ and\
  \bibinfo {author} {\bibfnamefont {A.}~\bibnamefont {Tesi}},\ }\bibfield
  {title} {\bibinfo {title} {The photo-philic qcd axion},\ }\href
  {https://doi.org/10.1007/JHEP01(2017)095} {\bibfield  {journal} {\bibinfo
  {journal} {Journal of High Energy Physics}\ }\textbf {\bibinfo {volume}
  {2017}},\ \bibinfo {pages} {95} (\bibinfo {year} {2017})}\BibitemShut
  {NoStop}%
\bibitem [{\citenamefont {Abbott}\ and\ \citenamefont
  {Sikivie}(1983)}]{ABBOTT1983133}%
  \BibitemOpen
  \bibfield  {author} {\bibinfo {author} {\bibfnamefont {L.}~\bibnamefont
  {Abbott}}\ and\ \bibinfo {author} {\bibfnamefont {P.}~\bibnamefont
  {Sikivie}},\ }\bibfield  {title} {\bibinfo {title} {A cosmological bound on
  the invisible axion},\ }\href
  {https://doi.org/https://doi.org/10.1016/0370-2693(83)90638-X} {\bibfield
  {journal} {\bibinfo  {journal} {Physics Letters B}\ }\textbf {\bibinfo
  {volume} {120}},\ \bibinfo {pages} {133} (\bibinfo {year}
  {1983})}\BibitemShut {NoStop}%
\bibitem [{\citenamefont {Preskill}\ \emph {et~al.}(1983)\citenamefont
  {Preskill}, \citenamefont {Wise},\ and\ \citenamefont
  {Wilczek}}]{PRESKILL1983127}%
  \BibitemOpen
  \bibfield  {author} {\bibinfo {author} {\bibfnamefont {J.}~\bibnamefont
  {Preskill}}, \bibinfo {author} {\bibfnamefont {M.~B.}\ \bibnamefont {Wise}},\
  and\ \bibinfo {author} {\bibfnamefont {F.}~\bibnamefont {Wilczek}},\
  }\bibfield  {title} {\bibinfo {title} {Cosmology of the invisible axion},\
  }\href {https://doi.org/https://doi.org/10.1016/0370-2693(83)90637-8}
  {\bibfield  {journal} {\bibinfo  {journal} {Physics Letters B}\ }\textbf
  {\bibinfo {volume} {120}},\ \bibinfo {pages} {127} (\bibinfo {year}
  {1983})}\BibitemShut {NoStop}%
\bibitem [{\citenamefont {Dine}\ and\ \citenamefont
  {Fischler}(1983)}]{DINE1983137}%
  \BibitemOpen
  \bibfield  {author} {\bibinfo {author} {\bibfnamefont {M.}~\bibnamefont
  {Dine}}\ and\ \bibinfo {author} {\bibfnamefont {W.}~\bibnamefont
  {Fischler}},\ }\bibfield  {title} {\bibinfo {title} {The not-so-harmless
  axion},\ }\href
  {https://doi.org/https://doi.org/10.1016/0370-2693(83)90639-1} {\bibfield
  {journal} {\bibinfo  {journal} {Physics Letters B}\ }\textbf {\bibinfo
  {volume} {120}},\ \bibinfo {pages} {137} (\bibinfo {year}
  {1983})}\BibitemShut {NoStop}%
\bibitem [{\citenamefont {Amruth}\ \emph {et~al.}(2023)\citenamefont {Amruth},
  \citenamefont {Broadhurst}, \citenamefont {Lim}, \citenamefont {Oguri},
  \citenamefont {Smoot}, \citenamefont {Diego}, \citenamefont {Leung},
  \citenamefont {Emami}, \citenamefont {Li}, \citenamefont {Chiueh},
  \citenamefont {Schive}, \citenamefont {Yeung},\ and\ \citenamefont
  {Li}}]{Amruth_2023}%
  \BibitemOpen
  \bibfield  {author} {\bibinfo {author} {\bibfnamefont {A.}~\bibnamefont
  {Amruth}}, \bibinfo {author} {\bibfnamefont {T.}~\bibnamefont {Broadhurst}},
  \bibinfo {author} {\bibfnamefont {J.}~\bibnamefont {Lim}}, \bibinfo {author}
  {\bibfnamefont {M.}~\bibnamefont {Oguri}}, \bibinfo {author} {\bibfnamefont
  {G.~F.}\ \bibnamefont {Smoot}}, \bibinfo {author} {\bibfnamefont {J.~M.}\
  \bibnamefont {Diego}}, \bibinfo {author} {\bibfnamefont {E.}~\bibnamefont
  {Leung}}, \bibinfo {author} {\bibfnamefont {R.}~\bibnamefont {Emami}},
  \bibinfo {author} {\bibfnamefont {J.}~\bibnamefont {Li}}, \bibinfo {author}
  {\bibfnamefont {T.}~\bibnamefont {Chiueh}}, \bibinfo {author} {\bibfnamefont
  {H.-Y.}\ \bibnamefont {Schive}}, \bibinfo {author} {\bibfnamefont {M.~C.~H.}\
  \bibnamefont {Yeung}},\ and\ \bibinfo {author} {\bibfnamefont {S.~K.}\
  \bibnamefont {Li}},\ }\bibfield  {title} {\bibinfo {title} {Einstein rings
  modulated by wavelike dark matter from anomalies in gravitationally lensed},\
  }\href {https://doi.org/10.1038/s41550-023-01943-9} {\bibfield  {journal}
  {\bibinfo  {journal} {Nat. Astron.}\ }\textbf {\bibinfo {volume} {7}},\
  \bibinfo {pages} {736} (\bibinfo {year} {2023})}\BibitemShut {NoStop}%
\bibitem [{\citenamefont {Akerib}\ \emph {et~al.}(2013)\citenamefont {Akerib},
  \citenamefont {Bai}, \citenamefont {Bedikian}, \citenamefont {Bernard},
  \citenamefont {Bernstein}, \citenamefont {Bolozdynya}, \citenamefont
  {Bradley}, \citenamefont {Byram}, \citenamefont {Cahn}, \citenamefont {Camp},
  \citenamefont {Carmona-Benitez}, \citenamefont {Carr}, \citenamefont
  {Chapman}, \citenamefont {Chiller}, \citenamefont {Chiller}, \citenamefont
  {Clark}, \citenamefont {Classen}, \citenamefont {Coffey}, \citenamefont
  {Curioni}, \citenamefont {Dahl}, \citenamefont {Dazeley}, \citenamefont
  {de~Viveiros}, \citenamefont {Dobi}, \citenamefont {Dragowsky}, \citenamefont
  {Druszkiewicz}, \citenamefont {Edwards}, \citenamefont {Faham}, \citenamefont
  {Fiorucci}, \citenamefont {Gaitskell}, \citenamefont {Gibson}, \citenamefont
  {Gilchriese}, \citenamefont {Hall}, \citenamefont {Hanhardt}, \citenamefont
  {Holbrook}, \citenamefont {Ihm}, \citenamefont {Jacobsen}, \citenamefont
  {Kastens}, \citenamefont {Kazkaz}, \citenamefont {Knoche}, \citenamefont
  {Kyre}, \citenamefont {Kwong}, \citenamefont {Lander}, \citenamefont
  {Larsen}, \citenamefont {Lee}, \citenamefont {Leonard}, \citenamefont
  {Lesko}, \citenamefont {Lindote}, \citenamefont {Lopes}, \citenamefont
  {Lyashenko}, \citenamefont {Malling}, \citenamefont {Mannino}, \citenamefont
  {Marquez}, \citenamefont {McKinsey}, \citenamefont {Mei}, \citenamefont
  {Mock}, \citenamefont {Moongweluwan}, \citenamefont {Morii}, \citenamefont
  {Nelson}, \citenamefont {Neves}, \citenamefont {Nikkel}, \citenamefont
  {Pangilinan}, \citenamefont {Parker}, \citenamefont {Pease}, \citenamefont
  {Pech}, \citenamefont {Phelps}, \citenamefont {Rodionov}, \citenamefont
  {Roberts}, \citenamefont {Shei}, \citenamefont {Shutt}, \citenamefont
  {Silva}, \citenamefont {Skulski}, \citenamefont {Solovov}, \citenamefont
  {Sofka}, \citenamefont {Sorensen}, \citenamefont {Spaans}, \citenamefont
  {Stiegler}, \citenamefont {Stolp}, \citenamefont {Svoboda}, \citenamefont
  {Sweany}, \citenamefont {Szydagis}, \citenamefont {Taylor}, \citenamefont
  {Thomson}, \citenamefont {Tripathi}, \citenamefont {Uvarov}, \citenamefont
  {Verbus}, \citenamefont {Walsh}, \citenamefont {Webb}, \citenamefont {White},
  \citenamefont {White}, \citenamefont {Whitis}, \citenamefont {Wlasenko},
  \citenamefont {Wolfs}, \citenamefont {Woods},\ and\ \citenamefont
  {Zhang}}]{Akerib_Lux_2013}%
  \BibitemOpen
  \bibfield  {author} {\bibinfo {author} {\bibfnamefont {D.}~\bibnamefont
  {Akerib}}, \bibinfo {author} {\bibfnamefont {X.}~\bibnamefont {Bai}},
  \bibinfo {author} {\bibfnamefont {S.}~\bibnamefont {Bedikian}}, \bibinfo
  {author} {\bibfnamefont {E.}~\bibnamefont {Bernard}}, \bibinfo {author}
  {\bibfnamefont {A.}~\bibnamefont {Bernstein}}, \bibinfo {author}
  {\bibfnamefont {A.}~\bibnamefont {Bolozdynya}}, \bibinfo {author}
  {\bibfnamefont {A.}~\bibnamefont {Bradley}}, \bibinfo {author} {\bibfnamefont
  {D.}~\bibnamefont {Byram}}, \bibinfo {author} {\bibfnamefont
  {S.}~\bibnamefont {Cahn}}, \bibinfo {author} {\bibfnamefont {C.}~\bibnamefont
  {Camp}}, \bibinfo {author} {\bibfnamefont {M.}~\bibnamefont
  {Carmona-Benitez}}, \bibinfo {author} {\bibfnamefont {D.}~\bibnamefont
  {Carr}}, \bibinfo {author} {\bibfnamefont {J.}~\bibnamefont {Chapman}},
  \bibinfo {author} {\bibfnamefont {A.}~\bibnamefont {Chiller}}, \bibinfo
  {author} {\bibfnamefont {C.}~\bibnamefont {Chiller}}, \bibinfo {author}
  {\bibfnamefont {K.}~\bibnamefont {Clark}}, \bibinfo {author} {\bibfnamefont
  {T.}~\bibnamefont {Classen}}, \bibinfo {author} {\bibfnamefont
  {T.}~\bibnamefont {Coffey}}, \bibinfo {author} {\bibfnamefont
  {A.}~\bibnamefont {Curioni}}, \bibinfo {author} {\bibfnamefont
  {E.}~\bibnamefont {Dahl}}, \bibinfo {author} {\bibfnamefont {S.}~\bibnamefont
  {Dazeley}}, \bibinfo {author} {\bibfnamefont {L.}~\bibnamefont
  {de~Viveiros}}, \bibinfo {author} {\bibfnamefont {A.}~\bibnamefont {Dobi}},
  \bibinfo {author} {\bibfnamefont {E.}~\bibnamefont {Dragowsky}}, \bibinfo
  {author} {\bibfnamefont {E.}~\bibnamefont {Druszkiewicz}}, \bibinfo {author}
  {\bibfnamefont {B.}~\bibnamefont {Edwards}}, \bibinfo {author} {\bibfnamefont
  {C.}~\bibnamefont {Faham}}, \bibinfo {author} {\bibfnamefont
  {S.}~\bibnamefont {Fiorucci}}, \bibinfo {author} {\bibfnamefont
  {R.}~\bibnamefont {Gaitskell}}, \bibinfo {author} {\bibfnamefont
  {K.}~\bibnamefont {Gibson}}, \bibinfo {author} {\bibfnamefont
  {M.}~\bibnamefont {Gilchriese}}, \bibinfo {author} {\bibfnamefont
  {C.}~\bibnamefont {Hall}}, \bibinfo {author} {\bibfnamefont {M.}~\bibnamefont
  {Hanhardt}}, \bibinfo {author} {\bibfnamefont {B.}~\bibnamefont {Holbrook}},
  \bibinfo {author} {\bibfnamefont {M.}~\bibnamefont {Ihm}}, \bibinfo {author}
  {\bibfnamefont {R.}~\bibnamefont {Jacobsen}}, \bibinfo {author}
  {\bibfnamefont {L.}~\bibnamefont {Kastens}}, \bibinfo {author} {\bibfnamefont
  {K.}~\bibnamefont {Kazkaz}}, \bibinfo {author} {\bibfnamefont
  {R.}~\bibnamefont {Knoche}}, \bibinfo {author} {\bibfnamefont
  {S.}~\bibnamefont {Kyre}}, \bibinfo {author} {\bibfnamefont {J.}~\bibnamefont
  {Kwong}}, \bibinfo {author} {\bibfnamefont {R.}~\bibnamefont {Lander}},
  \bibinfo {author} {\bibfnamefont {N.}~\bibnamefont {Larsen}}, \bibinfo
  {author} {\bibfnamefont {C.}~\bibnamefont {Lee}}, \bibinfo {author}
  {\bibfnamefont {D.}~\bibnamefont {Leonard}}, \bibinfo {author} {\bibfnamefont
  {K.}~\bibnamefont {Lesko}}, \bibinfo {author} {\bibfnamefont
  {A.}~\bibnamefont {Lindote}}, \bibinfo {author} {\bibfnamefont
  {M.}~\bibnamefont {Lopes}}, \bibinfo {author} {\bibfnamefont
  {A.}~\bibnamefont {Lyashenko}}, \bibinfo {author} {\bibfnamefont
  {D.}~\bibnamefont {Malling}}, \bibinfo {author} {\bibfnamefont
  {R.}~\bibnamefont {Mannino}}, \bibinfo {author} {\bibfnamefont
  {Z.}~\bibnamefont {Marquez}}, \bibinfo {author} {\bibfnamefont
  {D.}~\bibnamefont {McKinsey}}, \bibinfo {author} {\bibfnamefont {D.-M.}\
  \bibnamefont {Mei}}, \bibinfo {author} {\bibfnamefont {J.}~\bibnamefont
  {Mock}}, \bibinfo {author} {\bibfnamefont {M.}~\bibnamefont {Moongweluwan}},
  \bibinfo {author} {\bibfnamefont {M.}~\bibnamefont {Morii}}, \bibinfo
  {author} {\bibfnamefont {H.}~\bibnamefont {Nelson}}, \bibinfo {author}
  {\bibfnamefont {F.}~\bibnamefont {Neves}}, \bibinfo {author} {\bibfnamefont
  {J.}~\bibnamefont {Nikkel}}, \bibinfo {author} {\bibfnamefont
  {M.}~\bibnamefont {Pangilinan}}, \bibinfo {author} {\bibfnamefont
  {P.}~\bibnamefont {Parker}}, \bibinfo {author} {\bibfnamefont
  {E.}~\bibnamefont {Pease}}, \bibinfo {author} {\bibfnamefont
  {K.}~\bibnamefont {Pech}}, \bibinfo {author} {\bibfnamefont {P.}~\bibnamefont
  {Phelps}}, \bibinfo {author} {\bibfnamefont {A.}~\bibnamefont {Rodionov}},
  \bibinfo {author} {\bibfnamefont {P.}~\bibnamefont {Roberts}}, \bibinfo
  {author} {\bibfnamefont {A.}~\bibnamefont {Shei}}, \bibinfo {author}
  {\bibfnamefont {T.}~\bibnamefont {Shutt}}, \bibinfo {author} {\bibfnamefont
  {C.}~\bibnamefont {Silva}}, \bibinfo {author} {\bibfnamefont
  {W.}~\bibnamefont {Skulski}}, \bibinfo {author} {\bibfnamefont
  {V.}~\bibnamefont {Solovov}}, \bibinfo {author} {\bibfnamefont
  {C.}~\bibnamefont {Sofka}}, \bibinfo {author} {\bibfnamefont
  {P.}~\bibnamefont {Sorensen}}, \bibinfo {author} {\bibfnamefont
  {J.}~\bibnamefont {Spaans}}, \bibinfo {author} {\bibfnamefont
  {T.}~\bibnamefont {Stiegler}}, \bibinfo {author} {\bibfnamefont
  {D.}~\bibnamefont {Stolp}}, \bibinfo {author} {\bibfnamefont
  {R.}~\bibnamefont {Svoboda}}, \bibinfo {author} {\bibfnamefont
  {M.}~\bibnamefont {Sweany}}, \bibinfo {author} {\bibfnamefont
  {M.}~\bibnamefont {Szydagis}}, \bibinfo {author} {\bibfnamefont
  {D.}~\bibnamefont {Taylor}}, \bibinfo {author} {\bibfnamefont
  {J.}~\bibnamefont {Thomson}}, \bibinfo {author} {\bibfnamefont
  {M.}~\bibnamefont {Tripathi}}, \bibinfo {author} {\bibfnamefont
  {S.}~\bibnamefont {Uvarov}}, \bibinfo {author} {\bibfnamefont
  {J.}~\bibnamefont {Verbus}}, \bibinfo {author} {\bibfnamefont
  {N.}~\bibnamefont {Walsh}}, \bibinfo {author} {\bibfnamefont
  {R.}~\bibnamefont {Webb}}, \bibinfo {author} {\bibfnamefont {D.}~\bibnamefont
  {White}}, \bibinfo {author} {\bibfnamefont {J.}~\bibnamefont {White}},
  \bibinfo {author} {\bibfnamefont {T.}~\bibnamefont {Whitis}}, \bibinfo
  {author} {\bibfnamefont {M.}~\bibnamefont {Wlasenko}}, \bibinfo {author}
  {\bibfnamefont {F.}~\bibnamefont {Wolfs}}, \bibinfo {author} {\bibfnamefont
  {M.}~\bibnamefont {Woods}},\ and\ \bibinfo {author} {\bibfnamefont
  {C.}~\bibnamefont {Zhang}},\ }\bibfield  {title} {\bibinfo {title} {The large
  underground xenon (lux) experiment},\ }\href
  {https://doi.org/https://doi.org/10.1016/j.nima.2012.11.135} {\bibfield
  {journal} {\bibinfo  {journal} {Nuclear Instruments and Methods in Physics
  Research Section A: Accelerators, Spectrometers, Detectors and Associated
  Equipment}\ }\textbf {\bibinfo {volume} {704}},\ \bibinfo {pages} {111 }
  (\bibinfo {year} {2013})}\BibitemShut {NoStop}%
\bibitem [{\citenamefont {Aprile}\ \emph {et~al.}(2018)\citenamefont {Aprile},
  \citenamefont {Aalbers}, \citenamefont {Agostini}, \citenamefont {Alfonsi},
  \citenamefont {Althueser}, \citenamefont {Amaro}, \citenamefont {Anthony},
  \citenamefont {Arneodo}, \citenamefont {Baudis}, \citenamefont
  {Bauermeister}, \citenamefont {Benabderrahmane}, \citenamefont {Berger},
  \citenamefont {Breur}, \citenamefont {Brown}, \citenamefont {Brown},
  \citenamefont {Brown}, \citenamefont {Bruenner}, \citenamefont {Bruno},
  \citenamefont {Budnik}, \citenamefont {Capelli}, \citenamefont {Cardoso},
  \citenamefont {Cichon}, \citenamefont {Coderre}, \citenamefont {Colijn},
  \citenamefont {Conrad}, \citenamefont {Cussonneau}, \citenamefont {Decowski},
  \citenamefont {de~Perio}, \citenamefont {Di~Gangi}, \citenamefont
  {Di~Giovanni}, \citenamefont {Diglio}, \citenamefont {Elykov}, \citenamefont
  {Eurin}, \citenamefont {Fei}, \citenamefont {Ferella}, \citenamefont
  {Fieguth}, \citenamefont {Fulgione}, \citenamefont {Gallo~Rosso},
  \citenamefont {Galloway}, \citenamefont {Gao}, \citenamefont {Garbini},
  \citenamefont {Geis}, \citenamefont {Grandi}, \citenamefont {Greene},
  \citenamefont {Qiu}, \citenamefont {Hasterok}, \citenamefont {Hogenbirk},
  \citenamefont {Howlett}, \citenamefont {Itay}, \citenamefont {Joerg},
  \citenamefont {Kaminsky}, \citenamefont {Kazama}, \citenamefont {Kish},
  \citenamefont {Koltman}, \citenamefont {Landsman}, \citenamefont {Lang},
  \citenamefont {Levinson}, \citenamefont {Lin}, \citenamefont {Lindemann},
  \citenamefont {Lindner}, \citenamefont {Lombardi}, \citenamefont {Lopes},
  \citenamefont {Mahlstedt}, \citenamefont {Manfredini}, \citenamefont
  {Marrod\'an~Undagoitia}, \citenamefont {Masbou}, \citenamefont {Masson},
  \citenamefont {Messina}, \citenamefont {Micheneau}, \citenamefont {Miller},
  \citenamefont {Molinario}, \citenamefont {Mor\aa{}}, \citenamefont {Murra},
  \citenamefont {Naganoma}, \citenamefont {Ni}, \citenamefont {Oberlack},
  \citenamefont {Pelssers}, \citenamefont {Piastra}, \citenamefont {Pienaar},
  \citenamefont {Pizzella}, \citenamefont {Plante}, \citenamefont
  {Podviianiuk}, \citenamefont {Priel}, \citenamefont
  {Ram\'{\i}rez~Garc\'{\i}a}, \citenamefont {Rauch}, \citenamefont {Reichard},
  \citenamefont {Reuter}, \citenamefont {Riedel}, \citenamefont {Rizzo},
  \citenamefont {Rocchetti}, \citenamefont {Rupp}, \citenamefont {dos Santos},
  \citenamefont {Sartorelli}, \citenamefont {Scheibelhut}, \citenamefont
  {Schindler}, \citenamefont {Schreiner}, \citenamefont {Schulte},
  \citenamefont {Schumann}, \citenamefont {Scotto~Lavina}, \citenamefont
  {Selvi}, \citenamefont {Shagin}, \citenamefont {Shockley}, \citenamefont
  {Silva}, \citenamefont {Simgen}, \citenamefont {Thers}, \citenamefont
  {Toschi}, \citenamefont {Trinchero}, \citenamefont {Tunnell}, \citenamefont
  {Upole}, \citenamefont {Vargas}, \citenamefont {Wack}, \citenamefont {Wang},
  \citenamefont {Wang}, \citenamefont {Wei}, \citenamefont {Weinheimer},
  \citenamefont {Wittweg}, \citenamefont {Wulf}, \citenamefont {Ye},
  \citenamefont {Zhang},\ and\ \citenamefont {Zhu}}]{Aprile_2018}%
  \BibitemOpen
  \bibfield  {author} {\bibinfo {author} {\bibfnamefont {E.}~\bibnamefont
  {Aprile}}, \bibinfo {author} {\bibfnamefont {J.}~\bibnamefont {Aalbers}},
  \bibinfo {author} {\bibfnamefont {F.}~\bibnamefont {Agostini}}, \bibinfo
  {author} {\bibfnamefont {M.}~\bibnamefont {Alfonsi}}, \bibinfo {author}
  {\bibfnamefont {L.}~\bibnamefont {Althueser}}, \bibinfo {author}
  {\bibfnamefont {F.~D.}\ \bibnamefont {Amaro}}, \bibinfo {author}
  {\bibfnamefont {M.}~\bibnamefont {Anthony}}, \bibinfo {author} {\bibfnamefont
  {F.}~\bibnamefont {Arneodo}}, \bibinfo {author} {\bibfnamefont
  {L.}~\bibnamefont {Baudis}}, \bibinfo {author} {\bibfnamefont
  {B.}~\bibnamefont {Bauermeister}}, \bibinfo {author} {\bibfnamefont {M.~L.}\
  \bibnamefont {Benabderrahmane}}, \bibinfo {author} {\bibfnamefont
  {T.}~\bibnamefont {Berger}}, \bibinfo {author} {\bibfnamefont {P.~A.}\
  \bibnamefont {Breur}}, \bibinfo {author} {\bibfnamefont {A.}~\bibnamefont
  {Brown}}, \bibinfo {author} {\bibfnamefont {A.}~\bibnamefont {Brown}},
  \bibinfo {author} {\bibfnamefont {E.}~\bibnamefont {Brown}}, \bibinfo
  {author} {\bibfnamefont {S.}~\bibnamefont {Bruenner}}, \bibinfo {author}
  {\bibfnamefont {G.}~\bibnamefont {Bruno}}, \bibinfo {author} {\bibfnamefont
  {R.}~\bibnamefont {Budnik}}, \bibinfo {author} {\bibfnamefont
  {C.}~\bibnamefont {Capelli}}, \bibinfo {author} {\bibfnamefont {J.~M.~R.}\
  \bibnamefont {Cardoso}}, \bibinfo {author} {\bibfnamefont {D.}~\bibnamefont
  {Cichon}}, \bibinfo {author} {\bibfnamefont {D.}~\bibnamefont {Coderre}},
  \bibinfo {author} {\bibfnamefont {A.~P.}\ \bibnamefont {Colijn}}, \bibinfo
  {author} {\bibfnamefont {J.}~\bibnamefont {Conrad}}, \bibinfo {author}
  {\bibfnamefont {J.~P.}\ \bibnamefont {Cussonneau}}, \bibinfo {author}
  {\bibfnamefont {M.~P.}\ \bibnamefont {Decowski}}, \bibinfo {author}
  {\bibfnamefont {P.}~\bibnamefont {de~Perio}}, \bibinfo {author}
  {\bibfnamefont {P.}~\bibnamefont {Di~Gangi}}, \bibinfo {author}
  {\bibfnamefont {A.}~\bibnamefont {Di~Giovanni}}, \bibinfo {author}
  {\bibfnamefont {S.}~\bibnamefont {Diglio}}, \bibinfo {author} {\bibfnamefont
  {A.}~\bibnamefont {Elykov}}, \bibinfo {author} {\bibfnamefont
  {G.}~\bibnamefont {Eurin}}, \bibinfo {author} {\bibfnamefont
  {J.}~\bibnamefont {Fei}}, \bibinfo {author} {\bibfnamefont {A.~D.}\
  \bibnamefont {Ferella}}, \bibinfo {author} {\bibfnamefont {A.}~\bibnamefont
  {Fieguth}}, \bibinfo {author} {\bibfnamefont {W.}~\bibnamefont {Fulgione}},
  \bibinfo {author} {\bibfnamefont {A.}~\bibnamefont {Gallo~Rosso}}, \bibinfo
  {author} {\bibfnamefont {M.}~\bibnamefont {Galloway}}, \bibinfo {author}
  {\bibfnamefont {F.}~\bibnamefont {Gao}}, \bibinfo {author} {\bibfnamefont
  {M.}~\bibnamefont {Garbini}}, \bibinfo {author} {\bibfnamefont
  {C.}~\bibnamefont {Geis}}, \bibinfo {author} {\bibfnamefont {L.}~\bibnamefont
  {Grandi}}, \bibinfo {author} {\bibfnamefont {Z.}~\bibnamefont {Greene}},
  \bibinfo {author} {\bibfnamefont {H.}~\bibnamefont {Qiu}}, \bibinfo {author}
  {\bibfnamefont {C.}~\bibnamefont {Hasterok}}, \bibinfo {author}
  {\bibfnamefont {E.}~\bibnamefont {Hogenbirk}}, \bibinfo {author}
  {\bibfnamefont {J.}~\bibnamefont {Howlett}}, \bibinfo {author} {\bibfnamefont
  {R.}~\bibnamefont {Itay}}, \bibinfo {author} {\bibfnamefont {F.}~\bibnamefont
  {Joerg}}, \bibinfo {author} {\bibfnamefont {B.}~\bibnamefont {Kaminsky}},
  \bibinfo {author} {\bibfnamefont {S.}~\bibnamefont {Kazama}}, \bibinfo
  {author} {\bibfnamefont {A.}~\bibnamefont {Kish}}, \bibinfo {author}
  {\bibfnamefont {G.}~\bibnamefont {Koltman}}, \bibinfo {author} {\bibfnamefont
  {H.}~\bibnamefont {Landsman}}, \bibinfo {author} {\bibfnamefont {R.~F.}\
  \bibnamefont {Lang}}, \bibinfo {author} {\bibfnamefont {L.}~\bibnamefont
  {Levinson}}, \bibinfo {author} {\bibfnamefont {Q.}~\bibnamefont {Lin}},
  \bibinfo {author} {\bibfnamefont {S.}~\bibnamefont {Lindemann}}, \bibinfo
  {author} {\bibfnamefont {M.}~\bibnamefont {Lindner}}, \bibinfo {author}
  {\bibfnamefont {F.}~\bibnamefont {Lombardi}}, \bibinfo {author}
  {\bibfnamefont {J.~A.~M.}\ \bibnamefont {Lopes}}, \bibinfo {author}
  {\bibfnamefont {J.}~\bibnamefont {Mahlstedt}}, \bibinfo {author}
  {\bibfnamefont {A.}~\bibnamefont {Manfredini}}, \bibinfo {author}
  {\bibfnamefont {T.}~\bibnamefont {Marrod\'an~Undagoitia}}, \bibinfo {author}
  {\bibfnamefont {J.}~\bibnamefont {Masbou}}, \bibinfo {author} {\bibfnamefont
  {D.}~\bibnamefont {Masson}}, \bibinfo {author} {\bibfnamefont
  {M.}~\bibnamefont {Messina}}, \bibinfo {author} {\bibfnamefont
  {K.}~\bibnamefont {Micheneau}}, \bibinfo {author} {\bibfnamefont
  {K.}~\bibnamefont {Miller}}, \bibinfo {author} {\bibfnamefont
  {A.}~\bibnamefont {Molinario}}, \bibinfo {author} {\bibfnamefont
  {K.}~\bibnamefont {Mor\aa{}}}, \bibinfo {author} {\bibfnamefont
  {M.}~\bibnamefont {Murra}}, \bibinfo {author} {\bibfnamefont
  {J.}~\bibnamefont {Naganoma}}, \bibinfo {author} {\bibfnamefont
  {K.}~\bibnamefont {Ni}}, \bibinfo {author} {\bibfnamefont {U.}~\bibnamefont
  {Oberlack}}, \bibinfo {author} {\bibfnamefont {B.}~\bibnamefont {Pelssers}},
  \bibinfo {author} {\bibfnamefont {F.}~\bibnamefont {Piastra}}, \bibinfo
  {author} {\bibfnamefont {J.}~\bibnamefont {Pienaar}}, \bibinfo {author}
  {\bibfnamefont {V.}~\bibnamefont {Pizzella}}, \bibinfo {author}
  {\bibfnamefont {G.}~\bibnamefont {Plante}}, \bibinfo {author} {\bibfnamefont
  {R.}~\bibnamefont {Podviianiuk}}, \bibinfo {author} {\bibfnamefont
  {N.}~\bibnamefont {Priel}}, \bibinfo {author} {\bibfnamefont
  {D.}~\bibnamefont {Ram\'{\i}rez~Garc\'{\i}a}}, \bibinfo {author}
  {\bibfnamefont {L.}~\bibnamefont {Rauch}}, \bibinfo {author} {\bibfnamefont
  {S.}~\bibnamefont {Reichard}}, \bibinfo {author} {\bibfnamefont
  {C.}~\bibnamefont {Reuter}}, \bibinfo {author} {\bibfnamefont
  {B.}~\bibnamefont {Riedel}}, \bibinfo {author} {\bibfnamefont
  {A.}~\bibnamefont {Rizzo}}, \bibinfo {author} {\bibfnamefont
  {A.}~\bibnamefont {Rocchetti}}, \bibinfo {author} {\bibfnamefont
  {N.}~\bibnamefont {Rupp}}, \bibinfo {author} {\bibfnamefont {J.~M.~F.}\
  \bibnamefont {dos Santos}}, \bibinfo {author} {\bibfnamefont
  {G.}~\bibnamefont {Sartorelli}}, \bibinfo {author} {\bibfnamefont
  {M.}~\bibnamefont {Scheibelhut}}, \bibinfo {author} {\bibfnamefont
  {S.}~\bibnamefont {Schindler}}, \bibinfo {author} {\bibfnamefont
  {J.}~\bibnamefont {Schreiner}}, \bibinfo {author} {\bibfnamefont
  {D.}~\bibnamefont {Schulte}}, \bibinfo {author} {\bibfnamefont
  {M.}~\bibnamefont {Schumann}}, \bibinfo {author} {\bibfnamefont
  {L.}~\bibnamefont {Scotto~Lavina}}, \bibinfo {author} {\bibfnamefont
  {M.}~\bibnamefont {Selvi}}, \bibinfo {author} {\bibfnamefont
  {P.}~\bibnamefont {Shagin}}, \bibinfo {author} {\bibfnamefont
  {E.}~\bibnamefont {Shockley}}, \bibinfo {author} {\bibfnamefont
  {M.}~\bibnamefont {Silva}}, \bibinfo {author} {\bibfnamefont
  {H.}~\bibnamefont {Simgen}}, \bibinfo {author} {\bibfnamefont
  {D.}~\bibnamefont {Thers}}, \bibinfo {author} {\bibfnamefont
  {F.}~\bibnamefont {Toschi}}, \bibinfo {author} {\bibfnamefont
  {G.}~\bibnamefont {Trinchero}}, \bibinfo {author} {\bibfnamefont
  {C.}~\bibnamefont {Tunnell}}, \bibinfo {author} {\bibfnamefont
  {N.}~\bibnamefont {Upole}}, \bibinfo {author} {\bibfnamefont
  {M.}~\bibnamefont {Vargas}}, \bibinfo {author} {\bibfnamefont
  {O.}~\bibnamefont {Wack}}, \bibinfo {author} {\bibfnamefont {H.}~\bibnamefont
  {Wang}}, \bibinfo {author} {\bibfnamefont {Z.}~\bibnamefont {Wang}}, \bibinfo
  {author} {\bibfnamefont {Y.}~\bibnamefont {Wei}}, \bibinfo {author}
  {\bibfnamefont {C.}~\bibnamefont {Weinheimer}}, \bibinfo {author}
  {\bibfnamefont {C.}~\bibnamefont {Wittweg}}, \bibinfo {author} {\bibfnamefont
  {J.}~\bibnamefont {Wulf}}, \bibinfo {author} {\bibfnamefont {J.}~\bibnamefont
  {Ye}}, \bibinfo {author} {\bibfnamefont {Y.}~\bibnamefont {Zhang}},\ and\
  \bibinfo {author} {\bibfnamefont {T.}~\bibnamefont {Zhu}} (\bibinfo
  {collaboration} {XENON Collaboration}),\ }\bibfield  {title} {\bibinfo
  {title} {{Dark Matter Search Results from a One Ton-Year Exposure of
  XENON1T}},\ }\href {https://doi.org/10.1103/PhysRevLett.121.111302}
  {\bibfield  {journal} {\bibinfo  {journal} {Phys. Rev. Lett.}\ }\textbf
  {\bibinfo {volume} {121}},\ \bibinfo {pages} {111302} (\bibinfo {year}
  {2018})}\BibitemShut {NoStop}%
\bibitem [{\citenamefont {Zhang}\ \emph {et~al.}(2018)\citenamefont {Zhang},
  \citenamefont {Abdukerim}, \citenamefont {Chen}, \citenamefont {Chen},
  \citenamefont {Chen}, \citenamefont {Cui}, \citenamefont {Dong},
  \citenamefont {Fang}, \citenamefont {Fu}, \citenamefont {Giboni},
  \citenamefont {Giuliani}, \citenamefont {Gu}, \citenamefont {Guo},
  \citenamefont {Guo}, \citenamefont {Han}, \citenamefont {He}, \citenamefont
  {He}, \citenamefont {Huang}, \citenamefont {Huang}, \citenamefont {Huang},
  \citenamefont {Ji}, \citenamefont {Ji}, \citenamefont {Ju}, \citenamefont
  {Li}, \citenamefont {Li}, \citenamefont {Lin}, \citenamefont {Liu},
  \citenamefont {Liu}, \citenamefont {Ma}, \citenamefont {Mao}, \citenamefont
  {Ni}, \citenamefont {Ning}, \citenamefont {Ren}, \citenamefont {Shi},
  \citenamefont {Tan}, \citenamefont {Wang}, \citenamefont {Wang},
  \citenamefont {Wang}, \citenamefont {Wang}, \citenamefont {Wang},
  \citenamefont {Wang}, \citenamefont {Wang}, \citenamefont {Wang},
  \citenamefont {Wang}, \citenamefont {Wu}, \citenamefont {Wu}, \citenamefont
  {Xia}, \citenamefont {Xiao}, \citenamefont {Xie}, \citenamefont {Yan},
  \citenamefont {Yang}, \citenamefont {Yang}, \citenamefont {Yu}, \citenamefont
  {Yuan}, \citenamefont {Yue}, \citenamefont {Zhang}, \citenamefont {Zhang},
  \citenamefont {Zhao}, \citenamefont {Zheng}, \citenamefont {Zhou},
  \citenamefont {Zhou},\ and\ \citenamefont {Zhou}}]{Zhang_PandaX_2018}%
  \BibitemOpen
  \bibfield  {author} {\bibinfo {author} {\bibfnamefont {H.}~\bibnamefont
  {Zhang}}, \bibinfo {author} {\bibfnamefont {A.}~\bibnamefont {Abdukerim}},
  \bibinfo {author} {\bibfnamefont {W.}~\bibnamefont {Chen}}, \bibinfo {author}
  {\bibfnamefont {X.}~\bibnamefont {Chen}}, \bibinfo {author} {\bibfnamefont
  {Y.}~\bibnamefont {Chen}}, \bibinfo {author} {\bibfnamefont {X.}~\bibnamefont
  {Cui}}, \bibinfo {author} {\bibfnamefont {B.}~\bibnamefont {Dong}}, \bibinfo
  {author} {\bibfnamefont {D.}~\bibnamefont {Fang}}, \bibinfo {author}
  {\bibfnamefont {C.}~\bibnamefont {Fu}}, \bibinfo {author} {\bibfnamefont
  {K.}~\bibnamefont {Giboni}}, \bibinfo {author} {\bibfnamefont
  {F.}~\bibnamefont {Giuliani}}, \bibinfo {author} {\bibfnamefont
  {L.}~\bibnamefont {Gu}}, \bibinfo {author} {\bibfnamefont {X.}~\bibnamefont
  {Guo}}, \bibinfo {author} {\bibfnamefont {Z.}~\bibnamefont {Guo}}, \bibinfo
  {author} {\bibfnamefont {K.}~\bibnamefont {Han}}, \bibinfo {author}
  {\bibfnamefont {C.}~\bibnamefont {He}}, \bibinfo {author} {\bibfnamefont
  {S.}~\bibnamefont {He}}, \bibinfo {author} {\bibfnamefont {D.}~\bibnamefont
  {Huang}}, \bibinfo {author} {\bibfnamefont {X.}~\bibnamefont {Huang}},
  \bibinfo {author} {\bibfnamefont {Z.}~\bibnamefont {Huang}}, \bibinfo
  {author} {\bibfnamefont {P.}~\bibnamefont {Ji}}, \bibinfo {author}
  {\bibfnamefont {X.}~\bibnamefont {Ji}}, \bibinfo {author} {\bibfnamefont
  {Y.}~\bibnamefont {Ju}}, \bibinfo {author} {\bibfnamefont {S.}~\bibnamefont
  {Li}}, \bibinfo {author} {\bibfnamefont {Y.}~\bibnamefont {Li}}, \bibinfo
  {author} {\bibfnamefont {H.}~\bibnamefont {Lin}}, \bibinfo {author}
  {\bibfnamefont {H.}~\bibnamefont {Liu}}, \bibinfo {author} {\bibfnamefont
  {J.}~\bibnamefont {Liu}}, \bibinfo {author} {\bibfnamefont {Y.}~\bibnamefont
  {Ma}}, \bibinfo {author} {\bibfnamefont {Y.}~\bibnamefont {Mao}}, \bibinfo
  {author} {\bibfnamefont {K.}~\bibnamefont {Ni}}, \bibinfo {author}
  {\bibfnamefont {J.}~\bibnamefont {Ning}}, \bibinfo {author} {\bibfnamefont
  {X.}~\bibnamefont {Ren}}, \bibinfo {author} {\bibfnamefont {F.}~\bibnamefont
  {Shi}}, \bibinfo {author} {\bibfnamefont {A.}~\bibnamefont {Tan}}, \bibinfo
  {author} {\bibfnamefont {A.}~\bibnamefont {Wang}}, \bibinfo {author}
  {\bibfnamefont {C.}~\bibnamefont {Wang}}, \bibinfo {author} {\bibfnamefont
  {H.}~\bibnamefont {Wang}}, \bibinfo {author} {\bibfnamefont {M.}~\bibnamefont
  {Wang}}, \bibinfo {author} {\bibfnamefont {Q.}~\bibnamefont {Wang}}, \bibinfo
  {author} {\bibfnamefont {S.}~\bibnamefont {Wang}}, \bibinfo {author}
  {\bibfnamefont {X.}~\bibnamefont {Wang}}, \bibinfo {author} {\bibfnamefont
  {X.}~\bibnamefont {Wang}}, \bibinfo {author} {\bibfnamefont {Z.}~\bibnamefont
  {Wang}}, \bibinfo {author} {\bibfnamefont {M.}~\bibnamefont {Wu}}, \bibinfo
  {author} {\bibfnamefont {S.}~\bibnamefont {Wu}}, \bibinfo {author}
  {\bibfnamefont {J.}~\bibnamefont {Xia}}, \bibinfo {author} {\bibfnamefont
  {M.}~\bibnamefont {Xiao}}, \bibinfo {author} {\bibfnamefont {P.}~\bibnamefont
  {Xie}}, \bibinfo {author} {\bibfnamefont {B.}~\bibnamefont {Yan}}, \bibinfo
  {author} {\bibfnamefont {J.}~\bibnamefont {Yang}}, \bibinfo {author}
  {\bibfnamefont {Y.}~\bibnamefont {Yang}}, \bibinfo {author} {\bibfnamefont
  {C.}~\bibnamefont {Yu}}, \bibinfo {author} {\bibfnamefont {J.}~\bibnamefont
  {Yuan}}, \bibinfo {author} {\bibfnamefont {J.}~\bibnamefont {Yue}}, \bibinfo
  {author} {\bibfnamefont {D.}~\bibnamefont {Zhang}}, \bibinfo {author}
  {\bibfnamefont {T.}~\bibnamefont {Zhang}}, \bibinfo {author} {\bibfnamefont
  {L.}~\bibnamefont {Zhao}}, \bibinfo {author} {\bibfnamefont {Q.}~\bibnamefont
  {Zheng}}, \bibinfo {author} {\bibfnamefont {J.}~\bibnamefont {Zhou}},
  \bibinfo {author} {\bibfnamefont {N.}~\bibnamefont {Zhou}},\ and\ \bibinfo
  {author} {\bibfnamefont {X.}~\bibnamefont {Zhou}},\ }\bibfield  {title}
  {\bibinfo {title} {{Dark matter direct search sensitivity of the PandaX-4T
  experiment}},\ }\href {https://doi.org/10.1007/s11433-018-9259-0} {\bibfield
  {journal} {\bibinfo  {journal} {Science China Physics, Mechanics {\&}
  Astronomy}\ }\textbf {\bibinfo {volume} {62}},\ \bibinfo {pages} {31011}
  (\bibinfo {year} {2018})}\BibitemShut {NoStop}%
\bibitem [{\citenamefont {Caldwell}\ \emph {et~al.}(2017)\citenamefont
  {Caldwell}, \citenamefont {Dvali}, \citenamefont {Majorovits}, \citenamefont
  {Millar}, \citenamefont {Raffelt}, \citenamefont {Redondo}, \citenamefont
  {Reimann}, \citenamefont {Simon},\ and\ \citenamefont
  {Steffen}}]{Caldwell_2017}%
  \BibitemOpen
  \bibfield  {author} {\bibinfo {author} {\bibfnamefont {A.}~\bibnamefont
  {Caldwell}}, \bibinfo {author} {\bibfnamefont {G.}~\bibnamefont {Dvali}},
  \bibinfo {author} {\bibfnamefont {B.}~\bibnamefont {Majorovits}}, \bibinfo
  {author} {\bibfnamefont {A.}~\bibnamefont {Millar}}, \bibinfo {author}
  {\bibfnamefont {G.}~\bibnamefont {Raffelt}}, \bibinfo {author} {\bibfnamefont
  {J.}~\bibnamefont {Redondo}}, \bibinfo {author} {\bibfnamefont
  {O.}~\bibnamefont {Reimann}}, \bibinfo {author} {\bibfnamefont
  {F.}~\bibnamefont {Simon}},\ and\ \bibinfo {author} {\bibfnamefont
  {F.}~\bibnamefont {Steffen}} (\bibinfo {collaboration} {MADMAX Working
  Group}),\ }\bibfield  {title} {\bibinfo {title} {Dielectric haloscopes: A new
  way to detect axion dark matter},\ }\href
  {https://doi.org/10.1103/PhysRevLett.118.091801} {\bibfield  {journal}
  {\bibinfo  {journal} {Phys. Rev. Lett.}\ }\textbf {\bibinfo {volume} {118}},\
  \bibinfo {pages} {091801} (\bibinfo {year} {2017})}\BibitemShut {NoStop}%
\bibitem [{\citenamefont {Brouwer}\ \emph {et~al.}(2022)\citenamefont
  {Brouwer}, \citenamefont {Chaudhuri}, \citenamefont {Cho}, \citenamefont
  {Corbin}, \citenamefont {Craddock}, \citenamefont {Dawson}, \citenamefont
  {Droster}, \citenamefont {Foster}, \citenamefont {Fry}, \citenamefont
  {Graham}, \citenamefont {Henning}, \citenamefont {Irwin}, \citenamefont
  {Kadribasic}, \citenamefont {Kahn}, \citenamefont {Keller}, \citenamefont
  {Kolevatov}, \citenamefont {Kuenstner}, \citenamefont {Leder}, \citenamefont
  {Li}, \citenamefont {Ouellet}, \citenamefont {Pappas}, \citenamefont
  {Phipps}, \citenamefont {Rapidis}, \citenamefont {Safdi}, \citenamefont
  {Salemi}, \citenamefont {Simanovskaia}, \citenamefont {Singh}, \citenamefont
  {van Assendelft}, \citenamefont {van Bibber}, \citenamefont {Wells},
  \citenamefont {Winslow}, \citenamefont {Wisniewski},\ and\ \citenamefont
  {Young}}]{Brouwer_2022}%
  \BibitemOpen
  \bibfield  {author} {\bibinfo {author} {\bibfnamefont {L.}~\bibnamefont
  {Brouwer}}, \bibinfo {author} {\bibfnamefont {S.}~\bibnamefont {Chaudhuri}},
  \bibinfo {author} {\bibfnamefont {H.-M.}\ \bibnamefont {Cho}}, \bibinfo
  {author} {\bibfnamefont {J.}~\bibnamefont {Corbin}}, \bibinfo {author}
  {\bibfnamefont {W.}~\bibnamefont {Craddock}}, \bibinfo {author}
  {\bibfnamefont {C.~S.}\ \bibnamefont {Dawson}}, \bibinfo {author}
  {\bibfnamefont {A.}~\bibnamefont {Droster}}, \bibinfo {author} {\bibfnamefont
  {J.~W.}\ \bibnamefont {Foster}}, \bibinfo {author} {\bibfnamefont {J.~T.}\
  \bibnamefont {Fry}}, \bibinfo {author} {\bibfnamefont {P.~W.}\ \bibnamefont
  {Graham}}, \bibinfo {author} {\bibfnamefont {R.}~\bibnamefont {Henning}},
  \bibinfo {author} {\bibfnamefont {K.~D.}\ \bibnamefont {Irwin}}, \bibinfo
  {author} {\bibfnamefont {F.}~\bibnamefont {Kadribasic}}, \bibinfo {author}
  {\bibfnamefont {Y.}~\bibnamefont {Kahn}}, \bibinfo {author} {\bibfnamefont
  {A.}~\bibnamefont {Keller}}, \bibinfo {author} {\bibfnamefont
  {R.}~\bibnamefont {Kolevatov}}, \bibinfo {author} {\bibfnamefont
  {S.}~\bibnamefont {Kuenstner}}, \bibinfo {author} {\bibfnamefont {A.~F.}\
  \bibnamefont {Leder}}, \bibinfo {author} {\bibfnamefont {D.}~\bibnamefont
  {Li}}, \bibinfo {author} {\bibfnamefont {J.~L.}\ \bibnamefont {Ouellet}},
  \bibinfo {author} {\bibfnamefont {K.~M.~W.}\ \bibnamefont {Pappas}}, \bibinfo
  {author} {\bibfnamefont {A.}~\bibnamefont {Phipps}}, \bibinfo {author}
  {\bibfnamefont {N.~M.}\ \bibnamefont {Rapidis}}, \bibinfo {author}
  {\bibfnamefont {B.~R.}\ \bibnamefont {Safdi}}, \bibinfo {author}
  {\bibfnamefont {C.~P.}\ \bibnamefont {Salemi}}, \bibinfo {author}
  {\bibfnamefont {M.}~\bibnamefont {Simanovskaia}}, \bibinfo {author}
  {\bibfnamefont {J.}~\bibnamefont {Singh}}, \bibinfo {author} {\bibfnamefont
  {E.~C.}\ \bibnamefont {van Assendelft}}, \bibinfo {author} {\bibfnamefont
  {K.}~\bibnamefont {van Bibber}}, \bibinfo {author} {\bibfnamefont
  {K.}~\bibnamefont {Wells}}, \bibinfo {author} {\bibfnamefont
  {L.}~\bibnamefont {Winslow}}, \bibinfo {author} {\bibfnamefont {W.~J.}\
  \bibnamefont {Wisniewski}},\ and\ \bibinfo {author} {\bibfnamefont {B.~A.}\
  \bibnamefont {Young}} (\bibinfo {collaboration} {DMRadio Collaboration}),\
  }\bibfield  {title} {\bibinfo {title} {{Projected sensitivity of
  ${\text{DMRadio-m}}^{3}$: A search for the QCD axion below $1\text{ }\text{
  }\mathrm{\ensuremath{\mu}}\mathrm{eV}$}},\ }\href
  {https://doi.org/10.1103/PhysRevD.106.103008} {\bibfield  {journal} {\bibinfo
   {journal} {Phys. Rev. D}\ }\textbf {\bibinfo {volume} {106}},\ \bibinfo
  {pages} {103008} (\bibinfo {year} {2022})}\BibitemShut {NoStop}%
\bibitem [{\citenamefont {Anastassopoulos}\ \emph {et~al.}(2017)\citenamefont
  {Anastassopoulos} \emph {et~al.}}]{Anastassopoulos_2017}%
  \BibitemOpen
  \bibfield  {author} {\bibinfo {author} {\bibfnamefont {V.}~\bibnamefont
  {Anastassopoulos}} \emph {et~al.} (\bibinfo {collaboration} {CAST}),\
  }\bibfield  {title} {\bibinfo {title} {{New CAST Limit on the Axion-Photon
  Interaction}},\ }\href {https://doi.org/10.1038/nphys4109} {\bibfield
  {journal} {\bibinfo  {journal} {Nature Phys.}\ }\textbf {\bibinfo {volume}
  {13}},\ \bibinfo {pages} {584} (\bibinfo {year} {2017})},\ \Eprint
  {https://arxiv.org/abs/1705.02290} {arXiv:1705.02290 [hep-ex]} \BibitemShut
  {NoStop}%
\bibitem [{\citenamefont {Armengaud}\ \emph {et~al.}(2014)\citenamefont
  {Armengaud}, \citenamefont {Avignone}, \citenamefont {Betz}, \citenamefont
  {Brax}, \citenamefont {Brun}, \citenamefont {Cantatore}, \citenamefont
  {Carmona}, \citenamefont {Carosi}, \citenamefont {Caspers}, \citenamefont
  {Caspi}, \citenamefont {Cetin}, \citenamefont {Chelouche}, \citenamefont
  {Christensen}, \citenamefont {Dael}, \citenamefont {Dafni}, \citenamefont
  {Davenport}, \citenamefont {Derbin}, \citenamefont {Desch}, \citenamefont
  {Diago}, \citenamefont {Döbrich}, \citenamefont {Dratchnev}, \citenamefont
  {Dudarev}, \citenamefont {Eleftheriadis}, \citenamefont {Fanourakis},
  \citenamefont {Ferrer-Ribas}, \citenamefont {Galán}, \citenamefont
  {García}, \citenamefont {Garza}, \citenamefont {Geralis}, \citenamefont
  {Gimeno}, \citenamefont {Giomataris}, \citenamefont {Gninenko}, \citenamefont
  {Gómez}, \citenamefont {González-Díaz}, \citenamefont {Guendelman},
  \citenamefont {Hailey}, \citenamefont {Hiramatsu}, \citenamefont {Hoffmann},
  \citenamefont {Horns}, \citenamefont {Iguaz}, \citenamefont {Irastorza},
  \citenamefont {Isern}, \citenamefont {Imai}, \citenamefont {Jakobsen},
  \citenamefont {Jaeckel}, \citenamefont {Jakovčić}, \citenamefont
  {Kaminski}, \citenamefont {Kawasaki}, \citenamefont {Karuza}, \citenamefont
  {Krčmar}, \citenamefont {Kousouris}, \citenamefont {Krieger}, \citenamefont
  {Lakić}, \citenamefont {Limousin}, \citenamefont {Lindner}, \citenamefont
  {Liolios}, \citenamefont {Luzón}, \citenamefont {Matsuki}, \citenamefont
  {Muratova}, \citenamefont {Nones}, \citenamefont {Ortega}, \citenamefont
  {Papaevangelou}, \citenamefont {Pivovaroff}, \citenamefont {Raffelt},
  \citenamefont {Redondo}, \citenamefont {Ringwald}, \citenamefont
  {Russenschuck}, \citenamefont {Ruz}, \citenamefont {Saikawa}, \citenamefont
  {Savvidis}, \citenamefont {Sekiguchi}, \citenamefont {Semertzidis},
  \citenamefont {Shilon}, \citenamefont {Sikivie}, \citenamefont {Silva},
  \citenamefont {ten Kate}, \citenamefont {Tomas}, \citenamefont {Troitsky},
  \citenamefont {Vafeiadis}, \citenamefont {van Bibber}, \citenamefont
  {Vedrine}, \citenamefont {Villar}, \citenamefont {Vogel}, \citenamefont
  {Walckiers}, \citenamefont {Weltman}, \citenamefont {Wester}, \citenamefont
  {Yildiz},\ and\ \citenamefont {Zioutas}}]{Armengaud_2014}%
  \BibitemOpen
  \bibfield  {author} {\bibinfo {author} {\bibfnamefont {E.}~\bibnamefont
  {Armengaud}}, \bibinfo {author} {\bibfnamefont {F.~T.}\ \bibnamefont
  {Avignone}}, \bibinfo {author} {\bibfnamefont {M.}~\bibnamefont {Betz}},
  \bibinfo {author} {\bibfnamefont {P.}~\bibnamefont {Brax}}, \bibinfo {author}
  {\bibfnamefont {P.}~\bibnamefont {Brun}}, \bibinfo {author} {\bibfnamefont
  {G.}~\bibnamefont {Cantatore}}, \bibinfo {author} {\bibfnamefont {J.~M.}\
  \bibnamefont {Carmona}}, \bibinfo {author} {\bibfnamefont {G.~P.}\
  \bibnamefont {Carosi}}, \bibinfo {author} {\bibfnamefont {F.}~\bibnamefont
  {Caspers}}, \bibinfo {author} {\bibfnamefont {S.}~\bibnamefont {Caspi}},
  \bibinfo {author} {\bibfnamefont {S.~A.}\ \bibnamefont {Cetin}}, \bibinfo
  {author} {\bibfnamefont {D.}~\bibnamefont {Chelouche}}, \bibinfo {author}
  {\bibfnamefont {F.~E.}\ \bibnamefont {Christensen}}, \bibinfo {author}
  {\bibfnamefont {A.}~\bibnamefont {Dael}}, \bibinfo {author} {\bibfnamefont
  {T.}~\bibnamefont {Dafni}}, \bibinfo {author} {\bibfnamefont
  {M.}~\bibnamefont {Davenport}}, \bibinfo {author} {\bibfnamefont {A.~V.}\
  \bibnamefont {Derbin}}, \bibinfo {author} {\bibfnamefont {K.}~\bibnamefont
  {Desch}}, \bibinfo {author} {\bibfnamefont {A.}~\bibnamefont {Diago}},
  \bibinfo {author} {\bibfnamefont {B.}~\bibnamefont {Döbrich}}, \bibinfo
  {author} {\bibfnamefont {I.}~\bibnamefont {Dratchnev}}, \bibinfo {author}
  {\bibfnamefont {A.}~\bibnamefont {Dudarev}}, \bibinfo {author} {\bibfnamefont
  {C.}~\bibnamefont {Eleftheriadis}}, \bibinfo {author} {\bibfnamefont
  {G.}~\bibnamefont {Fanourakis}}, \bibinfo {author} {\bibfnamefont
  {E.}~\bibnamefont {Ferrer-Ribas}}, \bibinfo {author} {\bibfnamefont
  {J.}~\bibnamefont {Galán}}, \bibinfo {author} {\bibfnamefont {J.~A.}\
  \bibnamefont {García}}, \bibinfo {author} {\bibfnamefont {J.~G.}\
  \bibnamefont {Garza}}, \bibinfo {author} {\bibfnamefont {T.}~\bibnamefont
  {Geralis}}, \bibinfo {author} {\bibfnamefont {B.}~\bibnamefont {Gimeno}},
  \bibinfo {author} {\bibfnamefont {I.}~\bibnamefont {Giomataris}}, \bibinfo
  {author} {\bibfnamefont {S.}~\bibnamefont {Gninenko}}, \bibinfo {author}
  {\bibfnamefont {H.}~\bibnamefont {Gómez}}, \bibinfo {author} {\bibfnamefont
  {D.}~\bibnamefont {González-Díaz}}, \bibinfo {author} {\bibfnamefont
  {E.}~\bibnamefont {Guendelman}}, \bibinfo {author} {\bibfnamefont {C.~J.}\
  \bibnamefont {Hailey}}, \bibinfo {author} {\bibfnamefont {T.}~\bibnamefont
  {Hiramatsu}}, \bibinfo {author} {\bibfnamefont {D.~H.~H.}\ \bibnamefont
  {Hoffmann}}, \bibinfo {author} {\bibfnamefont {D.}~\bibnamefont {Horns}},
  \bibinfo {author} {\bibfnamefont {F.~J.}\ \bibnamefont {Iguaz}}, \bibinfo
  {author} {\bibfnamefont {I.~G.}\ \bibnamefont {Irastorza}}, \bibinfo {author}
  {\bibfnamefont {J.}~\bibnamefont {Isern}}, \bibinfo {author} {\bibfnamefont
  {K.}~\bibnamefont {Imai}}, \bibinfo {author} {\bibfnamefont {A.~C.}\
  \bibnamefont {Jakobsen}}, \bibinfo {author} {\bibfnamefont {J.}~\bibnamefont
  {Jaeckel}}, \bibinfo {author} {\bibfnamefont {K.}~\bibnamefont {Jakovčić}},
  \bibinfo {author} {\bibfnamefont {J.}~\bibnamefont {Kaminski}}, \bibinfo
  {author} {\bibfnamefont {M.}~\bibnamefont {Kawasaki}}, \bibinfo {author}
  {\bibfnamefont {M.}~\bibnamefont {Karuza}}, \bibinfo {author} {\bibfnamefont
  {M.}~\bibnamefont {Krčmar}}, \bibinfo {author} {\bibfnamefont
  {K.}~\bibnamefont {Kousouris}}, \bibinfo {author} {\bibfnamefont
  {C.}~\bibnamefont {Krieger}}, \bibinfo {author} {\bibfnamefont
  {B.}~\bibnamefont {Lakić}}, \bibinfo {author} {\bibfnamefont
  {O.}~\bibnamefont {Limousin}}, \bibinfo {author} {\bibfnamefont
  {A.}~\bibnamefont {Lindner}}, \bibinfo {author} {\bibfnamefont
  {A.}~\bibnamefont {Liolios}}, \bibinfo {author} {\bibfnamefont
  {G.}~\bibnamefont {Luzón}}, \bibinfo {author} {\bibfnamefont
  {S.}~\bibnamefont {Matsuki}}, \bibinfo {author} {\bibfnamefont {V.~N.}\
  \bibnamefont {Muratova}}, \bibinfo {author} {\bibfnamefont {C.}~\bibnamefont
  {Nones}}, \bibinfo {author} {\bibfnamefont {I.}~\bibnamefont {Ortega}},
  \bibinfo {author} {\bibfnamefont {T.}~\bibnamefont {Papaevangelou}}, \bibinfo
  {author} {\bibfnamefont {M.~J.}\ \bibnamefont {Pivovaroff}}, \bibinfo
  {author} {\bibfnamefont {G.}~\bibnamefont {Raffelt}}, \bibinfo {author}
  {\bibfnamefont {J.}~\bibnamefont {Redondo}}, \bibinfo {author} {\bibfnamefont
  {A.}~\bibnamefont {Ringwald}}, \bibinfo {author} {\bibfnamefont
  {S.}~\bibnamefont {Russenschuck}}, \bibinfo {author} {\bibfnamefont
  {J.}~\bibnamefont {Ruz}}, \bibinfo {author} {\bibfnamefont {K.}~\bibnamefont
  {Saikawa}}, \bibinfo {author} {\bibfnamefont {I.}~\bibnamefont {Savvidis}},
  \bibinfo {author} {\bibfnamefont {T.}~\bibnamefont {Sekiguchi}}, \bibinfo
  {author} {\bibfnamefont {Y.~K.}\ \bibnamefont {Semertzidis}}, \bibinfo
  {author} {\bibfnamefont {I.}~\bibnamefont {Shilon}}, \bibinfo {author}
  {\bibfnamefont {P.}~\bibnamefont {Sikivie}}, \bibinfo {author} {\bibfnamefont
  {H.}~\bibnamefont {Silva}}, \bibinfo {author} {\bibfnamefont
  {H.}~\bibnamefont {ten Kate}}, \bibinfo {author} {\bibfnamefont
  {A.}~\bibnamefont {Tomas}}, \bibinfo {author} {\bibfnamefont
  {S.}~\bibnamefont {Troitsky}}, \bibinfo {author} {\bibfnamefont
  {T.}~\bibnamefont {Vafeiadis}}, \bibinfo {author} {\bibfnamefont
  {K.}~\bibnamefont {van Bibber}}, \bibinfo {author} {\bibfnamefont
  {P.}~\bibnamefont {Vedrine}}, \bibinfo {author} {\bibfnamefont {J.~A.}\
  \bibnamefont {Villar}}, \bibinfo {author} {\bibfnamefont {J.~K.}\
  \bibnamefont {Vogel}}, \bibinfo {author} {\bibfnamefont {L.}~\bibnamefont
  {Walckiers}}, \bibinfo {author} {\bibfnamefont {A.}~\bibnamefont {Weltman}},
  \bibinfo {author} {\bibfnamefont {W.}~\bibnamefont {Wester}}, \bibinfo
  {author} {\bibfnamefont {S.~C.}\ \bibnamefont {Yildiz}},\ and\ \bibinfo
  {author} {\bibfnamefont {K.}~\bibnamefont {Zioutas}},\ }\bibfield  {title}
  {\bibinfo {title} {{Conceptual design of the International Axion Observatory
  (IAXO)}},\ }\href {https://doi.org/10.1088/1748-0221/9/05/T05002} {\bibfield
  {journal} {\bibinfo  {journal} {{JINST}}\ }\textbf {\bibinfo {volume} {9}},\
  \bibinfo {pages} {T05002} (\bibinfo {year} {2014})}\BibitemShut {NoStop}%
\bibitem [{\citenamefont {Bähre}\ \emph {et~al.}(2013)\citenamefont {Bähre},
  \citenamefont {Döbrich}, \citenamefont {Dreyling-Eschweiler}, \citenamefont
  {Ghazaryan}, \citenamefont {Hodajerdi}, \citenamefont {Horns}, \citenamefont
  {Januschek}, \citenamefont {Knabbe}, \citenamefont {Lindner}, \citenamefont
  {Notz}, \citenamefont {Ringwald}, \citenamefont {von Seggern}, \citenamefont
  {Stromhagen}, \citenamefont {Trines},\ and\ \citenamefont
  {Willke}}]{Baehre_2013}%
  \BibitemOpen
  \bibfield  {author} {\bibinfo {author} {\bibfnamefont {R.}~\bibnamefont
  {Bähre}}, \bibinfo {author} {\bibfnamefont {B.}~\bibnamefont {Döbrich}},
  \bibinfo {author} {\bibfnamefont {J.}~\bibnamefont {Dreyling-Eschweiler}},
  \bibinfo {author} {\bibfnamefont {S.}~\bibnamefont {Ghazaryan}}, \bibinfo
  {author} {\bibfnamefont {R.}~\bibnamefont {Hodajerdi}}, \bibinfo {author}
  {\bibfnamefont {D.}~\bibnamefont {Horns}}, \bibinfo {author} {\bibfnamefont
  {F.}~\bibnamefont {Januschek}}, \bibinfo {author} {\bibfnamefont {E.~A.}\
  \bibnamefont {Knabbe}}, \bibinfo {author} {\bibfnamefont {A.}~\bibnamefont
  {Lindner}}, \bibinfo {author} {\bibfnamefont {D.}~\bibnamefont {Notz}},
  \bibinfo {author} {\bibfnamefont {A.}~\bibnamefont {Ringwald}}, \bibinfo
  {author} {\bibfnamefont {J.~E.}\ \bibnamefont {von Seggern}}, \bibinfo
  {author} {\bibfnamefont {R.}~\bibnamefont {Stromhagen}}, \bibinfo {author}
  {\bibfnamefont {D.}~\bibnamefont {Trines}},\ and\ \bibinfo {author}
  {\bibfnamefont {B.}~\bibnamefont {Willke}},\ }\bibfield  {title} {\bibinfo
  {title} {{Any light particle search II — Technical Design Report}},\ }\href
  {https://doi.org/10.1088/1748-0221/8/09/T09001} {\bibfield  {journal}
  {\bibinfo  {journal} {Journal of Instrumentation}\ }\textbf {\bibinfo
  {volume} {8}}\bibinfo  {number} { (09)},\ \bibinfo {pages}
  {T09001}}\BibitemShut {NoStop}%
\bibitem [{\citenamefont {Betz}\ \emph {et~al.}(2013)\citenamefont {Betz},
  \citenamefont {Caspers}, \citenamefont {Gasior}, \citenamefont {Thumm},\ and\
  \citenamefont {Rieger}}]{Betz_2013}%
  \BibitemOpen
\bibfield  {number} {  }\bibfield  {author} {\bibinfo {author} {\bibfnamefont
  {M.}~\bibnamefont {Betz}}, \bibinfo {author} {\bibfnamefont {F.}~\bibnamefont
  {Caspers}}, \bibinfo {author} {\bibfnamefont {M.}~\bibnamefont {Gasior}},
  \bibinfo {author} {\bibfnamefont {M.}~\bibnamefont {Thumm}},\ and\ \bibinfo
  {author} {\bibfnamefont {S.~W.}\ \bibnamefont {Rieger}},\ }\bibfield  {title}
  {\bibinfo {title} {{First results of the CERN Resonant Weakly Interacting
  sub-eV Particle Search (CROWS)}},\ }\href
  {https://doi.org/10.1103/PhysRevD.88.075014} {\bibfield  {journal} {\bibinfo
  {journal} {Phys. Rev. D}\ }\textbf {\bibinfo {volume} {88}},\ \bibinfo
  {pages} {075014} (\bibinfo {year} {2013})}\BibitemShut {NoStop}%
\bibitem [{\citenamefont {Salemi}\ \emph {et~al.}(2021)\citenamefont {Salemi},
  \citenamefont {Foster}, \citenamefont {Ouellet}, \citenamefont {Gavin},
  \citenamefont {Pappas}, \citenamefont {Cheng}, \citenamefont {Richardson},
  \citenamefont {Henning}, \citenamefont {Kahn}, \citenamefont {Nguyen},
  \citenamefont {Rodd}, \citenamefont {Safdi},\ and\ \citenamefont
  {Winslow}}]{PhysRevLett.127.081801}%
  \BibitemOpen
  \bibfield  {author} {\bibinfo {author} {\bibfnamefont {C.~P.}\ \bibnamefont
  {Salemi}}, \bibinfo {author} {\bibfnamefont {J.~W.}\ \bibnamefont {Foster}},
  \bibinfo {author} {\bibfnamefont {J.~L.}\ \bibnamefont {Ouellet}}, \bibinfo
  {author} {\bibfnamefont {A.}~\bibnamefont {Gavin}}, \bibinfo {author}
  {\bibfnamefont {K.~M.~W.}\ \bibnamefont {Pappas}}, \bibinfo {author}
  {\bibfnamefont {S.}~\bibnamefont {Cheng}}, \bibinfo {author} {\bibfnamefont
  {K.~A.}\ \bibnamefont {Richardson}}, \bibinfo {author} {\bibfnamefont
  {R.}~\bibnamefont {Henning}}, \bibinfo {author} {\bibfnamefont
  {Y.}~\bibnamefont {Kahn}}, \bibinfo {author} {\bibfnamefont {R.}~\bibnamefont
  {Nguyen}}, \bibinfo {author} {\bibfnamefont {N.~L.}\ \bibnamefont {Rodd}},
  \bibinfo {author} {\bibfnamefont {B.~R.}\ \bibnamefont {Safdi}},\ and\
  \bibinfo {author} {\bibfnamefont {L.}~\bibnamefont {Winslow}},\ }\bibfield
  {title} {\bibinfo {title} {Search for low-mass axion dark matter with
  abracadabra-10 cm},\ }\href {https://doi.org/10.1103/PhysRevLett.127.081801}
  {\bibfield  {journal} {\bibinfo  {journal} {Phys. Rev. Lett.}\ }\textbf
  {\bibinfo {volume} {127}},\ \bibinfo {pages} {081801} (\bibinfo {year}
  {2021})}\BibitemShut {NoStop}%
\bibitem [{\citenamefont {Martynov}\ and\ \citenamefont
  {Miao}(2020)}]{Martynov_2020}%
  \BibitemOpen
  \bibfield  {author} {\bibinfo {author} {\bibfnamefont {D.}~\bibnamefont
  {Martynov}}\ and\ \bibinfo {author} {\bibfnamefont {H.}~\bibnamefont
  {Miao}},\ }\bibfield  {title} {\bibinfo {title} {Quantum-enhanced
  interferometry for axion searches},\ }\href
  {https://doi.org/10.1103/PhysRevD.101.095034} {\bibfield  {journal} {\bibinfo
   {journal} {Phys. Rev. D}\ }\textbf {\bibinfo {volume} {101}},\ \bibinfo
  {pages} {095034} (\bibinfo {year} {2020})}\BibitemShut {NoStop}%
\bibitem [{\citenamefont {DeRocco}\ and\ \citenamefont
  {Hook}(2018)}]{DeRocco_2018}%
  \BibitemOpen
  \bibfield  {author} {\bibinfo {author} {\bibfnamefont {W.}~\bibnamefont
  {DeRocco}}\ and\ \bibinfo {author} {\bibfnamefont {A.}~\bibnamefont {Hook}},\
  }\bibfield  {title} {\bibinfo {title} {Axion interferometry},\ }\href
  {https://doi.org/10.1103/PhysRevD.98.035021} {\bibfield  {journal} {\bibinfo
  {journal} {Phys. Rev. D}\ }\textbf {\bibinfo {volume} {98}},\ \bibinfo
  {pages} {035021} (\bibinfo {year} {2018})}\BibitemShut {NoStop}%
\bibitem [{\citenamefont {Obata}\ \emph {et~al.}(2018)\citenamefont {Obata},
  \citenamefont {Fujita},\ and\ \citenamefont {Michimura}}]{Obata_2018}%
  \BibitemOpen
  \bibfield  {author} {\bibinfo {author} {\bibfnamefont {I.}~\bibnamefont
  {Obata}}, \bibinfo {author} {\bibfnamefont {T.}~\bibnamefont {Fujita}},\ and\
  \bibinfo {author} {\bibfnamefont {Y.}~\bibnamefont {Michimura}},\ }\bibfield
  {title} {\bibinfo {title} {Optical ring cavity search for axion dark
  matter},\ }\href {https://doi.org/10.1103/PhysRevLett.121.161301} {\bibfield
  {journal} {\bibinfo  {journal} {Phys. Rev. Lett.}\ }\textbf {\bibinfo
  {volume} {121}},\ \bibinfo {pages} {161301} (\bibinfo {year}
  {2018})}\BibitemShut {NoStop}%
\bibitem [{\citenamefont {Liu}\ \emph {et~al.}(2019)\citenamefont {Liu},
  \citenamefont {Elwood}, \citenamefont {Evans},\ and\ \citenamefont
  {Thaler}}]{Liu_2019}%
  \BibitemOpen
  \bibfield  {author} {\bibinfo {author} {\bibfnamefont {H.}~\bibnamefont
  {Liu}}, \bibinfo {author} {\bibfnamefont {B.~D.}\ \bibnamefont {Elwood}},
  \bibinfo {author} {\bibfnamefont {M.}~\bibnamefont {Evans}},\ and\ \bibinfo
  {author} {\bibfnamefont {J.}~\bibnamefont {Thaler}},\ }\bibfield  {title}
  {\bibinfo {title} {Searching for axion dark matter with birefringent
  cavities},\ }\href {https://doi.org/10.1103/PhysRevD.100.023548} {\bibfield
  {journal} {\bibinfo  {journal} {Phys. Rev. D}\ }\textbf {\bibinfo {volume}
  {100}},\ \bibinfo {pages} {023548} (\bibinfo {year} {2019})}\BibitemShut
  {NoStop}%
\bibitem [{\citenamefont {Oshima}\ \emph {et~al.}(2023)\citenamefont {Oshima},
  \citenamefont {Fujimoto}, \citenamefont {Kume}, \citenamefont {Morisaki},
  \citenamefont {Nagano}, \citenamefont {Fujita}, \citenamefont {Obata},
  \citenamefont {Nishizawa}, \citenamefont {Michimura},\ and\ \citenamefont
  {Ando}}]{FirstDanceResults_2023}%
  \BibitemOpen
  \bibfield  {author} {\bibinfo {author} {\bibfnamefont {Y.}~\bibnamefont
  {Oshima}}, \bibinfo {author} {\bibfnamefont {H.}~\bibnamefont {Fujimoto}},
  \bibinfo {author} {\bibfnamefont {J.}~\bibnamefont {Kume}}, \bibinfo {author}
  {\bibfnamefont {S.}~\bibnamefont {Morisaki}}, \bibinfo {author}
  {\bibfnamefont {K.}~\bibnamefont {Nagano}}, \bibinfo {author} {\bibfnamefont
  {T.}~\bibnamefont {Fujita}}, \bibinfo {author} {\bibfnamefont
  {I.}~\bibnamefont {Obata}}, \bibinfo {author} {\bibfnamefont
  {A.}~\bibnamefont {Nishizawa}}, \bibinfo {author} {\bibfnamefont
  {Y.}~\bibnamefont {Michimura}},\ and\ \bibinfo {author} {\bibfnamefont
  {M.}~\bibnamefont {Ando}},\ }\bibfield  {title} {\bibinfo {title} {First
  results of axion dark matter search with dance},\ }\href
  {https://doi.org/10.1103/PhysRevD.108.072005} {\bibfield  {journal} {\bibinfo
   {journal} {Phys. Rev. D}\ }\textbf {\bibinfo {volume} {108}},\ \bibinfo
  {pages} {072005} (\bibinfo {year} {2023})}\BibitemShut {NoStop}%
\bibitem [{\citenamefont {Nagano}\ \emph {et~al.}(2019)\citenamefont {Nagano},
  \citenamefont {Fujita}, \citenamefont {Michimura},\ and\ \citenamefont
  {Obata}}]{Nagano_2019}%
  \BibitemOpen
  \bibfield  {author} {\bibinfo {author} {\bibfnamefont {K.}~\bibnamefont
  {Nagano}}, \bibinfo {author} {\bibfnamefont {T.}~\bibnamefont {Fujita}},
  \bibinfo {author} {\bibfnamefont {Y.}~\bibnamefont {Michimura}},\ and\
  \bibinfo {author} {\bibfnamefont {I.}~\bibnamefont {Obata}},\ }\bibfield
  {title} {\bibinfo {title} {Axion dark matter search with interferometric
  gravitational wave detectors},\ }\href
  {https://doi.org/10.1103/PhysRevLett.123.111301} {\bibfield  {journal}
  {\bibinfo  {journal} {Phys. Rev. Lett.}\ }\textbf {\bibinfo {volume} {123}},\
  \bibinfo {pages} {111301} (\bibinfo {year} {2019})}\BibitemShut {NoStop}%
\bibitem [{\citenamefont {Liu}\ \emph {et~al.}(2020)\citenamefont {Liu},
  \citenamefont {Smoot},\ and\ \citenamefont
  {Zhao}}]{Liu_linearlyPolarisedPulsarLight_2020}%
  \BibitemOpen
  \bibfield  {author} {\bibinfo {author} {\bibfnamefont {T.}~\bibnamefont
  {Liu}}, \bibinfo {author} {\bibfnamefont {G.}~\bibnamefont {Smoot}},\ and\
  \bibinfo {author} {\bibfnamefont {Y.}~\bibnamefont {Zhao}},\ }\bibfield
  {title} {\bibinfo {title} {Detecting axionlike dark matter with linearly
  polarized pulsar light},\ }\href
  {https://doi.org/10.1103/PhysRevD.101.063012} {\bibfield  {journal} {\bibinfo
   {journal} {Phys. Rev. D}\ }\textbf {\bibinfo {volume} {101}},\ \bibinfo
  {pages} {063012} (\bibinfo {year} {2020})}\BibitemShut {NoStop}%
\bibitem [{\citenamefont {Michimura}\ \emph {et~al.}(2020)\citenamefont
  {Michimura}, \citenamefont {Oshima}, \citenamefont {Watanabe}, \citenamefont
  {Kawasaki}, \citenamefont {Takeda}, \citenamefont {Ando}, \citenamefont
  {Nagano}, \citenamefont {Obata},\ and\ \citenamefont
  {Fujita}}]{Michimura_2020}%
  \BibitemOpen
  \bibfield  {author} {\bibinfo {author} {\bibfnamefont {Y.}~\bibnamefont
  {Michimura}}, \bibinfo {author} {\bibfnamefont {Y.}~\bibnamefont {Oshima}},
  \bibinfo {author} {\bibfnamefont {T.}~\bibnamefont {Watanabe}}, \bibinfo
  {author} {\bibfnamefont {T.}~\bibnamefont {Kawasaki}}, \bibinfo {author}
  {\bibfnamefont {H.}~\bibnamefont {Takeda}}, \bibinfo {author} {\bibfnamefont
  {M.}~\bibnamefont {Ando}}, \bibinfo {author} {\bibfnamefont {K.}~\bibnamefont
  {Nagano}}, \bibinfo {author} {\bibfnamefont {I.}~\bibnamefont {Obata}},\ and\
  \bibinfo {author} {\bibfnamefont {T.}~\bibnamefont {Fujita}},\ }\bibfield
  {title} {\bibinfo {title} {Dance: Dark matter axion search with ring cavity
  experiment},\ }\href {https://doi.org/10.1088/1742-6596/1468/1/012032}
  {\bibfield  {journal} {\bibinfo  {journal} {Journal of Physics: Conference
  Series}\ }\textbf {\bibinfo {volume} {1468}},\ \bibinfo {pages} {012032}
  (\bibinfo {year} {2020})}\BibitemShut {NoStop}%
\bibitem [{\citenamefont {Budker}\ \emph {et~al.}(2014)\citenamefont {Budker},
  \citenamefont {Graham}, \citenamefont {Ledbetter}, \citenamefont
  {Rajendran},\ and\ \citenamefont {Sushkov}}]{Budker_Casper_2014}%
  \BibitemOpen
  \bibfield  {author} {\bibinfo {author} {\bibfnamefont {D.}~\bibnamefont
  {Budker}}, \bibinfo {author} {\bibfnamefont {P.~W.}\ \bibnamefont {Graham}},
  \bibinfo {author} {\bibfnamefont {M.}~\bibnamefont {Ledbetter}}, \bibinfo
  {author} {\bibfnamefont {S.}~\bibnamefont {Rajendran}},\ and\ \bibinfo
  {author} {\bibfnamefont {A.~O.}\ \bibnamefont {Sushkov}},\ }\bibfield
  {title} {\bibinfo {title} {Proposal for a cosmic axion spin precession
  experiment (casper)},\ }\href {https://doi.org/10.1103/PhysRevX.4.021030}
  {\bibfield  {journal} {\bibinfo  {journal} {Phys. Rev. X}\ }\textbf {\bibinfo
  {volume} {4}},\ \bibinfo {pages} {021030} (\bibinfo {year}
  {2014})}\BibitemShut {NoStop}%
\bibitem [{\citenamefont {Black}(2001)}]{Black_PoundDreverHall_2001}%
  \BibitemOpen
  \bibfield  {author} {\bibinfo {author} {\bibfnamefont {E.~D.}\ \bibnamefont
  {Black}},\ }\bibfield  {title} {\bibinfo {title} {{An introduction to
  Pound-Drever-Hall laser frequency stabilization}},\ }\bibfield  {journal}
  {\bibinfo  {journal} {Am. J. Phys.}\ }\href
  {https://doi.org/10.1119/1.1286663} {10.1119/1.1286663} (\bibinfo {year}
  {2001})\BibitemShut {NoStop}%
\bibitem [{\citenamefont {Co}\ \emph {et~al.}(2016)\citenamefont {Co},
  \citenamefont {D'Eramo},\ and\ \citenamefont {Hall}}]{Co_2016}%
  \BibitemOpen
  \bibfield  {author} {\bibinfo {author} {\bibfnamefont {R.~T.}\ \bibnamefont
  {Co}}, \bibinfo {author} {\bibfnamefont {F.}~\bibnamefont {D'Eramo}},\ and\
  \bibinfo {author} {\bibfnamefont {L.~J.}\ \bibnamefont {Hall}},\ }\bibfield
  {title} {\bibinfo {title} {Supersymmetric axion grand unified theories and
  their predictions},\ }\href {https://doi.org/10.1103/PhysRevD.94.075001}
  {\bibfield  {journal} {\bibinfo  {journal} {Phys. Rev. D}\ }\textbf {\bibinfo
  {volume} {94}},\ \bibinfo {pages} {075001} (\bibinfo {year}
  {2016})}\BibitemShut {NoStop}%
\bibitem [{\citenamefont {Di~Luzio}\ \emph {et~al.}(2018)\citenamefont
  {Di~Luzio}, \citenamefont {Ringwald},\ and\ \citenamefont
  {Tamarit}}]{PhysRevD.98.095011}%
  \BibitemOpen
  \bibfield  {author} {\bibinfo {author} {\bibfnamefont {L.}~\bibnamefont
  {Di~Luzio}}, \bibinfo {author} {\bibfnamefont {A.}~\bibnamefont {Ringwald}},\
  and\ \bibinfo {author} {\bibfnamefont {C.}~\bibnamefont {Tamarit}},\
  }\bibfield  {title} {\bibinfo {title} {Axion mass prediction from minimal
  grand unification},\ }\href {https://doi.org/10.1103/PhysRevD.98.095011}
  {\bibfield  {journal} {\bibinfo  {journal} {Phys. Rev. D}\ }\textbf {\bibinfo
  {volume} {98}},\ \bibinfo {pages} {095011} (\bibinfo {year}
  {2018})}\BibitemShut {NoStop}%
\bibitem [{\citenamefont {Kohri}\ and\ \citenamefont
  {Kodama}(2017)}]{Kohri_2017}%
  \BibitemOpen
  \bibfield  {author} {\bibinfo {author} {\bibfnamefont {K.}~\bibnamefont
  {Kohri}}\ and\ \bibinfo {author} {\bibfnamefont {H.}~\bibnamefont {Kodama}},\
  }\bibfield  {title} {\bibinfo {title} {Axion-like particles and recent
  observations of the cosmic infrared background radiation},\ }\href
  {https://doi.org/10.1103/PhysRevD.96.051701} {\bibfield  {journal} {\bibinfo
  {journal} {Phys. Rev. D}\ }\textbf {\bibinfo {volume} {96}},\ \bibinfo
  {pages} {051701} (\bibinfo {year} {2017})}\BibitemShut {NoStop}%
\bibitem [{\citenamefont {Valle}\ \emph {et~al.}(2014)\citenamefont {Valle},
  \citenamefont {Milotti}, \citenamefont {Ejlli}, \citenamefont {Gastaldi},
  \citenamefont {Messineo}, \citenamefont {Piemontese}, \citenamefont
  {Zavattini}, \citenamefont {Pengo},\ and\ \citenamefont
  {Ruoso}}]{DellaValle_longDecayCavity_2014}%
  \BibitemOpen
  \bibfield  {author} {\bibinfo {author} {\bibfnamefont {F.~D.}\ \bibnamefont
  {Valle}}, \bibinfo {author} {\bibfnamefont {E.}~\bibnamefont {Milotti}},
  \bibinfo {author} {\bibfnamefont {A.}~\bibnamefont {Ejlli}}, \bibinfo
  {author} {\bibfnamefont {U.}~\bibnamefont {Gastaldi}}, \bibinfo {author}
  {\bibfnamefont {G.}~\bibnamefont {Messineo}}, \bibinfo {author}
  {\bibfnamefont {L.}~\bibnamefont {Piemontese}}, \bibinfo {author}
  {\bibfnamefont {G.}~\bibnamefont {Zavattini}}, \bibinfo {author}
  {\bibfnamefont {R.}~\bibnamefont {Pengo}},\ and\ \bibinfo {author}
  {\bibfnamefont {G.}~\bibnamefont {Ruoso}},\ }\bibfield  {title} {\bibinfo
  {title} {Extremely long decay time optical cavity},\ }\href
  {https://doi.org/10.1364/OE.22.011570} {\bibfield  {journal} {\bibinfo
  {journal} {Opt. Express}\ }\textbf {\bibinfo {volume} {22}},\ \bibinfo
  {pages} {11570} (\bibinfo {year} {2014})}\BibitemShut {NoStop}%
\bibitem [{\citenamefont {Ajello}\ \emph {et~al.}(2016)\citenamefont {Ajello},
  \citenamefont {Albert}, \citenamefont {Anderson}, \citenamefont {Baldini},
  \citenamefont {Barbiellini}, \citenamefont {Bastieri}, \citenamefont
  {Bellazzini}, \citenamefont {Bissaldi}, \citenamefont {Blandford},
  \citenamefont {Bloom}, \citenamefont {Bonino}, \citenamefont {Bottacini},
  \citenamefont {Bregeon}, \citenamefont {Bruel}, \citenamefont {Buehler},
  \citenamefont {Caliandro}, \citenamefont {Cameron}, \citenamefont
  {Caragiulo}, \citenamefont {Caraveo}, \citenamefont {Cecchi}, \citenamefont
  {Chekhtman}, \citenamefont {Ciprini}, \citenamefont {Cohen-Tanugi},
  \citenamefont {Conrad}, \citenamefont {Costanza}, \citenamefont {D'Ammando},
  \citenamefont {de~Angelis}, \citenamefont {de~Palma}, \citenamefont
  {Desiante}, \citenamefont {Di~Mauro}, \citenamefont {Di~Venere},
  \citenamefont {Dom\'{\i}nguez}, \citenamefont {Drell}, \citenamefont
  {Favuzzi}, \citenamefont {Focke}, \citenamefont {Franckowiak}, \citenamefont
  {Fukazawa}, \citenamefont {Funk}, \citenamefont {Fusco}, \citenamefont
  {Gargano}, \citenamefont {Gasparrini}, \citenamefont {Giglietto},
  \citenamefont {Glanzman}, \citenamefont {Godfrey}, \citenamefont {Guiriec},
  \citenamefont {Horan}, \citenamefont {J\'ohannesson}, \citenamefont
  {Katsuragawa}, \citenamefont {Kensei}, \citenamefont {Kuss}, \citenamefont
  {Larsson}, \citenamefont {Latronico}, \citenamefont {Li}, \citenamefont {Li},
  \citenamefont {Longo}, \citenamefont {Loparco}, \citenamefont {Lubrano},
  \citenamefont {Madejski}, \citenamefont {Maldera}, \citenamefont {Manfreda},
  \citenamefont {Mayer}, \citenamefont {Mazziotta}, \citenamefont {Meyer},
  \citenamefont {Michelson}, \citenamefont {Mirabal}, \citenamefont {Mizuno},
  \citenamefont {Monzani}, \citenamefont {Morselli}, \citenamefont
  {Moskalenko}, \citenamefont {Murgia}, \citenamefont {Negro}, \citenamefont
  {Nuss}, \citenamefont {Okada}, \citenamefont {Orlando}, \citenamefont
  {Ormes}, \citenamefont {Paneque}, \citenamefont {Perkins}, \citenamefont
  {Pesce-Rollins}, \citenamefont {Piron}, \citenamefont {Pivato}, \citenamefont
  {Porter}, \citenamefont {Rain\`o}, \citenamefont {Rando}, \citenamefont
  {Razzano}, \citenamefont {Reimer}, \citenamefont {S\'anchez-Conde},
  \citenamefont {Sgr\`o}, \citenamefont {Simone}, \citenamefont {Siskind},
  \citenamefont {Spada}, \citenamefont {Spandre}, \citenamefont {Spinelli},
  \citenamefont {Takahashi}, \citenamefont {Thayer}, \citenamefont {Torres},
  \citenamefont {Tosti}, \citenamefont {Troja}, \citenamefont {Uchiyama},
  \citenamefont {Wood}, \citenamefont {Wood}, \citenamefont {Zaharijas},\ and\
  \citenamefont {Zimmer}}]{Fermi-LAT_NGC1275_2016}%
  \BibitemOpen
  \bibfield  {author} {\bibinfo {author} {\bibfnamefont {M.}~\bibnamefont
  {Ajello}}, \bibinfo {author} {\bibfnamefont {A.}~\bibnamefont {Albert}},
  \bibinfo {author} {\bibfnamefont {B.}~\bibnamefont {Anderson}}, \bibinfo
  {author} {\bibfnamefont {L.}~\bibnamefont {Baldini}}, \bibinfo {author}
  {\bibfnamefont {G.}~\bibnamefont {Barbiellini}}, \bibinfo {author}
  {\bibfnamefont {D.}~\bibnamefont {Bastieri}}, \bibinfo {author}
  {\bibfnamefont {R.}~\bibnamefont {Bellazzini}}, \bibinfo {author}
  {\bibfnamefont {E.}~\bibnamefont {Bissaldi}}, \bibinfo {author}
  {\bibfnamefont {R.~D.}\ \bibnamefont {Blandford}}, \bibinfo {author}
  {\bibfnamefont {E.~D.}\ \bibnamefont {Bloom}}, \bibinfo {author}
  {\bibfnamefont {R.}~\bibnamefont {Bonino}}, \bibinfo {author} {\bibfnamefont
  {E.}~\bibnamefont {Bottacini}}, \bibinfo {author} {\bibfnamefont
  {J.}~\bibnamefont {Bregeon}}, \bibinfo {author} {\bibfnamefont
  {P.}~\bibnamefont {Bruel}}, \bibinfo {author} {\bibfnamefont
  {R.}~\bibnamefont {Buehler}}, \bibinfo {author} {\bibfnamefont {G.~A.}\
  \bibnamefont {Caliandro}}, \bibinfo {author} {\bibfnamefont {R.~A.}\
  \bibnamefont {Cameron}}, \bibinfo {author} {\bibfnamefont {M.}~\bibnamefont
  {Caragiulo}}, \bibinfo {author} {\bibfnamefont {P.~A.}\ \bibnamefont
  {Caraveo}}, \bibinfo {author} {\bibfnamefont {C.}~\bibnamefont {Cecchi}},
  \bibinfo {author} {\bibfnamefont {A.}~\bibnamefont {Chekhtman}}, \bibinfo
  {author} {\bibfnamefont {S.}~\bibnamefont {Ciprini}}, \bibinfo {author}
  {\bibfnamefont {J.}~\bibnamefont {Cohen-Tanugi}}, \bibinfo {author}
  {\bibfnamefont {J.}~\bibnamefont {Conrad}}, \bibinfo {author} {\bibfnamefont
  {F.}~\bibnamefont {Costanza}}, \bibinfo {author} {\bibfnamefont
  {F.}~\bibnamefont {D'Ammando}}, \bibinfo {author} {\bibfnamefont
  {A.}~\bibnamefont {de~Angelis}}, \bibinfo {author} {\bibfnamefont
  {F.}~\bibnamefont {de~Palma}}, \bibinfo {author} {\bibfnamefont
  {R.}~\bibnamefont {Desiante}}, \bibinfo {author} {\bibfnamefont
  {M.}~\bibnamefont {Di~Mauro}}, \bibinfo {author} {\bibfnamefont
  {L.}~\bibnamefont {Di~Venere}}, \bibinfo {author} {\bibfnamefont
  {A.}~\bibnamefont {Dom\'{\i}nguez}}, \bibinfo {author} {\bibfnamefont
  {P.~S.}\ \bibnamefont {Drell}}, \bibinfo {author} {\bibfnamefont
  {C.}~\bibnamefont {Favuzzi}}, \bibinfo {author} {\bibfnamefont {W.~B.}\
  \bibnamefont {Focke}}, \bibinfo {author} {\bibfnamefont {A.}~\bibnamefont
  {Franckowiak}}, \bibinfo {author} {\bibfnamefont {Y.}~\bibnamefont
  {Fukazawa}}, \bibinfo {author} {\bibfnamefont {S.}~\bibnamefont {Funk}},
  \bibinfo {author} {\bibfnamefont {P.}~\bibnamefont {Fusco}}, \bibinfo
  {author} {\bibfnamefont {F.}~\bibnamefont {Gargano}}, \bibinfo {author}
  {\bibfnamefont {D.}~\bibnamefont {Gasparrini}}, \bibinfo {author}
  {\bibfnamefont {N.}~\bibnamefont {Giglietto}}, \bibinfo {author}
  {\bibfnamefont {T.}~\bibnamefont {Glanzman}}, \bibinfo {author}
  {\bibfnamefont {G.}~\bibnamefont {Godfrey}}, \bibinfo {author} {\bibfnamefont
  {S.}~\bibnamefont {Guiriec}}, \bibinfo {author} {\bibfnamefont
  {D.}~\bibnamefont {Horan}}, \bibinfo {author} {\bibfnamefont
  {G.}~\bibnamefont {J\'ohannesson}}, \bibinfo {author} {\bibfnamefont
  {M.}~\bibnamefont {Katsuragawa}}, \bibinfo {author} {\bibfnamefont
  {S.}~\bibnamefont {Kensei}}, \bibinfo {author} {\bibfnamefont
  {M.}~\bibnamefont {Kuss}}, \bibinfo {author} {\bibfnamefont {S.}~\bibnamefont
  {Larsson}}, \bibinfo {author} {\bibfnamefont {L.}~\bibnamefont {Latronico}},
  \bibinfo {author} {\bibfnamefont {J.}~\bibnamefont {Li}}, \bibinfo {author}
  {\bibfnamefont {L.}~\bibnamefont {Li}}, \bibinfo {author} {\bibfnamefont
  {F.}~\bibnamefont {Longo}}, \bibinfo {author} {\bibfnamefont
  {F.}~\bibnamefont {Loparco}}, \bibinfo {author} {\bibfnamefont
  {P.}~\bibnamefont {Lubrano}}, \bibinfo {author} {\bibfnamefont {G.~M.}\
  \bibnamefont {Madejski}}, \bibinfo {author} {\bibfnamefont {S.}~\bibnamefont
  {Maldera}}, \bibinfo {author} {\bibfnamefont {A.}~\bibnamefont {Manfreda}},
  \bibinfo {author} {\bibfnamefont {M.}~\bibnamefont {Mayer}}, \bibinfo
  {author} {\bibfnamefont {M.~N.}\ \bibnamefont {Mazziotta}}, \bibinfo {author}
  {\bibfnamefont {M.}~\bibnamefont {Meyer}}, \bibinfo {author} {\bibfnamefont
  {P.~F.}\ \bibnamefont {Michelson}}, \bibinfo {author} {\bibfnamefont
  {N.}~\bibnamefont {Mirabal}}, \bibinfo {author} {\bibfnamefont
  {T.}~\bibnamefont {Mizuno}}, \bibinfo {author} {\bibfnamefont {M.~E.}\
  \bibnamefont {Monzani}}, \bibinfo {author} {\bibfnamefont {A.}~\bibnamefont
  {Morselli}}, \bibinfo {author} {\bibfnamefont {I.~V.}\ \bibnamefont
  {Moskalenko}}, \bibinfo {author} {\bibfnamefont {S.}~\bibnamefont {Murgia}},
  \bibinfo {author} {\bibfnamefont {M.}~\bibnamefont {Negro}}, \bibinfo
  {author} {\bibfnamefont {E.}~\bibnamefont {Nuss}}, \bibinfo {author}
  {\bibfnamefont {C.}~\bibnamefont {Okada}}, \bibinfo {author} {\bibfnamefont
  {E.}~\bibnamefont {Orlando}}, \bibinfo {author} {\bibfnamefont {J.~F.}\
  \bibnamefont {Ormes}}, \bibinfo {author} {\bibfnamefont {D.}~\bibnamefont
  {Paneque}}, \bibinfo {author} {\bibfnamefont {J.~S.}\ \bibnamefont
  {Perkins}}, \bibinfo {author} {\bibfnamefont {M.}~\bibnamefont
  {Pesce-Rollins}}, \bibinfo {author} {\bibfnamefont {F.}~\bibnamefont
  {Piron}}, \bibinfo {author} {\bibfnamefont {G.}~\bibnamefont {Pivato}},
  \bibinfo {author} {\bibfnamefont {T.~A.}\ \bibnamefont {Porter}}, \bibinfo
  {author} {\bibfnamefont {S.}~\bibnamefont {Rain\`o}}, \bibinfo {author}
  {\bibfnamefont {R.}~\bibnamefont {Rando}}, \bibinfo {author} {\bibfnamefont
  {M.}~\bibnamefont {Razzano}}, \bibinfo {author} {\bibfnamefont
  {A.}~\bibnamefont {Reimer}}, \bibinfo {author} {\bibfnamefont
  {M.}~\bibnamefont {S\'anchez-Conde}}, \bibinfo {author} {\bibfnamefont
  {C.}~\bibnamefont {Sgr\`o}}, \bibinfo {author} {\bibfnamefont
  {D.}~\bibnamefont {Simone}}, \bibinfo {author} {\bibfnamefont {E.~J.}\
  \bibnamefont {Siskind}}, \bibinfo {author} {\bibfnamefont {F.}~\bibnamefont
  {Spada}}, \bibinfo {author} {\bibfnamefont {G.}~\bibnamefont {Spandre}},
  \bibinfo {author} {\bibfnamefont {P.}~\bibnamefont {Spinelli}}, \bibinfo
  {author} {\bibfnamefont {H.}~\bibnamefont {Takahashi}}, \bibinfo {author}
  {\bibfnamefont {J.~B.}\ \bibnamefont {Thayer}}, \bibinfo {author}
  {\bibfnamefont {D.~F.}\ \bibnamefont {Torres}}, \bibinfo {author}
  {\bibfnamefont {G.}~\bibnamefont {Tosti}}, \bibinfo {author} {\bibfnamefont
  {E.}~\bibnamefont {Troja}}, \bibinfo {author} {\bibfnamefont
  {Y.}~\bibnamefont {Uchiyama}}, \bibinfo {author} {\bibfnamefont {K.~S.}\
  \bibnamefont {Wood}}, \bibinfo {author} {\bibfnamefont {M.}~\bibnamefont
  {Wood}}, \bibinfo {author} {\bibfnamefont {G.}~\bibnamefont {Zaharijas}},\
  and\ \bibinfo {author} {\bibfnamefont {S.}~\bibnamefont {Zimmer}} (\bibinfo
  {collaboration} {The Fermi-LAT Collaboration}),\ }\bibfield  {title}
  {\bibinfo {title} {Search for spectral irregularities due to
  photon--axionlike-particle oscillations with the fermi large area
  telescope},\ }\href {https://doi.org/10.1103/PhysRevLett.116.161101}
  {\bibfield  {journal} {\bibinfo  {journal} {Phys. Rev. Lett.}\ }\textbf
  {\bibinfo {volume} {116}},\ \bibinfo {pages} {161101} (\bibinfo {year}
  {2016})}\BibitemShut {NoStop}%
\bibitem [{\citenamefont {Dessert}\ \emph {et~al.}(2022)\citenamefont
  {Dessert}, \citenamefont {Dunsky},\ and\ \citenamefont
  {Safdi}}]{MWDpolarisation_2022}%
  \BibitemOpen
  \bibfield  {author} {\bibinfo {author} {\bibfnamefont {C.}~\bibnamefont
  {Dessert}}, \bibinfo {author} {\bibfnamefont {D.}~\bibnamefont {Dunsky}},\
  and\ \bibinfo {author} {\bibfnamefont {B.~R.}\ \bibnamefont {Safdi}},\
  }\bibfield  {title} {\bibinfo {title} {Upper limit on the axion-photon
  coupling from magnetic white dwarf polarization},\ }\href
  {https://doi.org/10.1103/PhysRevD.105.103034} {\bibfield  {journal} {\bibinfo
   {journal} {Phys. Rev. D}\ }\textbf {\bibinfo {volume} {105}},\ \bibinfo
  {pages} {103034} (\bibinfo {year} {2022})}\BibitemShut {NoStop}%
\bibitem [{\citenamefont {{The LIGO Scientific
  Collaboration}}(2013)}]{LSC_SQUEEZING_2013}%
  \BibitemOpen
  \bibfield  {author} {\bibinfo {author} {\bibnamefont {{The LIGO Scientific
  Collaboration}}},\ }\bibfield  {title} {\bibinfo {title} {Enhanced
  sensitivity of the {LIGO} gravitational wave detector by using squeezed
  states of light},\ }\href {https://doi.org/10.1038/nphoton.2013.177}
  {\bibfield  {journal} {\bibinfo  {journal} {Nature Photonics}\ }\textbf
  {\bibinfo {volume} {7}},\ \bibinfo {pages} {613} (\bibinfo {year}
  {2013})}\BibitemShut {NoStop}%
\bibitem [{\citenamefont {Buikema}\ \emph {et~al.}(2020)\citenamefont
  {Buikema}, \citenamefont {Cahillane}, \citenamefont {Mansell}, \citenamefont
  {Blair}, \citenamefont {Abbott}, \citenamefont {Adams}, \citenamefont
  {Adhikari}, \citenamefont {Ananyeva}, \citenamefont {Appert}, \citenamefont
  {Arai}, \citenamefont {Areeda}, \citenamefont {Asali}, \citenamefont {Aston},
  \citenamefont {Austin}, \citenamefont {Baer}, \citenamefont {Ball},
  \citenamefont {Ballmer}, \citenamefont {Banagiri}, \citenamefont {Barker},
  \citenamefont {Barsotti}, \citenamefont {Bartlett}, \citenamefont {Berger},
  \citenamefont {Betzwieser}, \citenamefont {Bhattacharjee}, \citenamefont
  {Billingsley}, \citenamefont {Biscans}, \citenamefont {Blair}, \citenamefont
  {Bode}, \citenamefont {Booker}, \citenamefont {Bork}, \citenamefont
  {Bramley}, \citenamefont {Brooks}, \citenamefont {Brown}, \citenamefont
  {Cannon}, \citenamefont {Chen}, \citenamefont {Ciobanu}, \citenamefont
  {Clara}, \citenamefont {Cooper}, \citenamefont {Corley}, \citenamefont
  {Countryman}, \citenamefont {Covas}, \citenamefont {Coyne}, \citenamefont
  {Datrier}, \citenamefont {Davis}, \citenamefont {Di~Fronzo}, \citenamefont
  {Dooley}, \citenamefont {Driggers}, \citenamefont {Dupej}, \citenamefont
  {Dwyer}, \citenamefont {Effler}, \citenamefont {Etzel}, \citenamefont
  {Evans}, \citenamefont {Evans}, \citenamefont {Feicht}, \citenamefont
  {Fernandez-Galiana}, \citenamefont {Fritschel}, \citenamefont {Frolov},
  \citenamefont {Fulda}, \citenamefont {Fyffe}, \citenamefont {Giaime},
  \citenamefont {Giardina}, \citenamefont {Godwin}, \citenamefont {Goetz},
  \citenamefont {Gras}, \citenamefont {Gray}, \citenamefont {Gray},
  \citenamefont {Green}, \citenamefont {Gustafson}, \citenamefont {Gustafson},
  \citenamefont {Hanks}, \citenamefont {Hanson}, \citenamefont {Hardwick},
  \citenamefont {Hasskew}, \citenamefont {Heintze}, \citenamefont
  {Helmling-Cornell}, \citenamefont {Holland}, \citenamefont {Jones},
  \citenamefont {Kandhasamy}, \citenamefont {Karki}, \citenamefont {Kasprzack},
  \citenamefont {Kawabe}, \citenamefont {Kijbunchoo}, \citenamefont {King},
  \citenamefont {Kissel}, \citenamefont {Kumar}, \citenamefont {Landry},
  \citenamefont {Lane}, \citenamefont {Lantz}, \citenamefont {Laxen},
  \citenamefont {Lecoeuche}, \citenamefont {Leviton}, \citenamefont {Liu},
  \citenamefont {Lormand}, \citenamefont {Lundgren}, \citenamefont {Macas},
  \citenamefont {MacInnis}, \citenamefont {Macleod}, \citenamefont {M\'arka},
  \citenamefont {M\'arka}, \citenamefont {Martynov}, \citenamefont {Mason},
  \citenamefont {Massinger}, \citenamefont {Matichard}, \citenamefont
  {Mavalvala}, \citenamefont {McCarthy}, \citenamefont {McClelland},
  \citenamefont {McCormick}, \citenamefont {McCuller}, \citenamefont {McIver},
  \citenamefont {McRae}, \citenamefont {Mendell}, \citenamefont {Merfeld},
  \citenamefont {Merilh}, \citenamefont {Meylahn}, \citenamefont {Mistry},
  \citenamefont {Mittleman}, \citenamefont {Moreno}, \citenamefont {Mow-Lowry},
  \citenamefont {Mozzon}, \citenamefont {Mullavey}, \citenamefont {Nelson},
  \citenamefont {Nguyen}, \citenamefont {Nuttall}, \citenamefont {Oberling},
  \citenamefont {Oram}, \citenamefont {O'Reilly}, \citenamefont {Osthelder},
  \citenamefont {Ottaway}, \citenamefont {Overmier}, \citenamefont {Palamos},
  \citenamefont {Parker}, \citenamefont {Payne}, \citenamefont {Pele},
  \citenamefont {Penhorwood}, \citenamefont {Perez}, \citenamefont {Pirello},
  \citenamefont {Radkins}, \citenamefont {Ramirez}, \citenamefont {Richardson},
  \citenamefont {Riles}, \citenamefont {Robertson}, \citenamefont {Rollins},
  \citenamefont {Romel}, \citenamefont {Romie}, \citenamefont {Ross},
  \citenamefont {Ryan}, \citenamefont {Sadecki}, \citenamefont {Sanchez},
  \citenamefont {Sanchez}, \citenamefont {Saravanan}, \citenamefont {Savage},
  \citenamefont {Schaetzl}, \citenamefont {Schnabel}, \citenamefont
  {Schofield}, \citenamefont {Schwartz}, \citenamefont {Sellers}, \citenamefont
  {Shaffer}, \citenamefont {Sigg}, \citenamefont {Slagmolen}, \citenamefont
  {Smith}, \citenamefont {Soni}, \citenamefont {Sorazu}, \citenamefont
  {Spencer}, \citenamefont {Strain}, \citenamefont {Sun}, \citenamefont
  {Szczepa\ifmmode~\acute{n}\else \'{n}\fi{}czyk}, \citenamefont {Thomas},
  \citenamefont {Thomas}, \citenamefont {Thorne}, \citenamefont {Toland},
  \citenamefont {Torrie}, \citenamefont {Traylor}, \citenamefont {Tse},
  \citenamefont {Urban}, \citenamefont {Vajente}, \citenamefont {Valdes},
  \citenamefont {Vander-Hyde}, \citenamefont {Veitch}, \citenamefont
  {Venkateswara}, \citenamefont {Venugopalan}, \citenamefont {Viets},
  \citenamefont {Vo}, \citenamefont {Vorvick}, \citenamefont {Wade},
  \citenamefont {Ward}, \citenamefont {Warner}, \citenamefont {Weaver},
  \citenamefont {Weiss}, \citenamefont {Whittle}, \citenamefont {Willke},
  \citenamefont {Wipf}, \citenamefont {Xiao}, \citenamefont {Yamamoto},
  \citenamefont {Yu}, \citenamefont {Yu}, \citenamefont {Zhang}, \citenamefont
  {Zucker},\ and\ \citenamefont {Zweizig}}]{Buikema_performanceOfLIGO_2020}%
  \BibitemOpen
  \bibfield  {author} {\bibinfo {author} {\bibfnamefont {A.}~\bibnamefont
  {Buikema}}, \bibinfo {author} {\bibfnamefont {C.}~\bibnamefont {Cahillane}},
  \bibinfo {author} {\bibfnamefont {G.~L.}\ \bibnamefont {Mansell}}, \bibinfo
  {author} {\bibfnamefont {C.~D.}\ \bibnamefont {Blair}}, \bibinfo {author}
  {\bibfnamefont {R.}~\bibnamefont {Abbott}}, \bibinfo {author} {\bibfnamefont
  {C.}~\bibnamefont {Adams}}, \bibinfo {author} {\bibfnamefont {R.~X.}\
  \bibnamefont {Adhikari}}, \bibinfo {author} {\bibfnamefont {A.}~\bibnamefont
  {Ananyeva}}, \bibinfo {author} {\bibfnamefont {S.}~\bibnamefont {Appert}},
  \bibinfo {author} {\bibfnamefont {K.}~\bibnamefont {Arai}}, \bibinfo {author}
  {\bibfnamefont {J.~S.}\ \bibnamefont {Areeda}}, \bibinfo {author}
  {\bibfnamefont {Y.}~\bibnamefont {Asali}}, \bibinfo {author} {\bibfnamefont
  {S.~M.}\ \bibnamefont {Aston}}, \bibinfo {author} {\bibfnamefont
  {C.}~\bibnamefont {Austin}}, \bibinfo {author} {\bibfnamefont {A.~M.}\
  \bibnamefont {Baer}}, \bibinfo {author} {\bibfnamefont {M.}~\bibnamefont
  {Ball}}, \bibinfo {author} {\bibfnamefont {S.~W.}\ \bibnamefont {Ballmer}},
  \bibinfo {author} {\bibfnamefont {S.}~\bibnamefont {Banagiri}}, \bibinfo
  {author} {\bibfnamefont {D.}~\bibnamefont {Barker}}, \bibinfo {author}
  {\bibfnamefont {L.}~\bibnamefont {Barsotti}}, \bibinfo {author}
  {\bibfnamefont {J.}~\bibnamefont {Bartlett}}, \bibinfo {author}
  {\bibfnamefont {B.~K.}\ \bibnamefont {Berger}}, \bibinfo {author}
  {\bibfnamefont {J.}~\bibnamefont {Betzwieser}}, \bibinfo {author}
  {\bibfnamefont {D.}~\bibnamefont {Bhattacharjee}}, \bibinfo {author}
  {\bibfnamefont {G.}~\bibnamefont {Billingsley}}, \bibinfo {author}
  {\bibfnamefont {S.}~\bibnamefont {Biscans}}, \bibinfo {author} {\bibfnamefont
  {R.~M.}\ \bibnamefont {Blair}}, \bibinfo {author} {\bibfnamefont
  {N.}~\bibnamefont {Bode}}, \bibinfo {author} {\bibfnamefont {P.}~\bibnamefont
  {Booker}}, \bibinfo {author} {\bibfnamefont {R.}~\bibnamefont {Bork}},
  \bibinfo {author} {\bibfnamefont {A.}~\bibnamefont {Bramley}}, \bibinfo
  {author} {\bibfnamefont {A.~F.}\ \bibnamefont {Brooks}}, \bibinfo {author}
  {\bibfnamefont {D.~D.}\ \bibnamefont {Brown}}, \bibinfo {author}
  {\bibfnamefont {K.~C.}\ \bibnamefont {Cannon}}, \bibinfo {author}
  {\bibfnamefont {X.}~\bibnamefont {Chen}}, \bibinfo {author} {\bibfnamefont
  {A.~A.}\ \bibnamefont {Ciobanu}}, \bibinfo {author} {\bibfnamefont
  {F.}~\bibnamefont {Clara}}, \bibinfo {author} {\bibfnamefont {S.~J.}\
  \bibnamefont {Cooper}}, \bibinfo {author} {\bibfnamefont {K.~R.}\
  \bibnamefont {Corley}}, \bibinfo {author} {\bibfnamefont {S.~T.}\
  \bibnamefont {Countryman}}, \bibinfo {author} {\bibfnamefont {P.~B.}\
  \bibnamefont {Covas}}, \bibinfo {author} {\bibfnamefont {D.~C.}\ \bibnamefont
  {Coyne}}, \bibinfo {author} {\bibfnamefont {L.~E.~H.}\ \bibnamefont
  {Datrier}}, \bibinfo {author} {\bibfnamefont {D.}~\bibnamefont {Davis}},
  \bibinfo {author} {\bibfnamefont {C.}~\bibnamefont {Di~Fronzo}}, \bibinfo
  {author} {\bibfnamefont {K.~L.}\ \bibnamefont {Dooley}}, \bibinfo {author}
  {\bibfnamefont {J.~C.}\ \bibnamefont {Driggers}}, \bibinfo {author}
  {\bibfnamefont {P.}~\bibnamefont {Dupej}}, \bibinfo {author} {\bibfnamefont
  {S.~E.}\ \bibnamefont {Dwyer}}, \bibinfo {author} {\bibfnamefont
  {A.}~\bibnamefont {Effler}}, \bibinfo {author} {\bibfnamefont
  {T.}~\bibnamefont {Etzel}}, \bibinfo {author} {\bibfnamefont
  {M.}~\bibnamefont {Evans}}, \bibinfo {author} {\bibfnamefont {T.~M.}\
  \bibnamefont {Evans}}, \bibinfo {author} {\bibfnamefont {J.}~\bibnamefont
  {Feicht}}, \bibinfo {author} {\bibfnamefont {A.}~\bibnamefont
  {Fernandez-Galiana}}, \bibinfo {author} {\bibfnamefont {P.}~\bibnamefont
  {Fritschel}}, \bibinfo {author} {\bibfnamefont {V.~V.}\ \bibnamefont
  {Frolov}}, \bibinfo {author} {\bibfnamefont {P.}~\bibnamefont {Fulda}},
  \bibinfo {author} {\bibfnamefont {M.}~\bibnamefont {Fyffe}}, \bibinfo
  {author} {\bibfnamefont {J.~A.}\ \bibnamefont {Giaime}}, \bibinfo {author}
  {\bibfnamefont {K.~D.}\ \bibnamefont {Giardina}}, \bibinfo {author}
  {\bibfnamefont {P.}~\bibnamefont {Godwin}}, \bibinfo {author} {\bibfnamefont
  {E.}~\bibnamefont {Goetz}}, \bibinfo {author} {\bibfnamefont
  {S.}~\bibnamefont {Gras}}, \bibinfo {author} {\bibfnamefont {C.}~\bibnamefont
  {Gray}}, \bibinfo {author} {\bibfnamefont {R.}~\bibnamefont {Gray}}, \bibinfo
  {author} {\bibfnamefont {A.~C.}\ \bibnamefont {Green}}, \bibinfo {author}
  {\bibfnamefont {E.~K.}\ \bibnamefont {Gustafson}}, \bibinfo {author}
  {\bibfnamefont {R.}~\bibnamefont {Gustafson}}, \bibinfo {author}
  {\bibfnamefont {J.}~\bibnamefont {Hanks}}, \bibinfo {author} {\bibfnamefont
  {J.}~\bibnamefont {Hanson}}, \bibinfo {author} {\bibfnamefont
  {T.}~\bibnamefont {Hardwick}}, \bibinfo {author} {\bibfnamefont {R.~K.}\
  \bibnamefont {Hasskew}}, \bibinfo {author} {\bibfnamefont {M.~C.}\
  \bibnamefont {Heintze}}, \bibinfo {author} {\bibfnamefont {A.~F.}\
  \bibnamefont {Helmling-Cornell}}, \bibinfo {author} {\bibfnamefont {N.~A.}\
  \bibnamefont {Holland}}, \bibinfo {author} {\bibfnamefont {J.~D.}\
  \bibnamefont {Jones}}, \bibinfo {author} {\bibfnamefont {S.}~\bibnamefont
  {Kandhasamy}}, \bibinfo {author} {\bibfnamefont {S.}~\bibnamefont {Karki}},
  \bibinfo {author} {\bibfnamefont {M.}~\bibnamefont {Kasprzack}}, \bibinfo
  {author} {\bibfnamefont {K.}~\bibnamefont {Kawabe}}, \bibinfo {author}
  {\bibfnamefont {N.}~\bibnamefont {Kijbunchoo}}, \bibinfo {author}
  {\bibfnamefont {P.~J.}\ \bibnamefont {King}}, \bibinfo {author}
  {\bibfnamefont {J.~S.}\ \bibnamefont {Kissel}}, \bibinfo {author}
  {\bibfnamefont {R.}~\bibnamefont {Kumar}}, \bibinfo {author} {\bibfnamefont
  {M.}~\bibnamefont {Landry}}, \bibinfo {author} {\bibfnamefont {B.~B.}\
  \bibnamefont {Lane}}, \bibinfo {author} {\bibfnamefont {B.}~\bibnamefont
  {Lantz}}, \bibinfo {author} {\bibfnamefont {M.}~\bibnamefont {Laxen}},
  \bibinfo {author} {\bibfnamefont {Y.~K.}\ \bibnamefont {Lecoeuche}}, \bibinfo
  {author} {\bibfnamefont {J.}~\bibnamefont {Leviton}}, \bibinfo {author}
  {\bibfnamefont {J.}~\bibnamefont {Liu}}, \bibinfo {author} {\bibfnamefont
  {M.}~\bibnamefont {Lormand}}, \bibinfo {author} {\bibfnamefont {A.~P.}\
  \bibnamefont {Lundgren}}, \bibinfo {author} {\bibfnamefont {R.}~\bibnamefont
  {Macas}}, \bibinfo {author} {\bibfnamefont {M.}~\bibnamefont {MacInnis}},
  \bibinfo {author} {\bibfnamefont {D.~M.}\ \bibnamefont {Macleod}}, \bibinfo
  {author} {\bibfnamefont {S.}~\bibnamefont {M\'arka}}, \bibinfo {author}
  {\bibfnamefont {Z.}~\bibnamefont {M\'arka}}, \bibinfo {author} {\bibfnamefont
  {D.~V.}\ \bibnamefont {Martynov}}, \bibinfo {author} {\bibfnamefont
  {K.}~\bibnamefont {Mason}}, \bibinfo {author} {\bibfnamefont {T.~J.}\
  \bibnamefont {Massinger}}, \bibinfo {author} {\bibfnamefont {F.}~\bibnamefont
  {Matichard}}, \bibinfo {author} {\bibfnamefont {N.}~\bibnamefont
  {Mavalvala}}, \bibinfo {author} {\bibfnamefont {R.}~\bibnamefont {McCarthy}},
  \bibinfo {author} {\bibfnamefont {D.~E.}\ \bibnamefont {McClelland}},
  \bibinfo {author} {\bibfnamefont {S.}~\bibnamefont {McCormick}}, \bibinfo
  {author} {\bibfnamefont {L.}~\bibnamefont {McCuller}}, \bibinfo {author}
  {\bibfnamefont {J.}~\bibnamefont {McIver}}, \bibinfo {author} {\bibfnamefont
  {T.}~\bibnamefont {McRae}}, \bibinfo {author} {\bibfnamefont
  {G.}~\bibnamefont {Mendell}}, \bibinfo {author} {\bibfnamefont
  {K.}~\bibnamefont {Merfeld}}, \bibinfo {author} {\bibfnamefont {E.~L.}\
  \bibnamefont {Merilh}}, \bibinfo {author} {\bibfnamefont {F.}~\bibnamefont
  {Meylahn}}, \bibinfo {author} {\bibfnamefont {T.}~\bibnamefont {Mistry}},
  \bibinfo {author} {\bibfnamefont {R.}~\bibnamefont {Mittleman}}, \bibinfo
  {author} {\bibfnamefont {G.}~\bibnamefont {Moreno}}, \bibinfo {author}
  {\bibfnamefont {C.~M.}\ \bibnamefont {Mow-Lowry}}, \bibinfo {author}
  {\bibfnamefont {S.}~\bibnamefont {Mozzon}}, \bibinfo {author} {\bibfnamefont
  {A.}~\bibnamefont {Mullavey}}, \bibinfo {author} {\bibfnamefont {T.~J.~N.}\
  \bibnamefont {Nelson}}, \bibinfo {author} {\bibfnamefont {P.}~\bibnamefont
  {Nguyen}}, \bibinfo {author} {\bibfnamefont {L.~K.}\ \bibnamefont {Nuttall}},
  \bibinfo {author} {\bibfnamefont {J.}~\bibnamefont {Oberling}}, \bibinfo
  {author} {\bibfnamefont {R.~J.}\ \bibnamefont {Oram}}, \bibinfo {author}
  {\bibfnamefont {B.}~\bibnamefont {O'Reilly}}, \bibinfo {author}
  {\bibfnamefont {C.}~\bibnamefont {Osthelder}}, \bibinfo {author}
  {\bibfnamefont {D.~J.}\ \bibnamefont {Ottaway}}, \bibinfo {author}
  {\bibfnamefont {H.}~\bibnamefont {Overmier}}, \bibinfo {author}
  {\bibfnamefont {J.~R.}\ \bibnamefont {Palamos}}, \bibinfo {author}
  {\bibfnamefont {W.}~\bibnamefont {Parker}}, \bibinfo {author} {\bibfnamefont
  {E.}~\bibnamefont {Payne}}, \bibinfo {author} {\bibfnamefont
  {A.}~\bibnamefont {Pele}}, \bibinfo {author} {\bibfnamefont {R.}~\bibnamefont
  {Penhorwood}}, \bibinfo {author} {\bibfnamefont {C.~J.}\ \bibnamefont
  {Perez}}, \bibinfo {author} {\bibfnamefont {M.}~\bibnamefont {Pirello}},
  \bibinfo {author} {\bibfnamefont {H.}~\bibnamefont {Radkins}}, \bibinfo
  {author} {\bibfnamefont {K.~E.}\ \bibnamefont {Ramirez}}, \bibinfo {author}
  {\bibfnamefont {J.~W.}\ \bibnamefont {Richardson}}, \bibinfo {author}
  {\bibfnamefont {K.}~\bibnamefont {Riles}}, \bibinfo {author} {\bibfnamefont
  {N.~A.}\ \bibnamefont {Robertson}}, \bibinfo {author} {\bibfnamefont {J.~G.}\
  \bibnamefont {Rollins}}, \bibinfo {author} {\bibfnamefont {C.~L.}\
  \bibnamefont {Romel}}, \bibinfo {author} {\bibfnamefont {J.~H.}\ \bibnamefont
  {Romie}}, \bibinfo {author} {\bibfnamefont {M.~P.}\ \bibnamefont {Ross}},
  \bibinfo {author} {\bibfnamefont {K.}~\bibnamefont {Ryan}}, \bibinfo {author}
  {\bibfnamefont {T.}~\bibnamefont {Sadecki}}, \bibinfo {author} {\bibfnamefont
  {E.~J.}\ \bibnamefont {Sanchez}}, \bibinfo {author} {\bibfnamefont {L.~E.}\
  \bibnamefont {Sanchez}}, \bibinfo {author} {\bibfnamefont {T.~R.}\
  \bibnamefont {Saravanan}}, \bibinfo {author} {\bibfnamefont {R.~L.}\
  \bibnamefont {Savage}}, \bibinfo {author} {\bibfnamefont {D.}~\bibnamefont
  {Schaetzl}}, \bibinfo {author} {\bibfnamefont {R.}~\bibnamefont {Schnabel}},
  \bibinfo {author} {\bibfnamefont {R.~M.~S.}\ \bibnamefont {Schofield}},
  \bibinfo {author} {\bibfnamefont {E.}~\bibnamefont {Schwartz}}, \bibinfo
  {author} {\bibfnamefont {D.}~\bibnamefont {Sellers}}, \bibinfo {author}
  {\bibfnamefont {T.}~\bibnamefont {Shaffer}}, \bibinfo {author} {\bibfnamefont
  {D.}~\bibnamefont {Sigg}}, \bibinfo {author} {\bibfnamefont {B.~J.~J.}\
  \bibnamefont {Slagmolen}}, \bibinfo {author} {\bibfnamefont {J.~R.}\
  \bibnamefont {Smith}}, \bibinfo {author} {\bibfnamefont {S.}~\bibnamefont
  {Soni}}, \bibinfo {author} {\bibfnamefont {B.}~\bibnamefont {Sorazu}},
  \bibinfo {author} {\bibfnamefont {A.~P.}\ \bibnamefont {Spencer}}, \bibinfo
  {author} {\bibfnamefont {K.~A.}\ \bibnamefont {Strain}}, \bibinfo {author}
  {\bibfnamefont {L.}~\bibnamefont {Sun}}, \bibinfo {author} {\bibfnamefont
  {M.~J.}\ \bibnamefont {Szczepa\ifmmode~\acute{n}\else \'{n}\fi{}czyk}},
  \bibinfo {author} {\bibfnamefont {M.}~\bibnamefont {Thomas}}, \bibinfo
  {author} {\bibfnamefont {P.}~\bibnamefont {Thomas}}, \bibinfo {author}
  {\bibfnamefont {K.~A.}\ \bibnamefont {Thorne}}, \bibinfo {author}
  {\bibfnamefont {K.}~\bibnamefont {Toland}}, \bibinfo {author} {\bibfnamefont
  {C.~I.}\ \bibnamefont {Torrie}}, \bibinfo {author} {\bibfnamefont
  {G.}~\bibnamefont {Traylor}}, \bibinfo {author} {\bibfnamefont
  {M.}~\bibnamefont {Tse}}, \bibinfo {author} {\bibfnamefont {A.~L.}\
  \bibnamefont {Urban}}, \bibinfo {author} {\bibfnamefont {G.}~\bibnamefont
  {Vajente}}, \bibinfo {author} {\bibfnamefont {G.}~\bibnamefont {Valdes}},
  \bibinfo {author} {\bibfnamefont {D.~C.}\ \bibnamefont {Vander-Hyde}},
  \bibinfo {author} {\bibfnamefont {P.~J.}\ \bibnamefont {Veitch}}, \bibinfo
  {author} {\bibfnamefont {K.}~\bibnamefont {Venkateswara}}, \bibinfo {author}
  {\bibfnamefont {G.}~\bibnamefont {Venugopalan}}, \bibinfo {author}
  {\bibfnamefont {A.~D.}\ \bibnamefont {Viets}}, \bibinfo {author}
  {\bibfnamefont {T.}~\bibnamefont {Vo}}, \bibinfo {author} {\bibfnamefont
  {C.}~\bibnamefont {Vorvick}}, \bibinfo {author} {\bibfnamefont
  {M.}~\bibnamefont {Wade}}, \bibinfo {author} {\bibfnamefont {R.~L.}\
  \bibnamefont {Ward}}, \bibinfo {author} {\bibfnamefont {J.}~\bibnamefont
  {Warner}}, \bibinfo {author} {\bibfnamefont {B.}~\bibnamefont {Weaver}},
  \bibinfo {author} {\bibfnamefont {R.}~\bibnamefont {Weiss}}, \bibinfo
  {author} {\bibfnamefont {C.}~\bibnamefont {Whittle}}, \bibinfo {author}
  {\bibfnamefont {B.}~\bibnamefont {Willke}}, \bibinfo {author} {\bibfnamefont
  {C.~C.}\ \bibnamefont {Wipf}}, \bibinfo {author} {\bibfnamefont
  {L.}~\bibnamefont {Xiao}}, \bibinfo {author} {\bibfnamefont {H.}~\bibnamefont
  {Yamamoto}}, \bibinfo {author} {\bibfnamefont {H.}~\bibnamefont {Yu}},
  \bibinfo {author} {\bibfnamefont {H.}~\bibnamefont {Yu}}, \bibinfo {author}
  {\bibfnamefont {L.}~\bibnamefont {Zhang}}, \bibinfo {author} {\bibfnamefont
  {M.~E.}\ \bibnamefont {Zucker}},\ and\ \bibinfo {author} {\bibfnamefont
  {J.}~\bibnamefont {Zweizig}},\ }\bibfield  {title} {\bibinfo {title}
  {Sensitivity and performance of the {A}dvanced {LIGO} detectors in the third
  observing run},\ }\href {https://doi.org/10.1103/PhysRevD.102.062003}
  {\bibfield  {journal} {\bibinfo  {journal} {Phys. Rev. D}\ }\textbf {\bibinfo
  {volume} {102}},\ \bibinfo {pages} {062003} (\bibinfo {year}
  {2020})}\BibitemShut {NoStop}%
\bibitem [{\citenamefont {Acernese}\ \emph {et~al.}(2019)\citenamefont
  {Acernese}, \citenamefont {Agathos}, \citenamefont {Aiello}, \citenamefont
  {Allocca}, \citenamefont {Amato}, \citenamefont {Ansoldi}, \citenamefont
  {Antier}, \citenamefont {Ar\`ene}, \citenamefont {Arnaud}, \citenamefont
  {Ascenzi}, \citenamefont {Astone}, \citenamefont {Aubin}, \citenamefont
  {Babak}, \citenamefont {Bacon}, \citenamefont {Badaracco}, \citenamefont
  {Bader}, \citenamefont {Baird}, \citenamefont {Baldaccini}, \citenamefont
  {Ballardin}, \citenamefont {Baltus}, \citenamefont {Barbieri}, \citenamefont
  {Barneo}, \citenamefont {Barone}, \citenamefont {Barsuglia}, \citenamefont
  {Barta}, \citenamefont {Basti}, \citenamefont {Bawaj}, \citenamefont
  {Bazzan}, \citenamefont {Bejger}, \citenamefont {Belahcene}, \citenamefont
  {Bernuzzi}, \citenamefont {Bersanetti}, \citenamefont {Bertolini},
  \citenamefont {Bischi}, \citenamefont {Bitossi}, \citenamefont {Bizouard},
  \citenamefont {Bobba}, \citenamefont {Boer}, \citenamefont {Bogaert},
  \citenamefont {Bondu}, \citenamefont {Bonnand}, \citenamefont {Boom},
  \citenamefont {Boschi}, \citenamefont {Bouffanais}, \citenamefont {Bozzi},
  \citenamefont {Bradaschia}, \citenamefont {Branchesi}, \citenamefont
  {Breschi}, \citenamefont {Briant}, \citenamefont {Brighenti}, \citenamefont
  {Brillet}, \citenamefont {Brooks}, \citenamefont {Bruno}, \citenamefont
  {Bulik}, \citenamefont {Bulten}, \citenamefont {Buskulic}, \citenamefont
  {Cagnoli}, \citenamefont {Calloni}, \citenamefont {Canepa}, \citenamefont
  {Carapella}, \citenamefont {Carbognani}, \citenamefont {Carullo},
  \citenamefont {Casanueva~Diaz}, \citenamefont {Casentini}, \citenamefont
  {Casta\~neda}, \citenamefont {Caudill}, \citenamefont {Cavalier},
  \citenamefont {Cavalieri}, \citenamefont {Cella}, \citenamefont
  {Cerd\'a-Dur\'an}, \citenamefont {Cesarini}, \citenamefont {Chaibi},
  \citenamefont {Chassande-Mottin}, \citenamefont {Chiadini}, \citenamefont
  {Chierici}, \citenamefont {Chincarini}, \citenamefont {Chiummo},
  \citenamefont {Christensen}, \citenamefont {Chua}, \citenamefont {Ciani},
  \citenamefont {Ciecielag}, \citenamefont {Cie\ifmmode~\acute{s}\else
  \'{s}\fi{}lar}, \citenamefont {Ciolfi}, \citenamefont {Cipriano},
  \citenamefont {Cirone}, \citenamefont {Clesse}, \citenamefont {Cleva},
  \citenamefont {Coccia}, \citenamefont {Cohadon}, \citenamefont {Cohen},
  \citenamefont {Colpi}, \citenamefont {Conti}, \citenamefont
  {Cordero-Carri\'on}, \citenamefont {Corezzi}, \citenamefont {Corre},
  \citenamefont {Cortese}, \citenamefont {Coulon}, \citenamefont {Croquette},
  \citenamefont {Cudell}, \citenamefont {Cuoco}, \citenamefont {Curylo},
  \citenamefont {D'Angelo}, \citenamefont {D'Antonio}, \citenamefont {Dattilo},
  \citenamefont {Davier}, \citenamefont {Degallaix}, \citenamefont
  {De~Laurentis}, \citenamefont {Del\'eglise}, \citenamefont {Del~Pozzo},
  \citenamefont {De~Pietri}, \citenamefont {De~Rosa}, \citenamefont {De~Rossi},
  \citenamefont {Dietrich}, \citenamefont {Di~Fiore}, \citenamefont
  {Di~Giorgio}, \citenamefont {Di~Giovanni}, \citenamefont {Di~Giovanni},
  \citenamefont {Di~Girolamo}, \citenamefont {Di~Lieto}, \citenamefont
  {Di~Pace}, \citenamefont {Di~Palma}, \citenamefont {Di~Renzo}, \citenamefont
  {Drago}, \citenamefont {Ducoin}, \citenamefont {Durante}, \citenamefont
  {D'Urso}, \citenamefont {Eisenmann}, \citenamefont {Errico}, \citenamefont
  {Estevez}, \citenamefont {Fafone}, \citenamefont {Farinon}, \citenamefont
  {Feng}, \citenamefont {Fenyvesi}, \citenamefont {Ferrante}, \citenamefont
  {Fidecaro}, \citenamefont {Fiori}, \citenamefont {Fiorucci}, \citenamefont
  {Fittipaldi}, \citenamefont {Fiumara}, \citenamefont {Flaminio},
  \citenamefont {Font}, \citenamefont {Fournier}, \citenamefont {Frasca},
  \citenamefont {Frasconi}, \citenamefont {Frey}, \citenamefont {Fronz\`e},
  \citenamefont {Garufi}, \citenamefont {Gemme}, \citenamefont {Genin},
  \citenamefont {Gennai}, \citenamefont {Ghosh}, \citenamefont {Giacomazzo},
  \citenamefont {Gosselin}, \citenamefont {Gouaty}, \citenamefont {Grado},
  \citenamefont {Granata}, \citenamefont {Greco}, \citenamefont {Grignani},
  \citenamefont {Grimaldi}, \citenamefont {Grimm}, \citenamefont {Gruning},
  \citenamefont {Guidi}, \citenamefont {Guix\'e}, \citenamefont {Guo},
  \citenamefont {Gupta}, \citenamefont {Halim}, \citenamefont {Harder},
  \citenamefont {Harms}, \citenamefont {Heidmann}, \citenamefont {Heitmann},
  \citenamefont {Hello}, \citenamefont {Hemming}, \citenamefont {Hennes},
  \citenamefont {Hinderer}, \citenamefont {Hofman}, \citenamefont {Huet},
  \citenamefont {Hui}, \citenamefont {Idzkowski}, \citenamefont {Iess},
  \citenamefont {Intini}, \citenamefont {Isac}, \citenamefont {Jacqmin},
  \citenamefont {Jaranowski}, \citenamefont {Jonker}, \citenamefont
  {Katsanevas}, \citenamefont {K\'ef\'elian}, \citenamefont {Khan},
  \citenamefont {Khetan}, \citenamefont {Koekoek}, \citenamefont {Koley},
  \citenamefont {Kr\'olak}, \citenamefont {Kutynia}, \citenamefont {Laghi},
  \citenamefont {Lamberts}, \citenamefont {La~Rosa}, \citenamefont
  {Lartaux-Vollard}, \citenamefont {Lazzaro}, \citenamefont {Leaci},
  \citenamefont {Leroy}, \citenamefont {Letendre}, \citenamefont {Linde},
  \citenamefont {Llorens-Monteagudo}, \citenamefont {Longo}, \citenamefont
  {Lorenzini}, \citenamefont {Loriette}, \citenamefont {Losurdo}, \citenamefont
  {Lumaca}, \citenamefont {Macquet}, \citenamefont {Majorana}, \citenamefont
  {Maksimovic}, \citenamefont {Man}, \citenamefont {Mangano}, \citenamefont
  {Mantovani}, \citenamefont {Mapelli}, \citenamefont {Marchesoni},
  \citenamefont {Marion}, \citenamefont {Marquina}, \citenamefont {Marsat},
  \citenamefont {Martelli}, \citenamefont {Martinez}, \citenamefont {Masserot},
  \citenamefont {Mastrogiovanni}, \citenamefont {Mejuto~Villa}, \citenamefont
  {Mereni}, \citenamefont {Merzougui}, \citenamefont {Metzdorff}, \citenamefont
  {Miani}, \citenamefont {Michel}, \citenamefont {Milano}, \citenamefont
  {Miller}, \citenamefont {Milotti}, \citenamefont {Minazzoli}, \citenamefont
  {Minenkov}, \citenamefont {Montani}, \citenamefont {Morawski}, \citenamefont
  {Mours}, \citenamefont {Muciaccia}, \citenamefont {Nagar}, \citenamefont
  {Nardecchia}, \citenamefont {Naticchioni}, \citenamefont {Neilson},
  \citenamefont {Nelemans}, \citenamefont {Nguyen}, \citenamefont {Nichols},
  \citenamefont {Nissanke}, \citenamefont {Nocera}, \citenamefont {Oganesyan},
  \citenamefont {Olivetto}, \citenamefont {Pagano}, \citenamefont {Pagliaroli},
  \citenamefont {Palomba}, \citenamefont {Pang}, \citenamefont {Pannarale},
  \citenamefont {Paoletti}, \citenamefont {Paoli}, \citenamefont {Pascucci},
  \citenamefont {Pasqualetti}, \citenamefont {Passaquieti}, \citenamefont
  {Passuello}, \citenamefont {Patricelli}, \citenamefont {Perego},
  \citenamefont {Pegoraro}, \citenamefont {P\'erigois}, \citenamefont
  {Perreca}, \citenamefont {Perri\`es}, \citenamefont {Phukon}, \citenamefont
  {Piccinni}, \citenamefont {Pichot}, \citenamefont {Piendibene}, \citenamefont
  {Piergiovanni}, \citenamefont {Pierro}, \citenamefont {Pillant},
  \citenamefont {Pinard}, \citenamefont {Pinto}, \citenamefont {Piotrzkowski},
  \citenamefont {Plastino}, \citenamefont {Poggiani}, \citenamefont
  {Popolizio}, \citenamefont {Porter}, \citenamefont {Prevedelli},
  \citenamefont {Principe}, \citenamefont {Prodi}, \citenamefont {Punturo},
  \citenamefont {Puppo}, \citenamefont {Raaijmakers}, \citenamefont
  {Radulesco}, \citenamefont {Rapagnani}, \citenamefont {Razzano},
  \citenamefont {Regimbau}, \citenamefont {Rei}, \citenamefont {Rettegno},
  \citenamefont {Ricci}, \citenamefont {Riemenschneider}, \citenamefont
  {Robinet}, \citenamefont {Rocchi}, \citenamefont {Rolland}, \citenamefont
  {Romanelli}, \citenamefont {Romano}, \citenamefont
  {Rosi\ifmmode~\acute{n}\else \'{n}\fi{}ska}, \citenamefont {Ruggi},
  \citenamefont {Salafia}, \citenamefont {Salconi}, \citenamefont {Samajdar},
  \citenamefont {Sanchis-Gual}, \citenamefont {Santos}, \citenamefont
  {Sassolas}, \citenamefont {Sauter}, \citenamefont {Sayah}, \citenamefont
  {Sentenac}, \citenamefont {Sequino}, \citenamefont {Sharma}, \citenamefont
  {Sieniawska}, \citenamefont {Singh}, \citenamefont {Singhal}, \citenamefont
  {Sipala}, \citenamefont {Sordini}, \citenamefont {Sorrentino}, \citenamefont
  {Spera}, \citenamefont {Stachie}, \citenamefont {Steer}, \citenamefont
  {Stratta}, \citenamefont {Sur}, \citenamefont {Swinkels}, \citenamefont
  {Tacca}, \citenamefont {Tanasijczuk}, \citenamefont {Tapia San~Martin},
  \citenamefont {Tiwari}, \citenamefont {Tonelli}, \citenamefont
  {Torres-Forn\'e}, \citenamefont {Tosta~e Melo}, \citenamefont {Travasso},
  \citenamefont {Tringali}, \citenamefont {Trovato}, \citenamefont {Tsang},
  \citenamefont {Turconi}, \citenamefont {Valentini}, \citenamefont {van
  Bakel}, \citenamefont {van Beuzekom}, \citenamefont {van~den Brand},
  \citenamefont {Van Den~Broeck}, \citenamefont {van~der Schaaf}, \citenamefont
  {Vardaro}, \citenamefont {Vas\'uth}, \citenamefont {Vedovato}, \citenamefont
  {Verkindt}, \citenamefont {Vetrano}, \citenamefont {Vicer\'e}, \citenamefont
  {Vinet}, \citenamefont {Vocca}, \citenamefont {Walet}, \citenamefont {Was},
  \citenamefont {Zadro\ifmmode~\dot{z}\else \.{z}\fi{}ny}, \citenamefont
  {Zelenova}, \citenamefont {Zendri}, \citenamefont {Vahlbruch}, \citenamefont
  {Mehmet}, \citenamefont {L\"uck},\ and\ \citenamefont
  {Danzmann}}]{Acernese_2019}%
  \BibitemOpen
  \bibfield  {author} {\bibinfo {author} {\bibfnamefont {F.}~\bibnamefont
  {Acernese}}, \bibinfo {author} {\bibfnamefont {M.}~\bibnamefont {Agathos}},
  \bibinfo {author} {\bibfnamefont {L.}~\bibnamefont {Aiello}}, \bibinfo
  {author} {\bibfnamefont {A.}~\bibnamefont {Allocca}}, \bibinfo {author}
  {\bibfnamefont {A.}~\bibnamefont {Amato}}, \bibinfo {author} {\bibfnamefont
  {S.}~\bibnamefont {Ansoldi}}, \bibinfo {author} {\bibfnamefont
  {S.}~\bibnamefont {Antier}}, \bibinfo {author} {\bibfnamefont
  {M.}~\bibnamefont {Ar\`ene}}, \bibinfo {author} {\bibfnamefont
  {N.}~\bibnamefont {Arnaud}}, \bibinfo {author} {\bibfnamefont
  {S.}~\bibnamefont {Ascenzi}}, \bibinfo {author} {\bibfnamefont
  {P.}~\bibnamefont {Astone}}, \bibinfo {author} {\bibfnamefont
  {F.}~\bibnamefont {Aubin}}, \bibinfo {author} {\bibfnamefont
  {S.}~\bibnamefont {Babak}}, \bibinfo {author} {\bibfnamefont
  {P.}~\bibnamefont {Bacon}}, \bibinfo {author} {\bibfnamefont
  {F.}~\bibnamefont {Badaracco}}, \bibinfo {author} {\bibfnamefont {M.~K.~M.}\
  \bibnamefont {Bader}}, \bibinfo {author} {\bibfnamefont {J.}~\bibnamefont
  {Baird}}, \bibinfo {author} {\bibfnamefont {F.}~\bibnamefont {Baldaccini}},
  \bibinfo {author} {\bibfnamefont {G.}~\bibnamefont {Ballardin}}, \bibinfo
  {author} {\bibfnamefont {G.}~\bibnamefont {Baltus}}, \bibinfo {author}
  {\bibfnamefont {C.}~\bibnamefont {Barbieri}}, \bibinfo {author}
  {\bibfnamefont {P.}~\bibnamefont {Barneo}}, \bibinfo {author} {\bibfnamefont
  {F.}~\bibnamefont {Barone}}, \bibinfo {author} {\bibfnamefont
  {M.}~\bibnamefont {Barsuglia}}, \bibinfo {author} {\bibfnamefont
  {D.}~\bibnamefont {Barta}}, \bibinfo {author} {\bibfnamefont
  {A.}~\bibnamefont {Basti}}, \bibinfo {author} {\bibfnamefont
  {M.}~\bibnamefont {Bawaj}}, \bibinfo {author} {\bibfnamefont
  {M.}~\bibnamefont {Bazzan}}, \bibinfo {author} {\bibfnamefont
  {M.}~\bibnamefont {Bejger}}, \bibinfo {author} {\bibfnamefont
  {I.}~\bibnamefont {Belahcene}}, \bibinfo {author} {\bibfnamefont
  {S.}~\bibnamefont {Bernuzzi}}, \bibinfo {author} {\bibfnamefont
  {D.}~\bibnamefont {Bersanetti}}, \bibinfo {author} {\bibfnamefont
  {A.}~\bibnamefont {Bertolini}}, \bibinfo {author} {\bibfnamefont
  {M.}~\bibnamefont {Bischi}}, \bibinfo {author} {\bibfnamefont
  {M.}~\bibnamefont {Bitossi}}, \bibinfo {author} {\bibfnamefont {M.~A.}\
  \bibnamefont {Bizouard}}, \bibinfo {author} {\bibfnamefont {F.}~\bibnamefont
  {Bobba}}, \bibinfo {author} {\bibfnamefont {M.}~\bibnamefont {Boer}},
  \bibinfo {author} {\bibfnamefont {G.}~\bibnamefont {Bogaert}}, \bibinfo
  {author} {\bibfnamefont {F.}~\bibnamefont {Bondu}}, \bibinfo {author}
  {\bibfnamefont {R.}~\bibnamefont {Bonnand}}, \bibinfo {author} {\bibfnamefont
  {B.~A.}\ \bibnamefont {Boom}}, \bibinfo {author} {\bibfnamefont
  {V.}~\bibnamefont {Boschi}}, \bibinfo {author} {\bibfnamefont
  {Y.}~\bibnamefont {Bouffanais}}, \bibinfo {author} {\bibfnamefont
  {A.}~\bibnamefont {Bozzi}}, \bibinfo {author} {\bibfnamefont
  {C.}~\bibnamefont {Bradaschia}}, \bibinfo {author} {\bibfnamefont
  {M.}~\bibnamefont {Branchesi}}, \bibinfo {author} {\bibfnamefont
  {M.}~\bibnamefont {Breschi}}, \bibinfo {author} {\bibfnamefont
  {T.}~\bibnamefont {Briant}}, \bibinfo {author} {\bibfnamefont
  {F.}~\bibnamefont {Brighenti}}, \bibinfo {author} {\bibfnamefont
  {A.}~\bibnamefont {Brillet}}, \bibinfo {author} {\bibfnamefont
  {J.}~\bibnamefont {Brooks}}, \bibinfo {author} {\bibfnamefont
  {G.}~\bibnamefont {Bruno}}, \bibinfo {author} {\bibfnamefont
  {T.}~\bibnamefont {Bulik}}, \bibinfo {author} {\bibfnamefont {H.~J.}\
  \bibnamefont {Bulten}}, \bibinfo {author} {\bibfnamefont {D.}~\bibnamefont
  {Buskulic}}, \bibinfo {author} {\bibfnamefont {G.}~\bibnamefont {Cagnoli}},
  \bibinfo {author} {\bibfnamefont {E.}~\bibnamefont {Calloni}}, \bibinfo
  {author} {\bibfnamefont {M.}~\bibnamefont {Canepa}}, \bibinfo {author}
  {\bibfnamefont {G.}~\bibnamefont {Carapella}}, \bibinfo {author}
  {\bibfnamefont {F.}~\bibnamefont {Carbognani}}, \bibinfo {author}
  {\bibfnamefont {G.}~\bibnamefont {Carullo}}, \bibinfo {author} {\bibfnamefont
  {J.}~\bibnamefont {Casanueva~Diaz}}, \bibinfo {author} {\bibfnamefont
  {C.}~\bibnamefont {Casentini}}, \bibinfo {author} {\bibfnamefont
  {J.}~\bibnamefont {Casta\~neda}}, \bibinfo {author} {\bibfnamefont
  {S.}~\bibnamefont {Caudill}}, \bibinfo {author} {\bibfnamefont
  {F.}~\bibnamefont {Cavalier}}, \bibinfo {author} {\bibfnamefont
  {R.}~\bibnamefont {Cavalieri}}, \bibinfo {author} {\bibfnamefont
  {G.}~\bibnamefont {Cella}}, \bibinfo {author} {\bibfnamefont
  {P.}~\bibnamefont {Cerd\'a-Dur\'an}}, \bibinfo {author} {\bibfnamefont
  {E.}~\bibnamefont {Cesarini}}, \bibinfo {author} {\bibfnamefont
  {O.}~\bibnamefont {Chaibi}}, \bibinfo {author} {\bibfnamefont
  {E.}~\bibnamefont {Chassande-Mottin}}, \bibinfo {author} {\bibfnamefont
  {F.}~\bibnamefont {Chiadini}}, \bibinfo {author} {\bibfnamefont
  {R.}~\bibnamefont {Chierici}}, \bibinfo {author} {\bibfnamefont
  {A.}~\bibnamefont {Chincarini}}, \bibinfo {author} {\bibfnamefont
  {A.}~\bibnamefont {Chiummo}}, \bibinfo {author} {\bibfnamefont
  {N.}~\bibnamefont {Christensen}}, \bibinfo {author} {\bibfnamefont
  {S.}~\bibnamefont {Chua}}, \bibinfo {author} {\bibfnamefont {G.}~\bibnamefont
  {Ciani}}, \bibinfo {author} {\bibfnamefont {P.}~\bibnamefont {Ciecielag}},
  \bibinfo {author} {\bibfnamefont {M.}~\bibnamefont
  {Cie\ifmmode~\acute{s}\else \'{s}\fi{}lar}}, \bibinfo {author} {\bibfnamefont
  {R.}~\bibnamefont {Ciolfi}}, \bibinfo {author} {\bibfnamefont
  {F.}~\bibnamefont {Cipriano}}, \bibinfo {author} {\bibfnamefont
  {A.}~\bibnamefont {Cirone}}, \bibinfo {author} {\bibfnamefont
  {S.}~\bibnamefont {Clesse}}, \bibinfo {author} {\bibfnamefont
  {F.}~\bibnamefont {Cleva}}, \bibinfo {author} {\bibfnamefont
  {E.}~\bibnamefont {Coccia}}, \bibinfo {author} {\bibfnamefont {P.-F.}\
  \bibnamefont {Cohadon}}, \bibinfo {author} {\bibfnamefont {D.}~\bibnamefont
  {Cohen}}, \bibinfo {author} {\bibfnamefont {M.}~\bibnamefont {Colpi}},
  \bibinfo {author} {\bibfnamefont {L.}~\bibnamefont {Conti}}, \bibinfo
  {author} {\bibfnamefont {I.}~\bibnamefont {Cordero-Carri\'on}}, \bibinfo
  {author} {\bibfnamefont {S.}~\bibnamefont {Corezzi}}, \bibinfo {author}
  {\bibfnamefont {D.}~\bibnamefont {Corre}}, \bibinfo {author} {\bibfnamefont
  {S.}~\bibnamefont {Cortese}}, \bibinfo {author} {\bibfnamefont {J.-P.}\
  \bibnamefont {Coulon}}, \bibinfo {author} {\bibfnamefont {M.}~\bibnamefont
  {Croquette}}, \bibinfo {author} {\bibfnamefont {J.-R.}\ \bibnamefont
  {Cudell}}, \bibinfo {author} {\bibfnamefont {E.}~\bibnamefont {Cuoco}},
  \bibinfo {author} {\bibfnamefont {M.}~\bibnamefont {Curylo}}, \bibinfo
  {author} {\bibfnamefont {B.}~\bibnamefont {D'Angelo}}, \bibinfo {author}
  {\bibfnamefont {S.}~\bibnamefont {D'Antonio}}, \bibinfo {author}
  {\bibfnamefont {V.}~\bibnamefont {Dattilo}}, \bibinfo {author} {\bibfnamefont
  {M.}~\bibnamefont {Davier}}, \bibinfo {author} {\bibfnamefont
  {J.}~\bibnamefont {Degallaix}}, \bibinfo {author} {\bibfnamefont
  {M.}~\bibnamefont {De~Laurentis}}, \bibinfo {author} {\bibfnamefont
  {S.}~\bibnamefont {Del\'eglise}}, \bibinfo {author} {\bibfnamefont
  {W.}~\bibnamefont {Del~Pozzo}}, \bibinfo {author} {\bibfnamefont
  {R.}~\bibnamefont {De~Pietri}}, \bibinfo {author} {\bibfnamefont
  {R.}~\bibnamefont {De~Rosa}}, \bibinfo {author} {\bibfnamefont
  {C.}~\bibnamefont {De~Rossi}}, \bibinfo {author} {\bibfnamefont
  {T.}~\bibnamefont {Dietrich}}, \bibinfo {author} {\bibfnamefont
  {L.}~\bibnamefont {Di~Fiore}}, \bibinfo {author} {\bibfnamefont
  {C.}~\bibnamefont {Di~Giorgio}}, \bibinfo {author} {\bibfnamefont
  {F.}~\bibnamefont {Di~Giovanni}}, \bibinfo {author} {\bibfnamefont
  {M.}~\bibnamefont {Di~Giovanni}}, \bibinfo {author} {\bibfnamefont
  {T.}~\bibnamefont {Di~Girolamo}}, \bibinfo {author} {\bibfnamefont
  {A.}~\bibnamefont {Di~Lieto}}, \bibinfo {author} {\bibfnamefont
  {S.}~\bibnamefont {Di~Pace}}, \bibinfo {author} {\bibfnamefont
  {I.}~\bibnamefont {Di~Palma}}, \bibinfo {author} {\bibfnamefont
  {F.}~\bibnamefont {Di~Renzo}}, \bibinfo {author} {\bibfnamefont
  {M.}~\bibnamefont {Drago}}, \bibinfo {author} {\bibfnamefont {J.-G.}\
  \bibnamefont {Ducoin}}, \bibinfo {author} {\bibfnamefont {O.}~\bibnamefont
  {Durante}}, \bibinfo {author} {\bibfnamefont {D.}~\bibnamefont {D'Urso}},
  \bibinfo {author} {\bibfnamefont {M.}~\bibnamefont {Eisenmann}}, \bibinfo
  {author} {\bibfnamefont {L.}~\bibnamefont {Errico}}, \bibinfo {author}
  {\bibfnamefont {D.}~\bibnamefont {Estevez}}, \bibinfo {author} {\bibfnamefont
  {V.}~\bibnamefont {Fafone}}, \bibinfo {author} {\bibfnamefont
  {S.}~\bibnamefont {Farinon}}, \bibinfo {author} {\bibfnamefont
  {F.}~\bibnamefont {Feng}}, \bibinfo {author} {\bibfnamefont {E.}~\bibnamefont
  {Fenyvesi}}, \bibinfo {author} {\bibfnamefont {I.}~\bibnamefont {Ferrante}},
  \bibinfo {author} {\bibfnamefont {F.}~\bibnamefont {Fidecaro}}, \bibinfo
  {author} {\bibfnamefont {I.}~\bibnamefont {Fiori}}, \bibinfo {author}
  {\bibfnamefont {D.}~\bibnamefont {Fiorucci}}, \bibinfo {author}
  {\bibfnamefont {R.}~\bibnamefont {Fittipaldi}}, \bibinfo {author}
  {\bibfnamefont {V.}~\bibnamefont {Fiumara}}, \bibinfo {author} {\bibfnamefont
  {R.}~\bibnamefont {Flaminio}}, \bibinfo {author} {\bibfnamefont {J.~A.}\
  \bibnamefont {Font}}, \bibinfo {author} {\bibfnamefont {J.-D.}\ \bibnamefont
  {Fournier}}, \bibinfo {author} {\bibfnamefont {S.}~\bibnamefont {Frasca}},
  \bibinfo {author} {\bibfnamefont {F.}~\bibnamefont {Frasconi}}, \bibinfo
  {author} {\bibfnamefont {V.}~\bibnamefont {Frey}}, \bibinfo {author}
  {\bibfnamefont {G.}~\bibnamefont {Fronz\`e}}, \bibinfo {author}
  {\bibfnamefont {F.}~\bibnamefont {Garufi}}, \bibinfo {author} {\bibfnamefont
  {G.}~\bibnamefont {Gemme}}, \bibinfo {author} {\bibfnamefont
  {E.}~\bibnamefont {Genin}}, \bibinfo {author} {\bibfnamefont
  {A.}~\bibnamefont {Gennai}}, \bibinfo {author} {\bibfnamefont
  {A.}~\bibnamefont {Ghosh}}, \bibinfo {author} {\bibfnamefont
  {B.}~\bibnamefont {Giacomazzo}}, \bibinfo {author} {\bibfnamefont
  {M.}~\bibnamefont {Gosselin}}, \bibinfo {author} {\bibfnamefont
  {R.}~\bibnamefont {Gouaty}}, \bibinfo {author} {\bibfnamefont
  {A.}~\bibnamefont {Grado}}, \bibinfo {author} {\bibfnamefont
  {M.}~\bibnamefont {Granata}}, \bibinfo {author} {\bibfnamefont
  {G.}~\bibnamefont {Greco}}, \bibinfo {author} {\bibfnamefont
  {G.}~\bibnamefont {Grignani}}, \bibinfo {author} {\bibfnamefont
  {A.}~\bibnamefont {Grimaldi}}, \bibinfo {author} {\bibfnamefont {S.~J.}\
  \bibnamefont {Grimm}}, \bibinfo {author} {\bibfnamefont {P.}~\bibnamefont
  {Gruning}}, \bibinfo {author} {\bibfnamefont {G.~M.}\ \bibnamefont {Guidi}},
  \bibinfo {author} {\bibfnamefont {G.}~\bibnamefont {Guix\'e}}, \bibinfo
  {author} {\bibfnamefont {Y.}~\bibnamefont {Guo}}, \bibinfo {author}
  {\bibfnamefont {P.}~\bibnamefont {Gupta}}, \bibinfo {author} {\bibfnamefont
  {O.}~\bibnamefont {Halim}}, \bibinfo {author} {\bibfnamefont
  {T.}~\bibnamefont {Harder}}, \bibinfo {author} {\bibfnamefont
  {J.}~\bibnamefont {Harms}}, \bibinfo {author} {\bibfnamefont
  {A.}~\bibnamefont {Heidmann}}, \bibinfo {author} {\bibfnamefont
  {H.}~\bibnamefont {Heitmann}}, \bibinfo {author} {\bibfnamefont
  {P.}~\bibnamefont {Hello}}, \bibinfo {author} {\bibfnamefont
  {G.}~\bibnamefont {Hemming}}, \bibinfo {author} {\bibfnamefont
  {E.}~\bibnamefont {Hennes}}, \bibinfo {author} {\bibfnamefont
  {T.}~\bibnamefont {Hinderer}}, \bibinfo {author} {\bibfnamefont
  {D.}~\bibnamefont {Hofman}}, \bibinfo {author} {\bibfnamefont
  {D.}~\bibnamefont {Huet}}, \bibinfo {author} {\bibfnamefont {V.}~\bibnamefont
  {Hui}}, \bibinfo {author} {\bibfnamefont {B.}~\bibnamefont {Idzkowski}},
  \bibinfo {author} {\bibfnamefont {A.}~\bibnamefont {Iess}}, \bibinfo {author}
  {\bibfnamefont {G.}~\bibnamefont {Intini}}, \bibinfo {author} {\bibfnamefont
  {J.-M.}\ \bibnamefont {Isac}}, \bibinfo {author} {\bibfnamefont
  {T.}~\bibnamefont {Jacqmin}}, \bibinfo {author} {\bibfnamefont
  {P.}~\bibnamefont {Jaranowski}}, \bibinfo {author} {\bibfnamefont {R.~J.~G.}\
  \bibnamefont {Jonker}}, \bibinfo {author} {\bibfnamefont {S.}~\bibnamefont
  {Katsanevas}}, \bibinfo {author} {\bibfnamefont {F.}~\bibnamefont
  {K\'ef\'elian}}, \bibinfo {author} {\bibfnamefont {I.}~\bibnamefont {Khan}},
  \bibinfo {author} {\bibfnamefont {N.}~\bibnamefont {Khetan}}, \bibinfo
  {author} {\bibfnamefont {G.}~\bibnamefont {Koekoek}}, \bibinfo {author}
  {\bibfnamefont {S.}~\bibnamefont {Koley}}, \bibinfo {author} {\bibfnamefont
  {A.}~\bibnamefont {Kr\'olak}}, \bibinfo {author} {\bibfnamefont
  {A.}~\bibnamefont {Kutynia}}, \bibinfo {author} {\bibfnamefont
  {D.}~\bibnamefont {Laghi}}, \bibinfo {author} {\bibfnamefont
  {A.}~\bibnamefont {Lamberts}}, \bibinfo {author} {\bibfnamefont
  {I.}~\bibnamefont {La~Rosa}}, \bibinfo {author} {\bibfnamefont
  {A.}~\bibnamefont {Lartaux-Vollard}}, \bibinfo {author} {\bibfnamefont
  {C.}~\bibnamefont {Lazzaro}}, \bibinfo {author} {\bibfnamefont
  {P.}~\bibnamefont {Leaci}}, \bibinfo {author} {\bibfnamefont
  {N.}~\bibnamefont {Leroy}}, \bibinfo {author} {\bibfnamefont
  {N.}~\bibnamefont {Letendre}}, \bibinfo {author} {\bibfnamefont
  {F.}~\bibnamefont {Linde}}, \bibinfo {author} {\bibfnamefont
  {M.}~\bibnamefont {Llorens-Monteagudo}}, \bibinfo {author} {\bibfnamefont
  {A.}~\bibnamefont {Longo}}, \bibinfo {author} {\bibfnamefont
  {M.}~\bibnamefont {Lorenzini}}, \bibinfo {author} {\bibfnamefont
  {V.}~\bibnamefont {Loriette}}, \bibinfo {author} {\bibfnamefont
  {G.}~\bibnamefont {Losurdo}}, \bibinfo {author} {\bibfnamefont
  {D.}~\bibnamefont {Lumaca}}, \bibinfo {author} {\bibfnamefont
  {A.}~\bibnamefont {Macquet}}, \bibinfo {author} {\bibfnamefont
  {E.}~\bibnamefont {Majorana}}, \bibinfo {author} {\bibfnamefont
  {I.}~\bibnamefont {Maksimovic}}, \bibinfo {author} {\bibfnamefont
  {N.}~\bibnamefont {Man}}, \bibinfo {author} {\bibfnamefont {V.}~\bibnamefont
  {Mangano}}, \bibinfo {author} {\bibfnamefont {M.}~\bibnamefont {Mantovani}},
  \bibinfo {author} {\bibfnamefont {M.}~\bibnamefont {Mapelli}}, \bibinfo
  {author} {\bibfnamefont {F.}~\bibnamefont {Marchesoni}}, \bibinfo {author}
  {\bibfnamefont {F.}~\bibnamefont {Marion}}, \bibinfo {author} {\bibfnamefont
  {A.}~\bibnamefont {Marquina}}, \bibinfo {author} {\bibfnamefont
  {S.}~\bibnamefont {Marsat}}, \bibinfo {author} {\bibfnamefont
  {F.}~\bibnamefont {Martelli}}, \bibinfo {author} {\bibfnamefont
  {V.}~\bibnamefont {Martinez}}, \bibinfo {author} {\bibfnamefont
  {A.}~\bibnamefont {Masserot}}, \bibinfo {author} {\bibfnamefont
  {S.}~\bibnamefont {Mastrogiovanni}}, \bibinfo {author} {\bibfnamefont
  {E.}~\bibnamefont {Mejuto~Villa}}, \bibinfo {author} {\bibfnamefont
  {L.}~\bibnamefont {Mereni}}, \bibinfo {author} {\bibfnamefont
  {M.}~\bibnamefont {Merzougui}}, \bibinfo {author} {\bibfnamefont
  {R.}~\bibnamefont {Metzdorff}}, \bibinfo {author} {\bibfnamefont
  {A.}~\bibnamefont {Miani}}, \bibinfo {author} {\bibfnamefont
  {C.}~\bibnamefont {Michel}}, \bibinfo {author} {\bibfnamefont
  {L.}~\bibnamefont {Milano}}, \bibinfo {author} {\bibfnamefont
  {A.}~\bibnamefont {Miller}}, \bibinfo {author} {\bibfnamefont
  {E.}~\bibnamefont {Milotti}}, \bibinfo {author} {\bibfnamefont
  {O.}~\bibnamefont {Minazzoli}}, \bibinfo {author} {\bibfnamefont
  {Y.}~\bibnamefont {Minenkov}}, \bibinfo {author} {\bibfnamefont
  {M.}~\bibnamefont {Montani}}, \bibinfo {author} {\bibfnamefont
  {F.}~\bibnamefont {Morawski}}, \bibinfo {author} {\bibfnamefont
  {B.}~\bibnamefont {Mours}}, \bibinfo {author} {\bibfnamefont
  {F.}~\bibnamefont {Muciaccia}}, \bibinfo {author} {\bibfnamefont
  {A.}~\bibnamefont {Nagar}}, \bibinfo {author} {\bibfnamefont
  {I.}~\bibnamefont {Nardecchia}}, \bibinfo {author} {\bibfnamefont
  {L.}~\bibnamefont {Naticchioni}}, \bibinfo {author} {\bibfnamefont
  {J.}~\bibnamefont {Neilson}}, \bibinfo {author} {\bibfnamefont
  {G.}~\bibnamefont {Nelemans}}, \bibinfo {author} {\bibfnamefont
  {C.}~\bibnamefont {Nguyen}}, \bibinfo {author} {\bibfnamefont
  {D.}~\bibnamefont {Nichols}}, \bibinfo {author} {\bibfnamefont
  {S.}~\bibnamefont {Nissanke}}, \bibinfo {author} {\bibfnamefont
  {F.}~\bibnamefont {Nocera}}, \bibinfo {author} {\bibfnamefont
  {G.}~\bibnamefont {Oganesyan}}, \bibinfo {author} {\bibfnamefont
  {C.}~\bibnamefont {Olivetto}}, \bibinfo {author} {\bibfnamefont
  {G.}~\bibnamefont {Pagano}}, \bibinfo {author} {\bibfnamefont
  {G.}~\bibnamefont {Pagliaroli}}, \bibinfo {author} {\bibfnamefont
  {C.}~\bibnamefont {Palomba}}, \bibinfo {author} {\bibfnamefont {P.~T.~H.}\
  \bibnamefont {Pang}}, \bibinfo {author} {\bibfnamefont {F.}~\bibnamefont
  {Pannarale}}, \bibinfo {author} {\bibfnamefont {F.}~\bibnamefont {Paoletti}},
  \bibinfo {author} {\bibfnamefont {A.}~\bibnamefont {Paoli}}, \bibinfo
  {author} {\bibfnamefont {D.}~\bibnamefont {Pascucci}}, \bibinfo {author}
  {\bibfnamefont {A.}~\bibnamefont {Pasqualetti}}, \bibinfo {author}
  {\bibfnamefont {R.}~\bibnamefont {Passaquieti}}, \bibinfo {author}
  {\bibfnamefont {D.}~\bibnamefont {Passuello}}, \bibinfo {author}
  {\bibfnamefont {B.}~\bibnamefont {Patricelli}}, \bibinfo {author}
  {\bibfnamefont {A.}~\bibnamefont {Perego}}, \bibinfo {author} {\bibfnamefont
  {M.}~\bibnamefont {Pegoraro}}, \bibinfo {author} {\bibfnamefont
  {C.}~\bibnamefont {P\'erigois}}, \bibinfo {author} {\bibfnamefont
  {A.}~\bibnamefont {Perreca}}, \bibinfo {author} {\bibfnamefont
  {S.}~\bibnamefont {Perri\`es}}, \bibinfo {author} {\bibfnamefont {K.~S.}\
  \bibnamefont {Phukon}}, \bibinfo {author} {\bibfnamefont {O.~J.}\
  \bibnamefont {Piccinni}}, \bibinfo {author} {\bibfnamefont {M.}~\bibnamefont
  {Pichot}}, \bibinfo {author} {\bibfnamefont {M.}~\bibnamefont {Piendibene}},
  \bibinfo {author} {\bibfnamefont {F.}~\bibnamefont {Piergiovanni}}, \bibinfo
  {author} {\bibfnamefont {V.}~\bibnamefont {Pierro}}, \bibinfo {author}
  {\bibfnamefont {G.}~\bibnamefont {Pillant}}, \bibinfo {author} {\bibfnamefont
  {L.}~\bibnamefont {Pinard}}, \bibinfo {author} {\bibfnamefont {I.~M.}\
  \bibnamefont {Pinto}}, \bibinfo {author} {\bibfnamefont {K.}~\bibnamefont
  {Piotrzkowski}}, \bibinfo {author} {\bibfnamefont {W.}~\bibnamefont
  {Plastino}}, \bibinfo {author} {\bibfnamefont {R.}~\bibnamefont {Poggiani}},
  \bibinfo {author} {\bibfnamefont {P.}~\bibnamefont {Popolizio}}, \bibinfo
  {author} {\bibfnamefont {E.~K.}\ \bibnamefont {Porter}}, \bibinfo {author}
  {\bibfnamefont {M.}~\bibnamefont {Prevedelli}}, \bibinfo {author}
  {\bibfnamefont {M.}~\bibnamefont {Principe}}, \bibinfo {author}
  {\bibfnamefont {G.~A.}\ \bibnamefont {Prodi}}, \bibinfo {author}
  {\bibfnamefont {M.}~\bibnamefont {Punturo}}, \bibinfo {author} {\bibfnamefont
  {P.}~\bibnamefont {Puppo}}, \bibinfo {author} {\bibfnamefont
  {G.}~\bibnamefont {Raaijmakers}}, \bibinfo {author} {\bibfnamefont
  {N.}~\bibnamefont {Radulesco}}, \bibinfo {author} {\bibfnamefont
  {P.}~\bibnamefont {Rapagnani}}, \bibinfo {author} {\bibfnamefont
  {M.}~\bibnamefont {Razzano}}, \bibinfo {author} {\bibfnamefont
  {T.}~\bibnamefont {Regimbau}}, \bibinfo {author} {\bibfnamefont
  {L.}~\bibnamefont {Rei}}, \bibinfo {author} {\bibfnamefont {P.}~\bibnamefont
  {Rettegno}}, \bibinfo {author} {\bibfnamefont {F.}~\bibnamefont {Ricci}},
  \bibinfo {author} {\bibfnamefont {G.}~\bibnamefont {Riemenschneider}},
  \bibinfo {author} {\bibfnamefont {F.}~\bibnamefont {Robinet}}, \bibinfo
  {author} {\bibfnamefont {A.}~\bibnamefont {Rocchi}}, \bibinfo {author}
  {\bibfnamefont {L.}~\bibnamefont {Rolland}}, \bibinfo {author} {\bibfnamefont
  {M.}~\bibnamefont {Romanelli}}, \bibinfo {author} {\bibfnamefont
  {R.}~\bibnamefont {Romano}}, \bibinfo {author} {\bibfnamefont
  {D.}~\bibnamefont {Rosi\ifmmode~\acute{n}\else \'{n}\fi{}ska}}, \bibinfo
  {author} {\bibfnamefont {P.}~\bibnamefont {Ruggi}}, \bibinfo {author}
  {\bibfnamefont {O.~S.}\ \bibnamefont {Salafia}}, \bibinfo {author}
  {\bibfnamefont {L.}~\bibnamefont {Salconi}}, \bibinfo {author} {\bibfnamefont
  {A.}~\bibnamefont {Samajdar}}, \bibinfo {author} {\bibfnamefont
  {N.}~\bibnamefont {Sanchis-Gual}}, \bibinfo {author} {\bibfnamefont
  {E.}~\bibnamefont {Santos}}, \bibinfo {author} {\bibfnamefont
  {B.}~\bibnamefont {Sassolas}}, \bibinfo {author} {\bibfnamefont
  {O.}~\bibnamefont {Sauter}}, \bibinfo {author} {\bibfnamefont
  {S.}~\bibnamefont {Sayah}}, \bibinfo {author} {\bibfnamefont
  {D.}~\bibnamefont {Sentenac}}, \bibinfo {author} {\bibfnamefont
  {V.}~\bibnamefont {Sequino}}, \bibinfo {author} {\bibfnamefont
  {A.}~\bibnamefont {Sharma}}, \bibinfo {author} {\bibfnamefont
  {M.}~\bibnamefont {Sieniawska}}, \bibinfo {author} {\bibfnamefont
  {N.}~\bibnamefont {Singh}}, \bibinfo {author} {\bibfnamefont
  {A.}~\bibnamefont {Singhal}}, \bibinfo {author} {\bibfnamefont
  {V.}~\bibnamefont {Sipala}}, \bibinfo {author} {\bibfnamefont
  {V.}~\bibnamefont {Sordini}}, \bibinfo {author} {\bibfnamefont
  {F.}~\bibnamefont {Sorrentino}}, \bibinfo {author} {\bibfnamefont
  {M.}~\bibnamefont {Spera}}, \bibinfo {author} {\bibfnamefont
  {C.}~\bibnamefont {Stachie}}, \bibinfo {author} {\bibfnamefont {D.~A.}\
  \bibnamefont {Steer}}, \bibinfo {author} {\bibfnamefont {G.}~\bibnamefont
  {Stratta}}, \bibinfo {author} {\bibfnamefont {A.}~\bibnamefont {Sur}},
  \bibinfo {author} {\bibfnamefont {B.~L.}\ \bibnamefont {Swinkels}}, \bibinfo
  {author} {\bibfnamefont {M.}~\bibnamefont {Tacca}}, \bibinfo {author}
  {\bibfnamefont {A.~J.}\ \bibnamefont {Tanasijczuk}}, \bibinfo {author}
  {\bibfnamefont {E.~N.}\ \bibnamefont {Tapia San~Martin}}, \bibinfo {author}
  {\bibfnamefont {S.}~\bibnamefont {Tiwari}}, \bibinfo {author} {\bibfnamefont
  {M.}~\bibnamefont {Tonelli}}, \bibinfo {author} {\bibfnamefont
  {A.}~\bibnamefont {Torres-Forn\'e}}, \bibinfo {author} {\bibfnamefont
  {I.}~\bibnamefont {Tosta~e Melo}}, \bibinfo {author} {\bibfnamefont
  {F.}~\bibnamefont {Travasso}}, \bibinfo {author} {\bibfnamefont {M.~C.}\
  \bibnamefont {Tringali}}, \bibinfo {author} {\bibfnamefont {A.}~\bibnamefont
  {Trovato}}, \bibinfo {author} {\bibfnamefont {K.~W.}\ \bibnamefont {Tsang}},
  \bibinfo {author} {\bibfnamefont {M.}~\bibnamefont {Turconi}}, \bibinfo
  {author} {\bibfnamefont {M.}~\bibnamefont {Valentini}}, \bibinfo {author}
  {\bibfnamefont {N.}~\bibnamefont {van Bakel}}, \bibinfo {author}
  {\bibfnamefont {M.}~\bibnamefont {van Beuzekom}}, \bibinfo {author}
  {\bibfnamefont {J.~F.~J.}\ \bibnamefont {van~den Brand}}, \bibinfo {author}
  {\bibfnamefont {C.}~\bibnamefont {Van Den~Broeck}}, \bibinfo {author}
  {\bibfnamefont {L.}~\bibnamefont {van~der Schaaf}}, \bibinfo {author}
  {\bibfnamefont {M.}~\bibnamefont {Vardaro}}, \bibinfo {author} {\bibfnamefont
  {M.}~\bibnamefont {Vas\'uth}}, \bibinfo {author} {\bibfnamefont
  {G.}~\bibnamefont {Vedovato}}, \bibinfo {author} {\bibfnamefont
  {D.}~\bibnamefont {Verkindt}}, \bibinfo {author} {\bibfnamefont
  {F.}~\bibnamefont {Vetrano}}, \bibinfo {author} {\bibfnamefont
  {A.}~\bibnamefont {Vicer\'e}}, \bibinfo {author} {\bibfnamefont {J.-Y.}\
  \bibnamefont {Vinet}}, \bibinfo {author} {\bibfnamefont {H.}~\bibnamefont
  {Vocca}}, \bibinfo {author} {\bibfnamefont {R.}~\bibnamefont {Walet}},
  \bibinfo {author} {\bibfnamefont {M.}~\bibnamefont {Was}}, \bibinfo {author}
  {\bibfnamefont {A.}~\bibnamefont {Zadro\ifmmode~\dot{z}\else \.{z}\fi{}ny}},
  \bibinfo {author} {\bibfnamefont {T.}~\bibnamefont {Zelenova}}, \bibinfo
  {author} {\bibfnamefont {J.-P.}\ \bibnamefont {Zendri}}, \bibinfo {author}
  {\bibfnamefont {H.}~\bibnamefont {Vahlbruch}}, \bibinfo {author}
  {\bibfnamefont {M.}~\bibnamefont {Mehmet}}, \bibinfo {author} {\bibfnamefont
  {H.}~\bibnamefont {L\"uck}},\ and\ \bibinfo {author} {\bibfnamefont
  {K.}~\bibnamefont {Danzmann}} (\bibinfo {collaboration} {Virgo
  Collaboration}),\ }\bibfield  {title} {\bibinfo {title} {Increasing the
  astrophysical reach of the advanced virgo detector via the application of
  squeezed vacuum states of light},\ }\href
  {https://doi.org/10.1103/PhysRevLett.123.231108} {\bibfield  {journal}
  {\bibinfo  {journal} {Phys. Rev. Lett.}\ }\textbf {\bibinfo {volume} {123}},\
  \bibinfo {pages} {231108} (\bibinfo {year} {2019})}\BibitemShut {NoStop}%
\bibitem [{\citenamefont {Bersanetti}\ \emph {et~al.}(2021)\citenamefont
  {Bersanetti}, \citenamefont {Patricelli}, \citenamefont {Piccinni},
  \citenamefont {Piergiovanni}, \citenamefont {Salemi},\ and\ \citenamefont
  {Sequino}}]{Bersanetti_VirgoStatus_2021}%
  \BibitemOpen
  \bibfield  {author} {\bibinfo {author} {\bibfnamefont {D.}~\bibnamefont
  {Bersanetti}}, \bibinfo {author} {\bibfnamefont {B.}~\bibnamefont
  {Patricelli}}, \bibinfo {author} {\bibfnamefont {O.~J.}\ \bibnamefont
  {Piccinni}}, \bibinfo {author} {\bibfnamefont {F.}~\bibnamefont
  {Piergiovanni}}, \bibinfo {author} {\bibfnamefont {F.}~\bibnamefont
  {Salemi}},\ and\ \bibinfo {author} {\bibfnamefont {V.}~\bibnamefont
  {Sequino}},\ }\bibfield  {title} {\bibinfo {title} {Advanced {V}irgo:
  {S}tatus of the {D}etector, {L}atest {R}esults and {F}uture {P}rospects},\
  }\href {https://doi.org/10.3390/universe7090322} {\bibfield  {journal}
  {\bibinfo  {journal} {Universe}\ }\textbf {\bibinfo {volume} {7}},\ \bibinfo
  {pages} {322} (\bibinfo {year} {2021})}\BibitemShut {NoStop}%
\bibitem [{\citenamefont {Lough}\ \emph {et~al.}(2021)\citenamefont {Lough},
  \citenamefont {Schreiber}, \citenamefont {Bergamin}, \citenamefont {Grote},
  \citenamefont {Mehmet}, \citenamefont {Vahlbruch}, \citenamefont {Affeldt},
  \citenamefont {Brinkmann}, \citenamefont {Bisht}, \citenamefont {Kringel},
  \citenamefont {L\"uck}, \citenamefont {Mukund}, \citenamefont {Nadji},
  \citenamefont {Sorazu}, \citenamefont {Strain}, \citenamefont {Weinert},\
  and\ \citenamefont {Danzmann}}]{Lough_6dBGEO600_2021}%
  \BibitemOpen
  \bibfield  {author} {\bibinfo {author} {\bibfnamefont {J.}~\bibnamefont
  {Lough}}, \bibinfo {author} {\bibfnamefont {E.}~\bibnamefont {Schreiber}},
  \bibinfo {author} {\bibfnamefont {F.}~\bibnamefont {Bergamin}}, \bibinfo
  {author} {\bibfnamefont {H.}~\bibnamefont {Grote}}, \bibinfo {author}
  {\bibfnamefont {M.}~\bibnamefont {Mehmet}}, \bibinfo {author} {\bibfnamefont
  {H.}~\bibnamefont {Vahlbruch}}, \bibinfo {author} {\bibfnamefont
  {C.}~\bibnamefont {Affeldt}}, \bibinfo {author} {\bibfnamefont
  {M.}~\bibnamefont {Brinkmann}}, \bibinfo {author} {\bibfnamefont
  {A.}~\bibnamefont {Bisht}}, \bibinfo {author} {\bibfnamefont
  {V.}~\bibnamefont {Kringel}}, \bibinfo {author} {\bibfnamefont
  {H.}~\bibnamefont {L\"uck}}, \bibinfo {author} {\bibfnamefont
  {N.}~\bibnamefont {Mukund}}, \bibinfo {author} {\bibfnamefont
  {S.}~\bibnamefont {Nadji}}, \bibinfo {author} {\bibfnamefont
  {B.}~\bibnamefont {Sorazu}}, \bibinfo {author} {\bibfnamefont
  {K.}~\bibnamefont {Strain}}, \bibinfo {author} {\bibfnamefont
  {M.}~\bibnamefont {Weinert}},\ and\ \bibinfo {author} {\bibfnamefont
  {K.}~\bibnamefont {Danzmann}},\ }\bibfield  {title} {\bibinfo {title} {First
  {D}emonstration of 6 d{B} {Q}uantum {N}oise {R}eduction in a {K}ilometer
  {S}cale {G}ravitational {W}ave {O}bservatory},\ }\href
  {https://doi.org/10.1103/PhysRevLett.126.041102} {\bibfield  {journal}
  {\bibinfo  {journal} {Phys. Rev. Lett.}\ }\textbf {\bibinfo {volume} {126}},\
  \bibinfo {pages} {041102} (\bibinfo {year} {2021})}\BibitemShut {NoStop}%
\bibitem [{\citenamefont {Evans}\ \emph {et~al.}(2010)\citenamefont {Evans},
  \citenamefont {Barsotti},\ and\ \citenamefont
  {Fritschel}}]{EVANS2010_parametricInstabilitiesInGeneral}%
  \BibitemOpen
  \bibfield  {author} {\bibinfo {author} {\bibfnamefont {M.}~\bibnamefont
  {Evans}}, \bibinfo {author} {\bibfnamefont {L.}~\bibnamefont {Barsotti}},\
  and\ \bibinfo {author} {\bibfnamefont {P.}~\bibnamefont {Fritschel}},\
  }\bibfield  {title} {\bibinfo {title} {A general approach to optomechanical
  parametric instabilities},\ }\href
  {https://doi.org/https://doi.org/10.1016/j.physleta.2009.11.023} {\bibfield
  {journal} {\bibinfo  {journal} {Physics Letters A}\ }\textbf {\bibinfo
  {volume} {374}},\ \bibinfo {pages} {665} (\bibinfo {year}
  {2010})}\BibitemShut {NoStop}%
\bibitem [{\citenamefont {Biscans}\ \emph {et~al.}(2019)\citenamefont
  {Biscans}, \citenamefont {Gras}, \citenamefont {Blair}, \citenamefont
  {Driggers}, \citenamefont {Evans}, \citenamefont {Fritschel}, \citenamefont
  {Hardwick},\ and\ \citenamefont {Mansell}}]{Biscans2019_acousticModeDampers}%
  \BibitemOpen
  \bibfield  {author} {\bibinfo {author} {\bibfnamefont {S.}~\bibnamefont
  {Biscans}}, \bibinfo {author} {\bibfnamefont {S.}~\bibnamefont {Gras}},
  \bibinfo {author} {\bibfnamefont {C.~D.}\ \bibnamefont {Blair}}, \bibinfo
  {author} {\bibfnamefont {J.}~\bibnamefont {Driggers}}, \bibinfo {author}
  {\bibfnamefont {M.}~\bibnamefont {Evans}}, \bibinfo {author} {\bibfnamefont
  {P.}~\bibnamefont {Fritschel}}, \bibinfo {author} {\bibfnamefont
  {T.}~\bibnamefont {Hardwick}},\ and\ \bibinfo {author} {\bibfnamefont
  {G.}~\bibnamefont {Mansell}},\ }\bibfield  {title} {\bibinfo {title}
  {Suppressing parametric instabilities in ligo using low-noise acoustic mode
  dampers},\ }\href {https://doi.org/10.1103/PhysRevD.100.122003} {\bibfield
  {journal} {\bibinfo  {journal} {Phys. Rev. D}\ }\textbf {\bibinfo {volume}
  {100}},\ \bibinfo {pages} {122003} (\bibinfo {year} {2019})}\BibitemShut
  {NoStop}%
\end{thebibliography}%

\end{document}